\documentclass[12pt]{article}
\newcommand{\be}{\begin{equation}}
\newcommand{\ee}{\end{equation}}
\usepackage{amsmath}
\usepackage{amssymb}
\usepackage{latexsym}
\usepackage{epsfig}

\newcommand{\Pexp}{{\rm Pexp}}
\def\<{\bigl\langle}
\def\>{\bigr\rangle}

\newcommand{\OO}{{\cal O}}
\newcommand{\BB}{{\cal B}}
\newcommand{\QQ}{{\cal Q}}

\newcommand{\PP}{{\cal P}}

\newcommand{\LL}{{\cal L}}
\newcommand{\Lrepp}{{\cal L}^{\rm rep'}}
\newcommand{\Brepp}{{\cal B}^{\rm rep'}}
\newcommand{\half}{\tfrac{1}{2}}
\newcommand{\ord}[1]{\mathcal{O}(#1)}

 \newcommand{\bra}[1]{\langle #1|}
\newcommand{\ket}[1]{|#1\rangle}
 
\newcommand{\del}{\partial}
 \newcommand{\oints}[1]{\oint_{#1}}

\newcommand{\LHB}{\overset{\,\,_{\frown}}{\cal L}}
\newcommand{\BHB}{\overset{\,\,_{\frown}}{\cal B}}
\newcommand{\QHB}{\overset{\,\,_{\frown}}{\cal Q}}
\newcommand{\LreppHB}{\overset{\,\,_{\frown}}{\cal L}{}^{\rm rep'}}
\newcommand{\BreppHB}{\overset{\,\,_{\frown}}{\cal B}{}^{\rm rep'}}
\newcommand{\LrepHB}{\overset{\,\,_{\frown}}{\cal L}{}^{\rm rep}}

\newcommand{\mP}{{\cal P}_\ast}
\newcommand{\mQ}{Q_\ast}
\newcommand{\PR}{{\cal P}}
\newcommand{\mB}{B_\ast}

\newcommand{\subs}{section~}

\begin{document}

\baselineskip=16pt

\begin{titlepage}
\rightline{\tt arXiv:0810.1737}
\rightline{\tt MIT-CTP-3990}
\rightline{\tt UT-Komaba/08-14}
\rightline{\tt IPMU 08-0074}
\begin{center}
\vskip .3cm
{\Large \bf {The boundary state from open string fields}}\\
\vskip .7cm
{\large {Michael Kiermaier${}^{1,2}$, Yuji Okawa${}^{3}$
and Barton Zwiebach${}^1$}}
\vskip 0.5cm
{\it {${}^1$ Center for Theoretical Physics}}\\
{\it {Massachusetts Institute of Technology}}\\
{\it {Cambridge, MA 02139, USA}}\\
mkiermai@mit.edu, zwiebach@mit.edu
\vskip 0.3cm
{\it {${}^2$ Institute for the Physics and Mathematics of the Universe}}\\
{\it {University of Tokyo}}\\
{\it {Kashiwa, Chiba 277-8582, Japan}}\\
\vskip 0.3cm
{\it {${}^3$ Institute of Physics, University of Tokyo}}\\
{\it {Komaba, Meguro-ku, Tokyo 153-8902, Japan}}\\
okawa@hep1.c.u-tokyo.ac.jp
\vskip 0.6cm
{\bf Abstract}
\end{center}

\noindent
We construct a class of BRST-invariant closed string states
for any classical solution of open string field theory.
The closed string state is a nonlinear functional
of the open string field and changes by a BRST-exact term
under a gauge transformation of the  solution.
As a result, its contraction with an on-shell closed string state
provides a gauge-invariant observable of open string field theory.
Unlike previously known observables,
however, the contraction with off-shell closed string states
in the Fock space is well defined and regular.
Moreover, we claim that the BRST-invariant closed string state
coincides, up to a possible BRST-exact term,
with the boundary state
of the boundary conformal field theory
which the solution is expected to describe.
Our construction requires a choice of a propagator strip.
If we choose the Schnabl propagator strip,
the BRST-invariant state becomes explicitly calculable.
We calculate it for various known analytic solutions
of open string field theory
and, remarkably, we find that it precisely coincides
with the boundary state without any additional BRST-exact term.
Our results imply, in particular, that
the wildly oscillatory rolling tachyon solution
of open string field theory actually describes
the regular closed string physics studied by
Sen
using the boundary state.

\medskip

\end{titlepage}

\baselineskip=17pt
\tableofcontents

\section{Introduction}
\setcounter{equation}{0}

The current formulation of open string field theory~\cite{wit1}
requires a choice of
a consistent
open string background
described by a
boundary conformal field
theory~(BCFT).\footnote{See \cite{Taylor:2003gn,
Rastelli:2005mz, Taylor:2006ye,Fuchs:2008cc}
for reviews.}
Other boundary conformal field theories are expected
to be described by classical solutions to the equation of motion
of the open string field theory based on the original
BCFT.
There have been remarkable developments
in open string field theory
since Schnabl's discovery
of an analytic solution for tachyon condensation~\cite{0511286},
and various analytic solutions have been constructed and studied~\cite{0603159}--\cite{Kishimoto:2008zj}.
It still remains difficult, however, to extract information
on the BCFT represented by a solution of open string field theory.

A useful
object that contains information on  a BCFT is
the boundary state $\ket{B}$.
The one-point function
of a closed string vertex operator
 $\phi_c$
inserted at the origin of a unit disk
can be written using the boundary state $\ket{B}$ as
\begin{equation}\label{disk1ptfct}
\langle \, ( c_0 - \tilde{c}_0 ) \, \phi_c (0) \, \rangle_{\rm disk} =
\bra{B} ( c_0 - \tilde{c}_0 ) \ket{\phi_c} \,,
\end{equation}
where $\ket{\phi_c}$ is the state corresponding to
 $\phi_c$
and the operator $c_0 - \tilde{c}_0$ is
associated with a conformal Killing vector on the disk.
A classical solution $\Psi$ of open string field theory is expected to describe a consistent open string background and thus a boundary conformal field theory, which we denote by BCFT$_*$.
If we can construct the boundary state $\ket{\mB}$ for BCFT$_*$
 from the solution $\Psi$,
we can extract
 all information contained in bulk
one-point functions
in the new background.
Interesting progress in that direction was recently reported
by Ellwood~\cite{Ellwood:2008jh}.
It was argued that
for \emph{on-shell} closed string vertex operators ${\cal V}$,
the one-point functions
on the disk with BCFT$_*$ boundary conditions
can be calculated from the gauge-invariant observables
$W ({\cal V}, \Psi)$ introduced in~\cite{Hashimoto:2001sm,Gaiotto:2001ji} as follows:
\be
\label{ellwood-claim}
\bra{B_*} ( c_0 - \tilde{c}_0 ) \ket{{\cal V}}  -
\bra{B} ( c_0 - \tilde{c}_0 ) \ket{{\cal V}}
= -4\pi i \, W ({\cal V} , \Psi) \,.
\ee
This remarkable observation
means that  the on-shell part of the information
encoded in the BCFT boundary state $\ket{\mB}$
can be extracted from the corresponding solution of open string field theory.

The restriction to on-shell closed string states arises
because the operator ${\cal V}$ in $W ({\cal V}, \Psi)$
is inserted at a point with a conical singularity
on a Riemann surface.
Therefore $W ({\cal V}, \Psi)$ is not well defined
when ${\cal V}$ is not a primary field of
weight~$(0,0)$.
Unfortunately, there are few
on-shell vertex operators
with nonvanishing one-point functions on a disk.
On the other hand, the boundary state  is well defined
when it is contracted with an arbitrary off-shell closed string state
and contains more information on the BCFT.
If we can relax the on-shell restriction on the closed string state
in $W ({\cal V}, \Psi)$,
we will be able to extract much more information
on the BCFT${}_*$ from the solution $\Psi$.
This is our motivation.

In this paper
we construct, for any open string field theory solution $\Psi$,
a class of
closed string states $\ket{\mB (\Psi)}$
of ghost number three.
Their contraction with
arbitrary off-shell closed string states is regular.
The states $\ket{\mB (\Psi)}$ are  BRST invariant,
namely,
\begin{equation}
Q \, \ket{\mB (\Psi)} = 0 \,.
\end{equation}
Under a gauge transformation
$\delta_{\chi} \Psi$
of the solution $\Psi$,
the states $\ket{\mB (\Psi)}$ change at most by a BRST-exact term:
\begin{equation}
     \ket{\mB (\Psi+\delta_\chi\Psi)}\,=\,\ket{\mB (\Psi)} \,
     +~ ( \, Q-\text{exact} \, ) \,.
\end{equation}
Therefore, a gauge-invariant observable can be constructed from $\ket{\mB (\Psi)}$ by its
contraction with an on-shell closed string
 state
${\cal V}$:
\begin{equation}
    \bra{{\cal V}} ( c_0 - \tilde{c}_0 ) \ket{\mB (\Psi+\delta_{\chi}\Psi)}=\bra{{\cal V}} ( c_0 - \tilde{c}_0 ) \ket{\mB (\Psi)}\,.
\end{equation}
The novelty in the construction of these observables is that they admit a perfectly regular off-shell extension
 and, as we will show,
the state $\ket{\mB (\Psi)}$ is explicitly calculable
in certain cases.
We claim that the state $\ket{\mB (\Psi)}$ coincides
with the boundary state $\ket{B_\ast}$
up to a possible BRST-exact term.
In fact, they precisely coincide
in all calculable examples that we examined.

Our construction of $\ket{\mB (\Psi)}$ was
inspired by Ellwood's paper~\cite{Ellwood:2008jh}
and by recent developments
in the calculation of Feynman diagrams
in Schnabl gauge~\cite{0708.2591,Kiermaier:2008jy}
and in a class of gauges called linear
$b$-gauges~\cite{Kiermaier:2007jg}.\footnote{For progress related to other gauge choices, see~\cite{Fuji:2006me,Asano:2006hk,Asano:2008iu}.}
In the construction of $\ket{\mB (\Psi)}$,
we first choose a propagator strip
associated with a linear $b$-gauge.
The shape of the strip is determined
by the operator $\BB$
used in the gauge-fixing condition
on the open string field
of ghost number one.
The length of the strip is determined
by a choice of a Schwinger parameter $s > 0$.
Then the chosen propagator strip can be represented
as the surface generated by the operator
$e^{-s{\LL}}$,
where $\LL$, defined by ${\cal L}=\{Q,\cal B\}$,
is the BRST transformation of $\BB$.

\begin{figure}[tb]
 \centerline{\hbox{\epsfig{figure=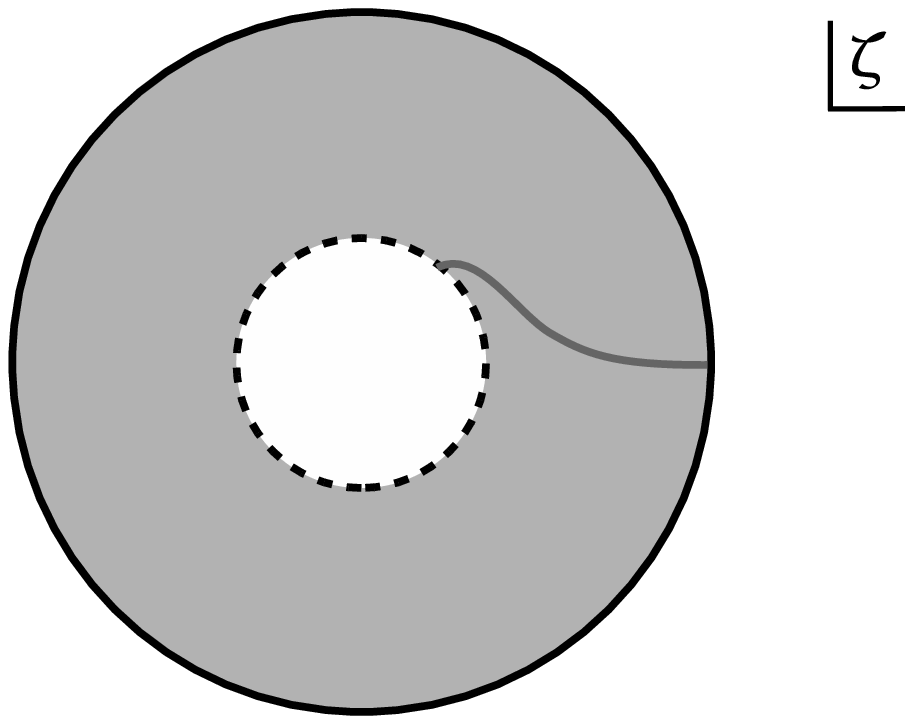, height=5.7cm}}
 \hskip2.8cm
 \hbox{\epsfig{figure=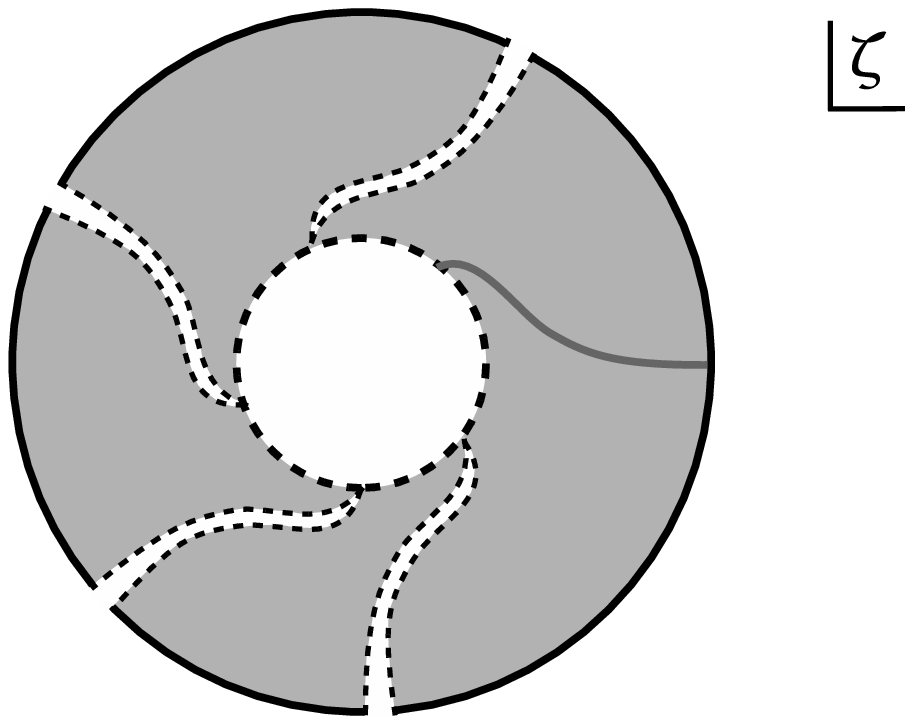, height=5.7cm}}}
 \smallskip
 \centerline{(a)\hskip9.4cm (b)\hskip1.5cm}
 \caption{(a)~An annulus constructed from a half-propagator strip.
The boundary
conditions
of the original BCFT
are
imposed
 on the outer boundary. The grey line in the figure represents the identified half-string edges of the half-propagator strip.
 The path integral over this annulus
 defines a closed string state at the inner boundary
 depicted as a dashed line in the figure.
 ~~(b)~An annulus with four slits. The classical solution $\Psi$ is glued to each slit.
 The path integral over this annulus
 after gluing classical solutions
 defines a closed string state at the inner boundary.}
 \label{intro-figure}
 \end{figure}

The main ingredient for the construction of $\ket{\mB (\Psi)}$
is a {\it half-propagator strip}.
We cut the chosen propagator strip in half along the line traced
by the open string midpoint.
We then take one of the resulting
half-propagator strips
and form an annulus by identifying its initial and final half-string edges. Imposing the original
BCFT boundary
conditions
at the open string boundary of this annulus, the path integral over the annulus defines a closed string state at the other boundary where we originally cut the propagator strip.
See Figure~\ref{intro-figure}(a). 
It is clear that this closed string state, after an appropriate
exponential
action of $L_0+\tilde{L}_0$,
reproduces
the boundary state $\ket{B}$ of the original BCFT.
We can thus construct the boundary state $\ket{B}$
for any choices of $\BB$ and~$s$.
It should be pointed out, however, that
the propagator for non-BPZ-even gauges
(${\cal B}^\star\neq {\cal B}$)
is a complicated object,
while our construction is based on $e^{-s{\cal L}}$
which in these cases is not the full propagator surface.

Let us now repeat the construction
with the above half-propagator strip
replaced by the one
for the background associated with $\Psi$.
The modified half-propagator strip can be
constructed by gluing the solution $\Psi$ to slits
which are inserted at various positions along the annulus.
See Figure~\ref{intro-figure}(b).
The shape of the slits is correlated with
the shape of the half-propagator strip before the identification
of the half-string edges
and is determined by the operator $\BB$.
The slits are accompanied by appropriate $b$-ghost
line integrals, and the positions of the slits are integrated over.
A closed string state is again defined
by the path integral over this annulus.
After
 an appropriate exponential action of $L_0+\tilde{L}_0$
 and summing over the number of solution insertions,
this defines the state $\ket{\mB (\Psi)}$.

The resulting state $\ket{\mB (\Psi)}$ depends
on $\BB$ and $s$,
but the gauge-invariant observables
$\bra{{\cal V}} ( c_0 - \tilde{c}_0 ) \ket{\mB (\Psi)}$
are independent of $\BB$ and $s$.
 Indeed, we
can show that as we vary $\BB$ and $s$,
the closed string state $\ket{\mB (\Psi)}$ changes
at most by a possible BRST-exact term.
While it is difficult to calculate the state $\ket{\mB (\Psi)}$
for generic choices of ${\cal B}$,
it is explicitly calculable
for solutions based on
the familiar wedge surfaces~\cite{Rastelli:2000iu,Schnabl:2002gg}
if we choose Schnabl's propagator strip.
 In fact, in this case the methods developed in~\cite{Kiermaier:2008jy} to map Riemann surfaces
for one-loop amplitudes in Schnabl gauge
to an annulus can be used to construct the Riemann surfaces which define $\ket{\mB (\Psi)}$.
We explicitly calculate
$\ket{\mB (\Psi)}$ based on the Schnabl propagator strip of arbitrary length $s$ for various known solutions of string field theory
such as
Schnabl's tachyon vacuum
solution
and the solutions for marginal deformations
with regular operator products constructed
in~\cite{0701248,0701249}
and in~\cite{0707.4472}.
We find that $\ket{\mB (\Psi)}$ vanishes identically for the tachyon vacuum
solution, which
is consistent with Sen's conjecture that the D-brane disappears at the tachyon vacuum.
For the marginal deformations, $\ket{\mB (\Psi)}$ \emph{precisely} reproduces the BCFT boundary state $\ket{B_*}$.
Both results hold independent of the length $s$ of the propagator strip used in the construction.
At least for these examples the exact BCFT
boundary state can
be obtained from the corresponding open string field solution!

Our results imply, in particular, that the boundary state $\ket{\mB (\Psi)}$
calculated from the known rolling tachyon solutions of open string field theory coincides with the BCFT
boundary state discussed in~\cite{Sen:2002nu,Sen:2002in,Sen:2002vv,Larsen:2002wc,Bagchi:2008et, Sen:2004nf}.
This boundary state
describes a regular behavior
of D-brane decay in the far future.
For example, the pressure decreases monotonically
and vanishes in the far future.
The rolling tachyon solution $\Psi$,
on the other hand, exhibits
ever-growing oscillations for the component fields
of the open string~\cite{0701248,0701249}.
It has been a long-standing puzzle
whether such
wildly oscillatory solutions
describe a regular time-dependent process
in the far future~\cite{Moeller:2002vx,Fujita:2003ex,Coletti:2005zj}.
Our explicit construction of the boundary state from the
rolling tachyon solution
confirms that the solution represents the
expected regular physics.
Our interpretation is that the wild oscillatory behavior
is due to the description of
the regular physics in the {\it closed} string channel
in terms of the {\it open} string degrees of freedom.

\medskip
The paper is organized as follows.
In section~\ref{sec2} we introduce the
 half-propagator
strips
and explain the construction of closed string states
using
 half-propagator
strips.
In section~\ref{sec3}
we define the closed string state $\ket{\mB(\Psi)}$. We show its BRST invariance and prove that it changes at most by a BRST-exact
term 
under a gauge transformation of $\Psi$.
We show  in section~\ref{sec4} that
as we vary $\BB$ and $s$,
the state $\ket{\mB(\Psi)}$ changes
at most by a BRST-exact term.

In section~\ref{secocsft}, we discuss the relation of our work
with an earlier approach to boundary states based on open-closed
string field theory~\cite{Zwiebach:1990qj,Zwiebach:1992bw,Zwiebach:1997fe}.  Indeed, a set of open-closed vertices can be used to
construct an alternative BRST-invariant closed string state
$\ket{B_*^{\rm oc}(\Psi)}$ associated with a solution $\Psi$. This state,
just like $\ket{\mB(\Psi)}$, changes
at most by
a BRST-exact term
under an
open string
gauge transformation.
We show that for choices of ${\cal B}$ that are invariant under
BPZ conjugation, $\ket{\mB(\Psi)}$ encodes
a set of consistent open-closed vertices.
Curiously, for general choices of ${\cal B}$, the open-closed
vertices encoded by $\ket{\mB(\Psi)}$ do not satisfy the
reality condition. Even so, the state $\ket{\mB(\Psi)}$
can be real for some classical solutions,
and we indeed find that it is the case
for all the explicit examples we discuss in section~\ref{secBCFT}.
If we assume the background independence of a certain
version of open-closed string field theory,
we can argue that
$\ket{\mB(\Psi)}$ and $\ket{B_*}$ coincide
up to a BRST-exact term.

In sections~\ref{secwedge} and~\ref{secBCFT}
we demonstrate that the state $\ket{\mB (\Psi)}$
is calculable for solutions based on wedge states
if we choose Schnabl's propagator strip.
We then explicitly
calculate $\ket{\mB(\Psi)}$ for various known solutions
and find that it coincides with the BCFT boundary state
for arbitrary~$s$.
In these cases, the state $\ket{\mB(\Psi)}$ factorizes
into matter and ghost sectors,
and the ghost sector
coincides with the boundary state of the $bc$ CFT.
It is important to understand when this factorization holds
because the state $\ket{\mB(\Psi)}$ factorized in this way
can be a consistent BCFT boundary state
without any BRST-exact term.
In section~\ref{sectrivialghost} we discuss
this factorization
and show that for solutions based on wedge states with
a certain class of
ghost insertions and arbitrary matter insertions,
the state $\ket{\mB(\Psi)}$ constructed from the
Schnabl propagator always factorizes in this way.
In section~\ref{secdisc} we end with concluding remarks.

\section{Half-propagator strips and closed string states}
\label{sec2}
\setcounter{equation}{0}

We begin this section
by reviewing the construction of propagator strips
in linear $b$-gauges~\cite{Kiermaier:2007jg}.
We then introduce half-propagator strips
by cutting the full strips along the line traced by the
open string
midpoint.
We further introduce various ingredients
to be used in section~\ref{sec3}
for the construction of $\ket{\mB (\Psi)}$,
such as the star multiplication of half-propagator strips
and operator insertions on the strips.
Finally, we construct closed string states from
half-propagator strips
by identifying the
 half-string edges.
The coordinate curve of the closed string
is the curve traced by
the open string midpoint.

\subsection{Half-propagator strips for regular linear $b$-gauges}

A large class of gauge choices for string perturbation theory was discussed in~\cite{Kiermaier:2007jg}.
These so-called \emph{linear $b$-gauges}
impose a gauge condition
\begin{equation}\label{classgaugecond}
   \BB\,\ket{\psi_{cl}}=0
\end{equation}
on the classical open string field
$\ket{\psi_{cl}}$ of ghost number one,\footnote{
String fields of different ghost numbers are introduced
in the process of gauge fixing.
See~\cite{Kiermaier:2007jg} for detailed discussions
about gauge conditions on such quantum string fields.
}
where the operator $\BB$ is a linear combination
of even-moded $b$-ghost oscillators:
\begin{equation}\label{Bandv}
      \BB=\sum_{j\in \mathbb{Z}} v_{2j} \, b_{2j}
      =\oint {d\xi\over 2\pi i} \, v(\xi) \, b(\xi)
      \qquad \textrm{with} \quad
      v(\xi)=\sum_{j\in\mathbb{Z}}v_{2j} \, \xi^{2j+1}\,, \quad v_{2j}\in\mathbb{R}\,.
\end{equation}
If the associated vector field $v(\xi)$ is analytic in a neighborhood of the unit circle $|\xi|=1$ and furthermore satisfies the condition
\begin{equation}\label{regularv}
     \Re\left(\bar\xi v(\xi)\right)>0\, \qquad  \textrm{for} \quad
     |\xi|=1\,,
\end{equation}
the gauge choice is called \emph{regular}. It was shown in~\cite{Kiermaier:2007jg} that regular linear $b$-gauges correctly reproduce open string on-shell amplitudes and that pure-gauge
external states decouple.

\begin{figure}[tb]
\centerline{
\hbox{\epsfig{figure=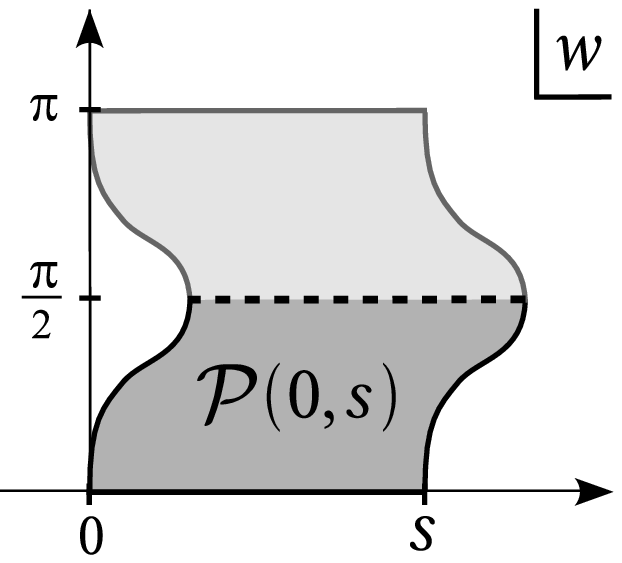, height=5.5cm}}
\hskip 2.5cm
\hbox{\epsfig{figure=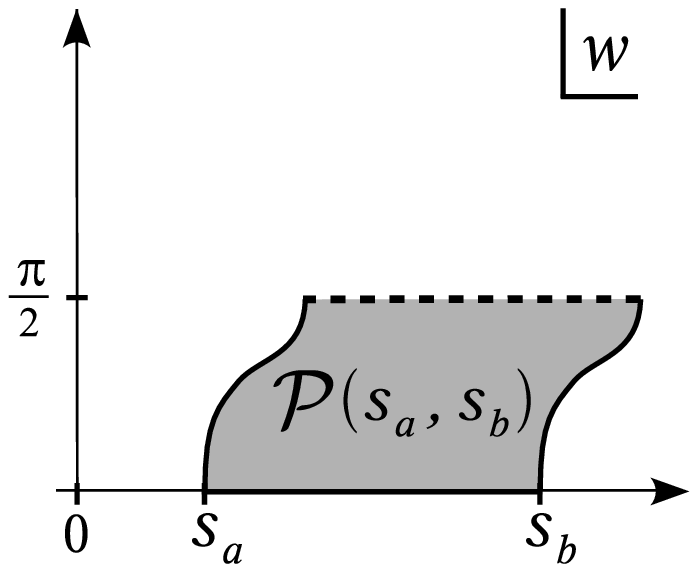, height=5.5cm}}}
\centerline{
(a)\hskip9cm (b)\hskip.5cm}
\caption{(a)~Illustration of the surface associated with $e^{-s\LL}$. It is generated by horizontal translations in the $w$ frame.
The half-propagator strip
$\PR (0,s)$ is obtained by cutting the surface $e^{-s\LL}$ along the line $\Im(w)=\frac{\pi}{2}$.
~~(b)~The surface $\PR (s_a,s_b)$ is
a horizontal translation of the surface $\PR (0,s_b-s_a)$ by $s_a$.
}\label{figPR}
\end{figure}

The propagator of a regular linear $b$-gauge is characterized by the strip surface generated by $e^{-s\LL}$, where
\begin{equation}
    \LL=\{ Q,\BB \}\,.
\end{equation}
In a certain conformal frame $w$, this strip surface is generated by  horizontal translations. See Figure~\ref{figPR}(a).
 The $w$ frame is obtained from
 the vector field  $v(\xi)$ through
 \begin{equation}
 \label{vmiahpcss}
 {dw(\xi)\over d\xi} = {1\over v(\xi)} \,,   \quad
 w(1) = 0\,.
 \end{equation}
 Normalizing $v(\xi)$ appropriately, we can impose the
 additional condition\footnote{This definition of the
  $w$ frame
 differs from the conventions of~\cite{Kiermaier:2008jy}. They are related by $w_{\rm here}=-w_{\rm there}+i\pi$.}
\begin{equation}\label{bndrcond}
       w(-1)= i \pi \,.
\end{equation}
The horizontal boundaries of the strip are then located at $\Im(w)=0$ and $\Im(w)=\pi$.
The left boundary is
the parameterized curve $\gamma(\theta) = w(e^{i\theta})$
for $0\leq\theta\leq\pi$.
 It follows from~(\ref{vmiahpcss}) and~(\ref{bndrcond}) that
 \begin{equation}
    \gamma(0)=0\,,\qquad\Im\bigl(\gamma(\tfrac{\pi}{2})\bigr)=\tfrac{\pi}{2}
    \,\,,\qquad\gamma(\pi)=i\pi\,,
 \end{equation}
where the middle equation holds
because of~(\ref{Bandv}).

When $e^{-s\LL}$ acts on an open string state $A$, the  parameterization on $\gamma$ is used to glue the strip associated with $e^{-s\LL}$ to the coordinate curve of $A$. In the $w$ frame, the right boundary of the strip $e^{-s\LL}$ is a horizontal translation by $s$ of its left boundary. It is therefore parameterized by the curve $s+\gamma(\theta)$ with
$0\leq\theta\leq\pi$.
This
fixes
the horizontal position of the strip surface $e^{-s\LL}$ in the
$w$ frame.  We will now consider the surfaces which arise when we cut the strip $e^{-s\LL}$ along the line $\Im(w)=\frac{\pi}{2}$. This line is generated by horizontal translations of $\gamma(\frac{\pi}{2})$ and is thus associated with the open string midpoint.
The resulting surfaces from this particular cut
are of interest for a number of reasons. In the annulus amplitude, we cut the propagator surface  along a closed string curve to read off the boundary state along the boundary generated by the cut. Choosing the open string midpoint for this cut is natural because of the special role of the midpoint in open string field theory.
Furthermore,
if one chooses this cut for the strip $e^{-s\LL}$ in Schnabl gauge, the resulting surfaces are the so-called slanted wedges
introduced
 in~\cite{Kiermaier:2008jy}. The remarkable algebraic properties of these slanted wedges under gluing allowed the explicit map of one-loop Riemann surfaces to an annulus frame, which is expected to facilitate  the explicit
calculation of off-shell amplitudes. Our analysis will experience a similarly drastic simplification in the Schnabl gauge limit.

The cutting of the strip $e^{-s\LL}$ along the line $\Im(w)=\frac{\pi}{2}$ yields two surfaces. We will denote the bottom one, located in the region
 $0\leq\Im(w)\leq\frac{\pi}{2}$,
by $\PR (0,s)$. See Figure~\ref{figPR}(a).
The arguments~$0$~and $s$ remind us that the open string boundary of $\PR (0,s)$ is located on the real axis between $w=0$ and $w=s$. More generally, we use the notation $\PR (s_a,s_b)$ with $s_b\geq s_a$ for the surface $\PR (0,s_b-s_a)$ shifted horizontally by $s_a$ in the $w$ frame. See Figure~\ref{figPR}(b). The left and right boundaries of $\PR (s_a,s_b)$ are parameterized by $s_a+\gamma(\theta)$ and $s_b+\gamma(\theta)$, respectively, where the range of $\theta$ is now restricted to $0\leq\theta\leq\frac{\pi}{2}$. Finally, $\PR (s_a,s_b)$ has a boundary induced by the cut. This boundary is neither an open string boundary nor the coordinate line of an open string state. For reasons that will become apparent later, we will refer to this boundary as the \emph{closed string boundary}.\footnote{For the particular case of Schnabl gauge, this boundary is the so-called \emph{hidden boundary} introduced in~\cite{Kiermaier:2008jy}.}

Naively, the surface $\PR (0,s)$
is  generated by the operator
$e^{-s\LL_R}$, where $\LL_R$ is the right half of $\LL$.
This notation, however,
is  misleading.
It suggests, incorrectly, that
the surface $\PR (0,s)$
with $\LL_R$ inserted at the left edge
is the same as $\PR (0,s)$
with $\LL_R$ inserted at the right edge
because $[ \, \LL_R, e^{-s\LL_R} \, ] = 0$.
This is not the case because
the line integral $\LL_R$ has an endpoint on the closed string boundary
and this endpoint cannot be moved by contour deformation.
Let us denote by $\LL_R(t)$ the line integral $\LL_R$
along the contour $t+ \gamma(\theta)$ with
$0\leq \theta \leq {\pi\over 2}$.
As $\LL_R$ generates translations in the $w$ frame, we have
\begin{equation}\label{LRt}
    \LL_R(t)\equiv
    \int_t^{\gamma(\frac{\pi}{2})+t}\biggl[\frac{dw}{2\pi i}\,T(w)+\frac{d\bar w}{2\pi i}\,\widetilde T(\bar w)\biggr]\,.
\end{equation}
The surface $\PR (s_a,s_b)$ can then be properly
expressed as the path-ordered exponential:
\begin{equation}\label{defPR}
    \PR (s_a,s_b)=\Pexp\biggl[-\int_{s_a}^{s_b} dt\,\LL_R(t) \biggr]\,.
\end{equation}
Our convention for the path-ordering is
$\LL_R (t_1) \, \LL_R (t_2)$ for $t_1 < t_2$.
It is now clear that
\begin{equation}\label{nocommutator}
  \LL_R(s_a)\,\PR (s_a,s_b)-\PR (s_a,s_b)\,\LL_R(s_b)\neq 0
\end{equation}
because the left-hand side represents a surface
with two disconnected contour integrals.
It is therefore natural to introduce an operator
that supplements the remaining line integral on $\PR (s_a,s_b)$
along the closed string boundary.
We thus define
\begin{equation}\label{LHB}
    \LHB = \int_{\gamma(\frac{\pi}{2})+s_a}^{\gamma(\frac{\pi}{2})+s_b}
    \biggl[\frac{dw}{2\pi i}\,T(w)+\frac{d\bar w}{2\pi i}\,\widetilde T(\bar w)\biggr]
\end{equation}
for $\LHB$ acting on $\PR (s_a,s_b)$.
We then have the identity
\begin{equation}\label{commLR}
 \LL_R(s_a)\,\PR (s_a,s_b)-\PR (s_a,s_b)\,\LL_R(s_b)  +\LHB\, \PR (s_a,s_b)=0\,,
\end{equation}
 which follows from first connecting the three line integrals in~(\ref{commLR})

 and then shrinking the resulting integral contour to zero size.
Furthermore, we have
\begin{equation}\label{dsPR}
\partial_{s_b} \, \PR (s_a, s_b)
 = -\PR (s_a, s_b) \, {\cal L}_R (s_b) \,, \\\qquad
\partial_{s_a} \, \PR (s_a, s_b)
 = {\cal L}_R (s_a) \, \PR (s_a, s_b) \,,
\end{equation}
which follow
from the definition~(\ref{defPR}).

Following the definitions~(\ref{LRt}) and~(\ref{LHB}) of
the line integrals
of the energy-momentum tensor,
we define the corresponding $b$-ghost line integrals
as follows:
\begin{equation}
\label{blineintegralsdef}
\begin{split}
    \BB_R(t) & =\int_t^{\gamma(\frac{\pi}{2})+t}
\biggl[\frac{dw}{2\pi i}\,b(w)+\frac{d\bar w}{2\pi i}\,\tilde b(\bar w)\biggr]\,,\\
    \BHB & = \int_{\gamma(\frac{\pi}{2})+s_a}^{\gamma(\frac{\pi}{2})+s_b}
    \biggl[\frac{dw}{2\pi i}\,b(w)+\frac{d\bar w}{2\pi i}\,
    \tilde b(\bar w)\biggr]
\end{split}
\end{equation}
for $\BHB$ acting on $\PR (s_a,s_b)$.
We also define the corresponding line integrals of the BRST current
\begin{equation}
\begin{split}
   \QQ_R(t) & =\int_t^{\gamma(\frac{\pi}{2})+t}
    \biggl[ \, \frac{dw}{2 \pi i} \, j_B (w)-\frac{d \bar{w}}{2 \pi i} \, \tilde{\jmath}_B (\bar{w}) \,\biggr]\, ,  \\
    \QHB & = \int_{\gamma(\frac{\pi}{2})+s_a}^{\gamma(\frac{\pi}{2})+s_b}
    \biggl[ \, \frac{dw}{2 \pi i} \, j_B (w)-
    \frac{d \bar{w}}{2 \pi i} \, \tilde{\jmath}_B (\bar{w}) \,\biggr]
        \end{split}
\end{equation}
for $\QHB$ acting on $\PR (s_a,s_b)$.
Just as in~(\ref{commLR}), we
connect
line integrals of the BRST current and the $b$ ghost to find
\begin{equation}\label{commBRQ}
\begin{split}
&
\BB_R(s_a)\PR (s_a,s_b)-\PR (s_a,s_b)\BB_R(s_b)\,+\,\BHB\, \PR (s_a,s_b)=0\,,\\
& \QQ_R(s_a)\PR (s_a,s_b)-\PR (s_a,s_b)\QQ_R(s_b)+\QHB\, \PR (s_a,s_b)=0\,.
\end{split}
\end{equation}

\subsection{Star multiplication of half-propagator strips}
\begin{figure}[tb]
\centerline{\hbox{\epsfig{figure=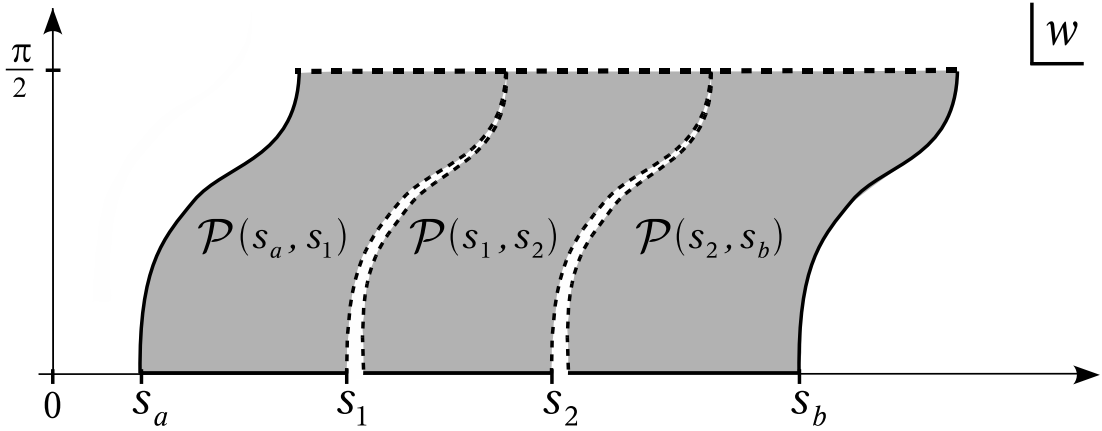, height=6cm}}}
\caption{
Illustration of the surface $\Sigma(s_a,s_b)$ for $k=2$. It is obtained from the surface $\PR(s_a,s_b)$ by inserting $k$ open string states along the parameterized slits $s_i+\gamma(\theta)$.
}\label{figSigma}
\end{figure}

The surface $\PR (s_a,s_b)$ is equipped with two parameterized boundaries which can be glued to open string states $A$ or
to other surfaces $\PR (s_a',s_b')$. For the latter, we require that the glued boundaries in the $w$ frame match. As for regular star multiplication of open string states, we use the symbol $\ast$ to denote this gluing.\footnote{We will suppress
explicit $\ast$ symbols in later sections.}
For the special case of Schnabl gauge, this type of gluing operation was discussed extensively in~\cite{Kiermaier:2008jy}.
 From the definition of $\PR (s_a,s_b)$
it immediately follows that
\begin{equation}
    \PR (s_a,s_b)\ast\PR (s_b,s_c)=\PR (s_a,s_c)\,.
\end{equation}
Open string states $A$ do not carry a closed string boundary. Therefore the gluing operation
\begin{equation}\label{PRAPR}
    \PR (s_a,s_b)\ast A\ast\PR (s_b,s_c)\,
\end{equation}
is well defined and yields a surface with one connected closed string boundary between $\gamma(\frac{\pi}{2})+s_a$ and
$\gamma(\frac{\pi}{2})+s_c$.
It can be thought of as
the surface $\PR (s_a,s_c)$ with a \emph{slit} along the curve $s_b+\gamma(\theta)$, where the open string state $A$
 is to be inserted.
See Figure~\ref{figSigma}.
It follows from~(\ref{dsPR})
that a change in $s_b$ on the surface~(\ref{PRAPR}) is generated by
\begin{equation}\label{dPAP}
\begin{split}
& \partial_{s_b} \, \bigl[ \,
\PR (s_a, s_b) \ast A \ast \PR (s_b, s_c) \, \bigr] \\
& = {}-\PR (s_a, s_b) \, {\cal L}_R (s_b) \ast A \ast \PR (s_b, s_c)
+ \PR (s_a, s_b) \ast A \ast {\cal L}_R (s_b) \, \PR (s_b, s_c) \\
& \equiv {}-\PR (s_a, s_b) \ast [ \, {\cal L}_R (s_b), A \, ] \ast
\PR (s_b, s_c) \,.
\end{split}
\end{equation}
 Note the location of the star symbols in the second line, which is implicit in the commutator defined in the last line.
This definition applies to any commutator or anticommutator
of a line integral up to the closed string boundary
with an open string state.
For example,
\begin{equation}
\begin{split}
& \PR (s_a, s_b) \ast \, \{ \, \BB_R (s_b), A \, \} \, \ast
\PR (s_b, s_c) \\
& \equiv
\PR (s_a, s_b) \, \BB_R (s_b) \ast A \ast \PR (s_b, s_c)
+ \PR (s_a, s_b) \ast A \ast \BB_R (s_b) \, \PR (s_b, s_c) \,.
\end{split}
\end{equation}
In the case of $\QQ_R (t)$, we have
\begin{equation}
\begin{split}
& \PR (s_a, s_b) \ast \, \{ \, \QQ_R (s_b), A \, \} \, \ast
\PR (s_b, s_c) \\
& \equiv
\PR (s_a, s_b) \, \QQ_R (s_b) \ast A \ast \PR (s_b, s_c)
+ \PR (s_a, s_b) \ast A \ast \QQ_R (s_b) \, \PR (s_b, s_c) \\
& = {}- \PR (s_a, s_b)  \ast ( Q A ) \ast \PR (s_b, s_c)
\end{split}
\end{equation}
for any Grassmann-odd state $A$
because the BRST current is a primary field of
weight
one
so that its integral in the $w$ frame can be deformed
and easily mapped to the frame for $A$.
Similar relations do not hold for $\LL_R (t)$
and $\BB_R (t)$ because the energy-momentum tensor
and the $b$ ghost are not
 primary fields
of weight one:
\begin{equation}
\begin{split}
\PR (s_a, s_b) \ast [ \, \LL_R (s_b), A \, ] \ast
\PR (s_b, s_c)
& \ne -\PR (s_a, s_b) \ast ( \LL A ) \ast \PR (s_b, s_c) \,, \\
\PR (s_a, s_b) \ast \{ \, \BB_R (s_b), A \, \} \ast
\PR (s_b, s_c)
& \ne -\PR (s_a, s_b) \ast ( \BB A ) \ast \PR (s_b, s_c) \,.
\end{split}
\end{equation}

\medskip
More generally, we will consider
surfaces $\Sigma(s_a,s_b)$
resulting from multiple insertions of
open string states
$A_1$, $A_2, \ldots, A_k$
into $\PR (s_a,s_b)$
in the following form:
\begin{equation}\label{Sigma}
   \Sigma(s_a,s_b)=\PR (s_a,s_1)\ast A_1\,
    \ldots\,\PR (s_{i-1},s_i)\ast A_i \ast \PR (s_{i},s_{i+1})\,\ldots
    \, A_k\ast\PR (s_{k},s_{b})
\end{equation}
with
 $s_a\leq s_1$,~ $s_i\leq s_{i+1}$,\,
and $s_k\leq s_b$. The surface $\Sigma(s_a,s_b)$ is illustrated in Figure~\ref{figSigma} for $k=2$.
The surface $\Sigma(s_a,s_b)$ is $\PR (s_a,s_b)$
with $k$ parameterized slits
along the curves $s_i+\gamma(\theta)$ where the states $A_i$ are to be glued. We denote
the Grassmann property
of $\Sigma$ by $(-)^\Sigma$:
\begin{equation}
    (-)^\Sigma=\prod_{i=1}^k \, (-)^{A_i}\,.
\end{equation}
The operators $\LHB$, $\BHB$, and $\QHB$ are derivations when acting on products of the form~(\ref{Sigma}). For example, we have  \begin{equation}\label{LHBderiv}
\begin{split}
    &\LHB\bigl[ \, \PR (s_a, s_b)\ast A\ast\PR (s_b, s_c) \, \bigl]\\
    &=\bigl[\LHB\PR (s_a, s_b)\bigr]\ast A\ast\PR (s_b, s_c)
    +\PR (s_a, s_b)\ast \bigl[ \, \LHB A \, \bigr]\ast\PR (s_b, s_c)
    +\PR (s_a, s_b)\ast A\ast\bigl[\LHB\PR (s_b, s_c)\bigr]\\
    &=\bigl[\LHB\PR (s_a, s_b)\bigr]\ast A\ast\PR (s_b, s_c)
    +\PR (s_a, s_b)\ast A\ast\bigl[\LHB\PR (s_b, s_c)\bigr]\,.
\end{split}
\end{equation}
Here we used the fact that
an open string state $A$ does not have a closed string boundary
and it is therefore annihilated by $\LHB$, $\BHB$, and $\QHB$:
\begin{equation}\label{LHBstate}
    \LHB A=0\,,\qquad \BHB A=0\,,\qquad \QHB A=0\,.
\end{equation}

We define the BRST operator $\QQ$ acting on a surface $\Sigma(s_a,s_b)$ of the form~(\ref{Sigma})
by
\begin{equation}\label{Qdef}
    \QQ \, \Sigma(s_a,s_b)
\equiv (-)^\Sigma \, \Sigma(s_a,s_b) \, \QQ_R(s_b)
-\QHB \, \Sigma(s_a,s_b)
-\QQ_R(s_a) \, \Sigma(s_a,s_b)\,.
\end{equation}
Note that the three integral contours can be connected.
We have
\begin{equation}
\{ \QQ, \BB_R(t) \} = \LL_R(t) \,, \qquad
\{ \QQ, \BHB \} = \LHB \,.
\end{equation}
{}From~(\ref{commBRQ})
we know that $\PR (s_a,s_b)$ is annihilated by this operator:
\begin{equation}\label{QPR}
    \QQ\,\PR (s_a,s_b)=0\,.
\end{equation}
On an open string state $A$, inserted along a slit $s_a+\gamma(\theta)$ in the $w$ frame, the definition of $\QQ$ in~(\ref{Qdef}) reduces to the usual BRST transformation $Q$:
\begin{equation}\label{QA}
    \QQ\,A=(-)^A A \,\QQ_R(s_a)-\QHB A-\QQ_R(s_a)\,A
    =Q\,A\,.
\end{equation}
Here we have used $\QHB A=0$.
Combining~(\ref{QPR}),~(\ref{QA}),
and the fact that the BRST current is a primary field
of weight one,
 we find that the BRST
transformation
of a product $\Sigma(s_a,s_b)$ of the form~(\ref{Sigma})
reduces to BRST transformations of the Fock-space states $A_i$.
We have
\begin{equation}\label{QSigma}
    \QQ\,\Sigma(s_a,s_b)
     =\sum_{i=1}^k(-)^{\sum_{j=1}^{i-1}A_j}
    \PR (s_a,s_1)\ast A_1 \,\ldots\,
    \PR (s_{i-1},s_i)\ast (Q A_i) \ast \PR (s_{i},s_{i+1})\,\ldots
    \,  A_k \ast \PR (s_{k},s_{b})\,.
\end{equation}
Similarly, the properties~(\ref{commLR}),~(\ref{LHBderiv})
and~(\ref{LHBstate}) can be used to
show
\begin{equation}\label{commLRSigma}
\begin{split}
&   \LL_R(s_a) \,\Sigma(s_a,s_b) -  \Sigma(s_a,s_b) \, \LL_R(s_b)
 +  \LHB \, \Sigma(s_a,s_b)\\
    &=\sum_{i=1}^k\PR (s_a,s_1)\ast A_1 \,\ldots\,
    \PR (s_{i-1},s_i)\ast [\LL_R(s_i), A_i] \ast \PR (s_{i},s_{i+1})\,\ldots
    \,  A_k \ast \PR (s_{k},s_{b})\\
    &=-\sum_{i=1}^k\del_{s_i}\Sigma(s_a,s_b)\,,
\end{split}
\end{equation}
where we used~(\ref{dPAP}) in the last step.
The corresponding identity for $b$-ghost line integrals
is
\begin{equation}
\begin{split}
  &\BB_R(s_a) \, \Sigma(s_a,s_b)-(-)^\Sigma \, \Sigma(s_a,s_b)\,\BB_R(s_b)\,+\,
  \BHB \, \Sigma(s_a,s_b)\\
    &=\sum_{i=1}^k \, (-)^{\sum_{j=1}^{i-1}A_j}
    \PR (s_a,s_1)\ast A_1 \,\ldots\,
    \bigl(\BB_R(s_i) A_i-(-)^{A_i}A_i\BB_R(s_i)\bigr) \,\ldots
    \,  A_k \ast \PR (s_{k},s_{b})\,.
\end{split}
\end{equation}

\subsection{Closed string states from half-propagator strips}

A surface $\Sigma(s_a,s_b)$ of the form~(\ref{Sigma})
can be used to construct a
{\em  closed string} surface state.
To do this, we first introduce the identification $w\sim w+(s_b-s_a)$
in the $w$ frame.
This identifies the left boundary $s_a+\gamma(\theta)$ with
the right boundary $s_b+\gamma(\theta)$ of $\Sigma(s_a,s_b)$.
We are left with the closed string boundary at $\Im(w)=\frac{\pi}{2}$, whose name we are now doing justice by gluing it to the coordinate line $0\leq \sigma\leq2\pi$ of a closed string coordinate patch. The map from $\sigma$ to the closed string boundary $\Im(w)=\frac{\pi}{2}$
of $\Sigma(s_a,s_b)$
is given by
\begin{equation}\label{wsigma}
    \sigma\,\to\,w=i\,\frac{\pi}{2}+(s_b-s_a)\,\frac{\sigma}{2\pi}\,.
\end{equation}
This map is consistent with the identifications $w\sim w+(s_b-s_a)$ and $\sigma\sim \sigma+2\pi$. The resulting surface is a closed string surface state with its coordinate line at $\Im(w)=\frac{\pi}{2}$ parameterized by~(\ref{wsigma}).
We denote this closed string surface state by
\begin{equation}
    \oints{s_b-s_a}\hskip-.2cm\Sigma(s_b,s_a)\,.
\end{equation}
We have represented the operation that  turns the surface $\Sigma(s_b,s_a)$
into a closed string surface state
by the symbol $\oints{s_b-s_a}$.
This notation is somewhat reminiscent of
the notation $\int A$,
 often used in open string field theory,
which glues the left and right parts
of the open string state $A$.
The subscript $s_b-s_a$
is a reminder that the width $s_b-s_a$ of the strip $\Sigma(s_a,s_b)$ explicitly enters the gluing prescription~(\ref{wsigma}).

A natural
representation of $\oints{s_b-s_a}\Sigma(s_a,s_b)$ can be
obtained in the $\zeta$ frame defined by
\begin{equation}\label{zetaw}
    \zeta=\exp\Bigl(\frac{2\pi i}{s_b-s_a}\,w\Bigr)\,.
\end{equation}
This maps the surface $\Sigma$ to an annulus
with the open string boundary placed at $|\zeta|=1$ and the closed
string coordinate line located at $|\zeta|=e^{-\frac{\pi^2}{s_b-s_a}}$. The surface state $\oints{s_b-s_a}\Sigma(s_a,s_b)$
is then defined through inner products with arbitrary
closed string states
$\ket{\phi_c}$
in the Fock space
by
\begin{equation}\label{defointSigma}
    \< \, \phi_c \,,
\,\oints{s_b-s_a}\hskip-.3cm\Sigma(s_a,s_b) \, \>
\,=\, \< \, d_{s_b-s_a}\circ \phi_c (0) \,
 [\ldots] \, \>_{
 \text{disk}
}\,,
\end{equation}
where the operator $\phi_c (0)$
corresponding to $\ket{\phi_c}$
is mapped from its canonical coordinate patch $|\xi|\leq 1$ to
the shrunk coordinate patch $|\zeta|\leq e^{-\frac{\pi^2}{s_b-s_a}}$ by the retraction
\begin{equation}
    d_{s_b-s_a}(\xi)=e^{-\frac{\pi^2}{s_b-s_a}}\,\xi\,.
\end{equation}
The dots $[\dots]$
in~(\ref{defointSigma}) represent the slits where the open string states $A_i$ are
inserted.  They are also
mapped from the $w$ frame to the $\zeta$ frame via~(\ref{zetaw}).
 For the case of $k=4$ slits, the $\zeta$-frame representation of the closed string surface state
 $\oints{s_b-s_a}\Sigma(s_a,s_b)$ was illustrated
in Figure~\ref{intro-figure}(b).

\medskip
The identification $w\sim w+(s_b-s_a)$ allows us to move line integrals cyclically in $\oints{s_b-s_a}$. We have
\begin{equation}\label{cyclLBQ}
\begin{split}
    \oints{s_b-s_a}\hskip-.2cm \LL_R(s_a)\,\Sigma(s_a,s_b)&=\oints{s_b-s_a}\hskip-.2cm \Sigma(s_a,s_b)\,\LL_R(s_b)\,,\\
    \oints{s_b-s_a}\hskip-.2cm \BB_R(s_a)\,\Sigma(s_a,s_b)&=(-)^\Sigma\oints{s_b-s_a}\hskip-.2cm \Sigma(s_a,s_b)\,\BB_R(s_b)\,,\\
    \oints{s_b-s_a}\hskip-.2cm \QQ_R(s_a)\,\Sigma(s_a,s_b)&=(-)^\Sigma\oints{s_b-s_a}\hskip-.2cm \Sigma(s_a,s_b)\,\QQ_R(s_b)\,.
\end{split}
\end{equation}
Let us examine how operators acting on the closed string state $\oints{s_b-s_a}\Sigma(s_a,s_b)$ translate into line integrals on $\Sigma(s_a,s_b)$.
The BRST operator
is invariant under conformal transformations, and we find
\begin{equation}\label{QonSigma}
\begin{split}
    Q\oints{s_b-s_a}\hskip-.2cm \Sigma(s_a,s_b)
    &=-\oints{s_b-s_a}\hskip-.2cm  \QHB\,\Sigma(s_a,s_b)\\
    &=-\oints{s_b-s_a}\hskip-.2cm  \Bigl(\QHB\,\Sigma(s_a,s_b)+\QQ_R(s_a)\,\Sigma(s_a,s_b)-(-)^\Sigma\,\Sigma(s_a,s_b)\,\QQ_R(s_b)\Bigr)\\
    &=\oints{s_b-s_a}\hskip-.2cm  \QQ\,\Sigma(s_a,s_b)\,,
\end{split}
\end{equation}
where we used~(\ref{cyclLBQ}) in the second step
and~(\ref{Qdef})
in the last step.
Let us now consider the action of $L_0-\tilde L_0$ on $\oints{s_b-s_a}\Sigma(s_a,s_b)$. As $L_0-\tilde L_0$
generates rotations in the $\zeta$ frame, we expect it to generate horizontal translations in the $w$ frame.
As a first step we note that
\begin{equation}
\begin{split}
  L_0-\tilde L_0
  &= \int_{|\zeta|=\exp(-\frac{\pi^2}{s_b-s_a})}
  \biggl[\frac{d\zeta}{2\pi i}\,\zeta\, T(\zeta)-\frac{d\bar \zeta}{2\pi i}\,\bar \zeta\,\widetilde T(\bar \zeta)\biggr]\\
  &= \frac{s_b-s_a}{2\pi i}\int_{\gamma(\frac{\pi}{2})+s_a}^{\gamma(\frac{\pi}{2})+s_b}
  \biggl[\frac{dw}{2\pi i}\, T(w)+\frac{d\bar w}{2\pi i}\,\widetilde T(\bar w)\biggr]\,,
\end{split}
\end{equation}
and therefore
\begin{equation}
\begin{split}
    (L_0-\tilde L_0)\oints{s_b-s_a}\hskip-.2cm \Sigma(s_a,s_b)
    &=\frac{s_b-s_a}{2\pi i}\oints{s_b-s_a}\hskip-.2cm  \LHB\,\Sigma(s_a,s_b)\,.
\end{split}
\end{equation}
For $\Sigma(s_a,s_b)$ of the form~(\ref{Sigma}) we can use~(\ref{cyclLBQ}) and~(\ref{commLRSigma}) to find
\begin{equation}\label{L0-}
\begin{split}
    (L_0-\tilde L_0)\oints{s_b-s_a}\hskip-.2cm \Sigma(s_a,s_b)
    &=\frac{s_b-s_a}{2\pi i}\oints{s_b-s_a}\hskip-.2cm  \Bigl([\,\LL_R,\Sigma(s_a,s_b)\,]+\LHB\,\Sigma(s_a,s_b)\Bigr)\\
    &=-\frac{s_b-s_a}{2\pi i}\oints{s_b-s_a} \Biggl[\,\sum_{i=1}^k\del_{s_i}\Sigma(s_a,s_b)\,\Biggr]\,.
\end{split}
\end{equation}
This result is consistent with the intuition that the generator of rotations in the $\zeta$ frame, $L_0-\tilde L_0$, should generate horizontal translations on the positions $s_i$ of the
 slits where the open string states $A_i$ are glued.
The translation in the $w$ frame
is proportional to the total length of the strip $s_b-s_a$, as expected.
The corresponding identity for
$b_0-\tilde b_0$
reads
\begin{equation}\label{b0-}
\begin{split}
    &(b_0-\tilde b_0)\oints{s_b-s_a}\hskip-.2cm \Sigma(s_a,s_b)
    =\frac{s_b-s_a}{2\pi i}\oints{s_b-s_a}\hskip-.2cm  \BHB\,\Sigma(s_a,s_b)\\
    &=\frac{s_b-s_a}{2\pi i}\oints{s_b-s_a}
     \sum_{i=1}^k(-)^{\sum_{j=1}^{i-1}A_j}
    \PR (s_a,s_1)\ast A_1 \,\ldots\\
    & \qquad \qquad \qquad \qquad \times
    \bigl(\BB_R(s_i) A_i-(-)^{A_i}A_i\,\BB_R(s_i)\bigr) \,\ldots \,
    A_k \ast \PR (s_{k},s_{b})\,.
\end{split}
\end{equation}

\medskip
We can move an open string state $A$ cyclically within $\oints{s_b-s_a}$ just as we did for line integrals in~(\ref{cyclLBQ}). We have
\begin{equation}\label{Acycl}
       \oints{s_b-s_a}\hskip-.2cm A\ast\Sigma(s_a,s_b)=(-)^{A\Sigma}\oints{s_b-s_a}\hskip-.2cm \Sigma(s_a,s_b)\ast A\,.
\end{equation}
Similarly, we can cyclically move half-propagator
strips in $\oints{s_b-s_a}$, but all surfaces must attach to the same segment of the closed string boundary after using the cyclicity.  We conclude that
\begin{equation}
\begin{split}
& \oints{s_b-s_a}\hskip-.2cm\PR (s_a,s_1)\ast A_1\ast \PR (s_1,s_2) \,\ldots\, A_k\ast\PR (s_{k},s_{b})\\[.5ex]
&=\oints{s_b-s_a}\hskip-.2cm A_1\ast\PR (s_1,s_2) \,\ldots\, A_k\ast\PR (s_{k},s_{b})\ast \PR (s_b,s_1+s_b-s_a) \\
&=\oints{s_b-s_a}\hskip-.2cm A_1\ast\PR (s_1,s_2) \,\ldots\, A_k\ast\PR (s_{k},s_1+s_b-s_a)\,,
\end{split}
\end{equation}
where
the position of $\PR (s_a,s_1)$ was translated
by $s_b-s_a$,
which is consistent with the periodicity
$w\sim w+(s_b-s_a)$ in the $w$ frame.

\section{Construction of BRST-invariant closed string states}
\label{sec3}
\setcounter{equation}{0}

In this section we construct a class of
closed string states from a solution
of open string field theory
using the half-propagator strips we discussed in section~\ref{sec2}.
We then show that the closed string states are BRST invariant
and change by BRST-exact terms under
gauge transformations of the classical solution.

\subsection{The boundary state from the half-propagator strip}

The surface $\PR(0,s)$ is closely related
to the BCFT boundary state $\ket{B}$.
Recall
 from~(\ref{disk1ptfct})
that
a one-point function of
a closed string vertex operator
at the origin on a unit disk
can be expressed in terms of
$\ket{B}$ as follows:
\begin{equation}
\bra{B} ( c_0 - \tilde{c}_0 ) \ket{\phi_c} \,,
\end{equation}
where $\ket{\phi_c}$ is
the closed string state
corresponding to the vertex operator.
When we cut the unit disk along a circle
of radius $e^{-\frac{\pi^2}{s}}$, the one-point function
can be thought of as an inner product of
$\bra{B} \, e^{-\frac{\pi^2}{s} ( L_0 + \tilde{L}_0 )}$
and
$e^{\frac{\pi^2}{s} ( L_0 + \tilde{L}_0 )}
( c_0 - \tilde{c}_0 ) \ket{\phi_c}$:
\begin{equation}
\bra{B} ( c_0 - \tilde{c}_0 ) \ket{\phi_c}
= \bra{B} \,
e^{-\frac{\pi^2}{s} ( L_0 + \tilde{L}_0 )}
e^{\frac{\pi^2}{s} ( L_0 + \tilde{L}_0 )}
( c_0 - \tilde{c}_0 ) \ket{\phi_c} \,.
\end{equation}
The half-propagator strip of
 length $s$
 with the initial and final half-string boundaries identified can be mapped
 to the annulus region on the unit disk bounded by the unit circle and the circle of radius $e^{-\frac{\pi^2}{s}}$.
We therefore have
\begin{equation}
\oint_s \PR (0,s)
= e^{-\frac{\pi^2}{s} ( L_0 + \tilde{L}_0 )} \, \ket{B} \,.
\end{equation}
The boundary state $\ket{B}$ can thus be expressed
in terms of the half-propagator strip as follows:
\begin{equation}\label{defB}
\ket{B} = e^{\frac{\pi^2}{s} ( L_0 + \tilde{L}_0 )} \,
\oint_s \PR (0,s) \,.
\end{equation}
This definition reproduces the BCFT boundary state for any value of $s$. In particular, we conclude
\begin{equation}\label{dsB}
    \del_s\biggl[e^{\frac{\pi^2}{s} ( L_0 + \tilde{L}_0 )} \,
    \oint_s \PR (0,s)\biggr]
    \,=\,\del_s \ket{B}\,=\,0\,.
\end{equation}
Later we will confirm this explicitly
in \subs\ref{sec4.2}.

\subsection{Construction of the closed string state $\ket{\mB (\Psi)}$}

 We now define a closed string state that is expected
 to be a generalization of $\ket{B}$ to
the background
 associated with
a solution
to the equation of motion of open string field theory.
 In~(\ref{defB}) the boundary state $\ket{B}$  was expressed  in terms of the surface $\PR (0,s)$
 which is the right half of the propagator strip generated by
 $e^{-s {\cal L}}$.
Since
$\{Q, {\cal B}\} = {\cal L}$, we can write
\begin{equation}
e^{-s {\cal L}} = e^{-s \, \{ Q,\, {\cal B} \} } \,.
\end{equation}
We generalize $e^{-s{\cal L}}$
by replacing $Q$ in this expression
by the BRST operator associated with the new background.

When we expand the open string field theory action
around a solution $\Psi$ of the equation of motion
\begin{equation}
\label{orijdfg98}
Q \Psi + \Psi^2 = 0 \,,
\end{equation}
the BRST operator $\mQ$ associated with the
new background
is given by
\begin{equation}\label{defmQ}
\mQ A \equiv
Q A + \Psi A - (-)^A A \, \Psi
\end{equation}
for any state $A$.
Thus
the operator $e^{-s {\cal L}}$ should be modified as
\begin{equation}
    e^{-s {\cal L}}~\to~e^{-s \{ \mQ, \, {\cal B} \} }\,.
\end{equation}
To define a modified half-propagator strip $\mP(0,s)$,
we have to extract the right half of the surface
associated with $e^{-s \{ \mQ, \, {\cal B} \} }$.
To do this, we first examine the action of $\{ \mQ, {\cal B} \}$
on an arbitrary state $A$.  Making use of (\ref{defmQ}),
we readily find
\begin{equation}
\{ \mQ, {\cal B} \} \, A
= {\cal L} A
+ \Psi \, ( {\cal B} A )
+ (-)^A \, ( {\cal B} A ) \, \Psi
+ {\cal B} \, ( \Psi A)
- (-)^A {\cal B} \, ( A \Psi ) \,.
\end{equation}
If we write ${\cal L} = {\cal L}_R + {\cal L}_L$
and ${\cal B} = {\cal B}_R + {\cal B}_L$,
we find that the action of $\{ \mQ, {\cal B} \}$ on $A$
decomposes into right and left pieces
as
\begin{equation}
\label{splitactionqb*}
\{ \mQ, {\cal B} \} A
= \Bigl[ \, {\cal L}_R A
+ (-)^A ( {\cal B}_R A ) \Psi
- (-)^A {\cal B}_R ( A \Psi ) \, \Bigr]
+ \Bigl[ \, {\cal L}_L A
+ \Psi ( {\cal B}_L A )
+ {\cal B}_L ( \Psi A) \, \Bigr] \,,
\end{equation}
where terms with a mixed action
on both right and left halves of the state $A$
have canceled as follows:
\begin{equation}
\Psi ( {\cal B}_R A )
+ {\cal B}_R ( \Psi A) = 0 \,, \quad
(-)^A ( {\cal B}_L A ) \Psi
- (-)^A {\cal B}_L ( A \Psi ) = 0 \,.
\end{equation}
Therefore the operator ${\cal L}_R(t)$
in the half-propagator strip ${\cal P}(s_a, s_b)$
defined in~(\ref{defPR}) should be modified as
\begin{equation}
{\cal L}_R (t) ~\to~ {\cal L}_R (t) + \{ {\cal B}_R (t), \Psi \} \,.
\end{equation}
The sign factors of $(-)^A$ in~(\ref{splitactionqb*})
have disappeared because of our path-ordering convention
stated after~(\ref{defPR}).
We thus
define the modified
half-propagator strip by
\begin{equation}
\label{defhalf-prop-strip}
 \mP (s_a, s_b)
\equiv \Pexp \biggl[ \, - \int_{s_a}^{s_b} dt \,
\Bigl[ \, {\cal L}_R (t) + \{ {\cal B}_R (t), \Psi \} \,
\Bigr] \, \biggr]\,.
 \end{equation}
It is useful to explicitly expand $\mP (s_a, s_b)$
in powers of the classical solution.
We obtain
 \begin{equation}\label{explicitmP}
 \begin{split}
 \mP (s_a, s_b)
& = \PR (s_a, s_b)
- \int_{s_a}^{s_b} ds_1 \, \PR (s_a, s_1)
\, \{ {\cal B}_R (s_1), \Psi \} \, \PR (s_1, s_b) \\
& \quad ~
+ \int_{s_a}^{s_b} ds_1 \, \int_{s_1}^{s_b} ds_2 \,
\PR (s_a, s_1)
\, \{ {\cal B}_R (s_1), \Psi \} \, \PR (s_1, s_2)
\, \{ {\cal B}_R (s_2), \Psi \} \, \PR (s_2, s_b)
+ \ldots \\
& = \sum_{k=0}^\infty \, (-1)^k
\int_{s_a}^{s_b} ds_1 \ldots \int_{s_{i-1}}^{s_b} ds_i \ldots
\int_{s_{k-1}}^{s_b} ds_k \,
\PR (s_a, s_1)
\, \{ {\cal B}_R (s_1), \Psi \} \, \PR (s_1, s_2) \ldots \\
& \qquad \times \ldots
\PR (s_{i-1}, s_i) \,
\{ {\cal B}_R (s_i), \Psi \} \, \PR (s_i, s_{i+1}) \ldots
\PR (s_{k-1}, s_k) \,
\{ {\cal B}_R (s_k), \Psi \} \, \PR (s_k, s_b) \,.
\end{split}
\end{equation}
The modified half-propagator strip
obeys the following relations:
\begin{equation}
\begin{split}
\partial_{s_b} \mP (s_a, s_b)
& = - \mP (s_a, s_b) \,
( \, {\cal L}_R (s_b) + \{ {\cal B}_R (s_b), \Psi \} \, ) \,, \\
\partial_{s_a} \mP (s_a, s_b)
& = ( \, {\cal L}_R (s_a) + \{ {\cal B}_R (s_a), \Psi \} \, ) \,
\mP (s_a, s_b) \,.
\end{split}
\end{equation}
The formula~(\ref{dPAP}) is generalized as follows:
\begin{equation}
\partial_t \bigl[ \, \mP (s_a, t) \, A \, \mP (t, s_b) \, \bigr]
= - \mP (s_a, t) \,
\bigl[ \, {\cal L}_R (t) + \{ {\cal B}_R (t), \Psi \}\,,\, A \, \bigr] \,
\mP (t, s_b) \,.
\label{dtmP}
\end{equation}

By analogy with the expression~(\ref{defB})
of the original boundary state $\ket{B}$,
we now introduce the
following
background-dependent state:
\begin{equation}\label{defmB}
\boxed{\phantom{\Biggl\{}
\ket{\mB (\Psi)} \equiv
e^{\frac{\pi^2}{s} ( L_0 + \tilde{L}_0 )} \,
\oint_s \mP (0,s) \,.
~}
\end{equation}
For future use we
expand
$\ket{\mB (\Psi)}$ in powers of the solution:
\begin{equation}\label{mBPsiexpansion}
\ket{\mB (\Psi)} = \sum_{k=0}^\infty \ket{\mB^{(k)} (\Psi)} \,,
\end{equation}
where
\begin{equation}\label{defmBexp}
\begin{split}
\ket{\mB^{(0)} (\Psi)} & = \ket{B} \,, \\
\ket{\mB^{(1)} (\Psi)} & =
{}- e^{\frac{\pi^2}{s} ( L_0 + \tilde{L}_0 )}
\oint_s\,  \int_{0}^{s} ds_1 \, \PR (0, s_1)
\, \{ {\cal B}_R (s_1), \Psi \} \, \PR (s_1, s) \,, \\
\ket{\mB^{(2)} (\Psi)} & =
e^{\frac{\pi^2}{s} ( L_0 + \tilde{L}_0 )}
\oint_s\,   \int_{0}^{s}\hskip-4pt ds_1  \int_{s_1}^{s} \hskip-4pt ds_2 \,
\PR (0, s_1)
\, \{ {\cal B}_R (s_1), \Psi \} \, \PR (s_1, s_2)
\, \{ {\cal B}_R (s_2), \Psi \} \, \PR (s_2, s) \,, \\
& \vdots \, \\
\ket{\mB^{(k)} (\Psi)} & =
(-1)^k \, e^{\frac{\pi^2}{s} ( L_0 + \tilde{L}_0 )}
\oint_s \, \int_0^s ds_1 \ldots \int_{s_{i-1}}^s ds_i \ldots
\int_{s_{k-1}}^s ds_k \,
\PR (0, s_1)
\, \{ {\cal B}_R (s_1), \Psi \} \, \PR (s_1, s_2) \ldots \\
& \qquad \times \ldots
\PR (s_{i-1}, s_i) \,
\{ {\cal B}_R (s_i), \Psi \} \, \PR (s_i, s_{i+1}) \ldots
\PR (s_{k-1}, s_k) \,
\{ {\cal B}_R (s_k), \Psi \} \, \PR (s_k, s) \,.
\end{split}
\end{equation}
We expect that  $\ket{\mB (\Psi)}$
is related
to the boundary state of the BCFT
described by the  solution~$\Psi$.
In the following we study various properties of $\ket{\mB (\Psi)}$
and in section~\ref{secBCFT}
 we explicitly calculate it for analytic solutions.

\subsection{BRST invariance of $\ket{\mB (\Psi)}$}

We show that the
closed string
state $\ket{\mB (\Psi)}$
is BRST closed when $\Psi$ satisfies the equation of motion
of open string field theory.
The BRST transformation of $\{ {\cal B}_R (t), \Psi \}$ is

\begin{equation}\label{QBPsi}
\begin{split}
\QQ \, \{ {\cal B}_R (t), \Psi \}
& = \QQ \, ( {\cal B}_R (t) \, \Psi + \Psi \, {\cal B}_R (t) )
= {\cal L}_R (t) \, \Psi + {\cal B}_R (t) \, \Psi^2
- \Psi^2 \, {\cal B}_R (t) - \Psi \, {\cal L}_R (t) \\
& = [ \, {\cal L}_R (t), \Psi \, ]
+ \{ {\cal B}_R (t), \Psi \} \, \Psi
- \Psi \, \{ {\cal B}_R (t), \Psi \}
= [ \, {\cal L}_R (t)+\{ {\cal B}_R (t), \Psi \}, \Psi \, ] \,,
\end{split}
\end{equation}
where we have used the equation of motion
$Q \Psi + \Psi^2 = 0 \,$.
Using~(\ref{dtmP}), we conclude
\begin{equation}\label{QmP1}
\begin{split}
  \QQ \, \mP(s_a,s_b)&=-\int_{s_a}^{s_b}dt\,\mP(s_a,t) \, \bigl(\QQ \, \{ {\cal B}_R (t), \Psi \}\bigr) \, \mP(t,s_b)\\
    &=-\int_{s_a}^{s_b}dt\,\mP(s_a,t)\,[ \, {\cal L}_R (t)+\{ {\cal B}_R (t), \Psi \}, \Psi \, ]\,\mP(t,s_b)\\
    &=\int_{s_a}^{s_b}dt\,\del_{t}\bigl[\mP(s_a,t)\, \Psi\,\mP(t,s_b)\bigr]\\[.5ex]
    &=-\bigl[ \, \Psi,\mP(s_a,s_b) \, \bigr] \,.
\end{split}
\end{equation}
It is instructive to derive this relation explicitly from the expansion~(\ref{explicitmP}) of the path-ordered exponential which
defines $\mP(s_a,s_b)$. It follows from~(\ref{QBPsi}) that
\begin{equation}\label{QBPsiagain}
\begin{split}
& \QQ \Bigl[ \, \PR (s_{i-1}, s_i) \,
\{ {\cal B}_R (s_i), \Psi \} \, \PR (s_i, s_{i+1}) \,
\Bigr]
= \PR (s_{i-1}, s_i) \,
( \QQ \{ {\cal B}_R (s_i), \Psi \} ) \, \PR (s_i, s_{i+1}) \\
& =
 -\partial_{s_i} \, \Bigl[ \, \PR (s_{i-1}, s_i) \,
\Psi \, \PR (s_i, s_{i+1})  \, \Bigr]
+ \PR (s_{i-1}, s_i) \,
[ \, \{ {\cal B}_R (s_i), \Psi \}, \Psi \, ] \,
\PR (s_i, s_{i+1}) \,,
\end{split}
\end{equation}
where we have used~(\ref{dPAP}) and~(\ref{QSigma}).
Let us consider the term in the expansion of $\mP(s_a,s_b)$
with $k$ insertions
of the classical solution. We need to
calculate its BRST
transformation
\begin{equation}\label{termm}
\begin{split}
(-1)^k \QQ
\int_{s_a}^{s_b} ds_1 \,.\,.\!\int_{s_{i-1}}^{s_b} ds_i \,.\,.\!
\int_{s_{k-1}}^{s_b} ds_k \,\PR (s_a, s_1)
\,.\,. \,\PR (s_{i-1}, s_i)\, \{ {\cal B}_R (s_i), \Psi \} \, \PR (s_i, s_{i+1}) \,.\,.
\, \PR (s_k, s_b)\,.
\end{split}
\end{equation}
Using~(\ref{QBPsiagain}) and the formula
 \begin{equation}
 \begin{split}
 &\int_{s_{i-1}}^{s_b} ds_i
 \int_{s_i}^{s_b} ds_{i+1} \,
 \partial_{s_i}
 f( s_{i-1}, s_i, s_{i+1}) \\
 &=  \int_{s_{i-1}}^{s_b} ds_i \,
 \partial_{s_i} \biggl[ \,
 \int_{s_i}^{s_b} ds_{i+1}  \,
 f( s_{i-1}, s_i, s_{i+1}) \,
 \biggr]
  + \int_{s_{i-1}}^{s_b} ds_i\,
 f( s_{i-1}, s_i, s_i) \\
 &=  {}- \,
 \int_{s_{i-1}}^{s_b} ds_{i+1} \,
 f( s_{i-1}, s_i, s_{i+1})
 \biggr|_{s_i = s_{i-1}}
  + \int_{s_{i-1}}^{s_b} ds_i \,
 f( s_{i-1}, s_i, s_{i+1})
 \biggr|_{s_{i+1} = s_i} \,,
 \end{split}
 \end{equation}
we calculate~(\ref{termm}) as
\begin{equation}\label{}
\begin{split}
&(-1)^k
\sum_{i=1}^k\int_{s_a}^{s_b} ds_1 \,.\,.\int_{s_{i-1}}^{s_b} ds_{i}\,.\,.
\int_{s_{k-1}}^{s_b} ds_k \,\\
&\hskip2cm\times\,\PR (s_a, s_1)\ldots \PR (s_{i-1}, s_{i})\, \bigl[\{ {\cal B}_R (s_{i}), \Psi \}\,, \,\Psi\bigr] \PR (s_{i}, s_{i+1}) \ldots
\, \PR (s_k, s_b)\\
&-(-1)^{k-1}
\int_{s_a}^{s_b} ds_1 \,.\,.\int_{s_{i-1}}^{s_b} ds_{i}\,.\,.
\int_{s_{k-2}}^{s_b} ds_{k-1} \,\\
&\hskip2cm\times\, \biggl[ \, \sum_{i=1}^{k-1}\PR (s_a, s_1)\ldots\PR (s_{i-1}, s_{i})\, \bigl[\{ {\cal B}_R (s_{i}), \Psi \}\,, \,\Psi\bigr] \PR (s_{i}, s_{i+1}) \ldots\, \PR (s_{k-1}, s_b)\\
&\hskip3cm+\Bigl[ \, \Psi\,,\,\PR (s_a, s_1)\ldots \PR (s_{i-1}, s_{i})\,\{ {\cal B}_R (s_{i}), \Psi \}\,\PR (s_{i}, s_{i+1}) \ldots
\, \PR (s_{k-1}, s_b) \, \Bigr] \, \biggr]\,.
\end{split}
\end{equation}
Here we relabeled the indices  in the last two terms so that the $k-1$ integration variables are $s_1,\ldots,s_{k-1}$.
We see that after summing over all $k$, the first term at order $k-1$ cancels the second term at order $k$, and we are left with contributions from the last term. As in~(\ref{QmP1}), we conclude that
\begin{equation}
\QQ \, \mP (s_a, s_b)
= {}- [ \, \Psi, \mP (s_a, s_b) \, ] \,.
\label{mQmP}
\end{equation}
Recalling the definition of the modified BRST operator $\mQ$ in~(\ref{defmQ}), this is in fact a natural modification of~(\ref{QPR}).
We then find
\begin{equation}
Q \oint_{s_b - s_a} \mP (s_a, s_b)
= \oint_{s_b - s_a} \QQ \, \mP (s_a, s_b)
= - \oint_{s_b - s_a} [ \, \Psi, \mP (s_a, s_b) \, ] = 0 \,,
\end{equation}
where we used~(\ref{QonSigma}) in the first step and~(\ref{Acycl}) in the last step.
Since the BRST operator commutes with $L_0 + \tilde{L}_0$,
we obtain
\begin{equation}
Q \, \ket{\mB (\Psi)} = 0 \,.
\end{equation}
We have thus constructed a BRST-invariant closed string state
for a given solution $\Psi$.

It is easy to see that the closed string state $\ket{\mB (\Psi)}$
is annihilated by $b_0 - \tilde{b}_0$.
It follows from~(\ref{b0-})
and $[ \, {\cal B}_R (t), \{ {\cal B}_R (t), \Psi \} \, ] = 0$ that
\begin{equation}
( b_0 - \tilde{b}_0 ) \oint_{s_b - s_a} \mP (s_a, s_b) = 0 \,.
\end{equation}
Since $b_0 - \tilde{b}_0$ commutes with $L_0 + \tilde{L}_0$,
we conclude that
\begin{equation}
( b_0 - \tilde{b}_0 ) \, \ket{\mB (\Psi)} = 0 \,.
\end{equation}
The state $\ket{\mB (\Psi)}$ is also annihilated by
$L_0 - \tilde{L}_0$ because
\begin{equation}
( L_0 - \tilde{L}_0 ) \, \ket{\mB (\Psi)}
= \{ Q, \, b_0 - \tilde{b}_0 \} \, \ket{\mB (\Psi)} = 0 \,.
\end{equation}

 \medskip

In summary, we have found
that the state $\ket{\mB (\Psi)}$ satisfies three
consistency requirements for its interpretation as  a boundary state,
namely,
\begin{equation}
\boxed{\phantom{\biggl(}
Q \, \ket{\mB (\Psi)} = 0\,,\qquad
( b_0 - \tilde{b}_0 ) \, \ket{\mB (\Psi)}=0\,,\qquad
( L_0 - \tilde{L}_0 ) \, \ket{\mB (\Psi)}=0\,.
~~}
\end{equation}

\subsection{Variation of $\ket{\mB (\Psi)}$
under open string gauge transformations}\label{secgaugeinv}

Consider an infinitesimal gauge transformation of the solution:
\begin{equation}
\delta_\chi \Psi = Q \, \chi + [ \, \Psi, \chi \, ] \,.
\end{equation}
It follows from the path-ordered expression
of $\mP (s_a, s_b)$
that it changes
under the gauge transformation as follows:
\begin{equation}
\begin{split}
& \delta_\chi \mP (s_a, s_b)
= - \int_{s_a}^{s_b} dt \, \mP (s_a, t) \,
\{ {\cal B}_R (t), \delta_\chi \Psi \} \, \mP (t, s_b) \\
& = - \int_{s_a}^{s_b} dt \, \mP (s_a, t) \,
\{ {\cal B}_R (t), Q \chi \} \, \mP (t, s_b)
- \int_{s_a}^{s_b} dt \, \mP (s_a, t) \,
\{ {\cal B}_R (t), [ \, \Psi, \chi \, ]  \} \, \mP (t, s_b) \,.
\end{split}
\end{equation}
The first term in the second line can be written as
\begin{equation}\label{thefirstterm}
\begin{split}
& - \int_{s_a}^{s_b} dt \, \mP (s_a, t) \,
\{ {\cal B}_R (t), Q \chi \} \, \mP (t, s_b) \\
& = \QQ \int_{s_a}^{s_b} dt \, \mP (s_a, t) \,
[ \, {\cal B}_R (t), \chi \, ] \, \mP (t, s_b)
- \int_{s_a}^{s_b} dt \,
( \QQ \, \mP (s_a, t) ) \,
[ \, {\cal B}_R (t), \chi \, ] \, \mP (t, s_b) \\
& \quad ~ {}- \int_{s_a}^{s_b} dt \, \mP (s_a, t) \,
[ \, {\cal L}_R (t), \chi \, ] \, \mP (t, s_b)
+ \int_{s_a}^{s_b} dt \, \mP (s_a, t) \,
[ \, {\cal B}_R (t), \chi \, ] \,
( \QQ \, \mP (t, s_b) ) \,.
\end{split}
\end{equation}
The first term on the right-hand side
 of~(\ref{thefirstterm})
is BRST exact.
The second and fourth terms on the right-hand side
can be written using~(\ref{mQmP}) as follows:
\begin{equation}
\begin{split}
& - \int_{s_a}^{s_b} dt \,
( \QQ \, \mP (s_a, t) ) \,
[ \, {\cal B}_R (t), \chi \, ] \, \mP (t, s_b)
+ \int_{s_a}^{s_b} dt \, \mP (s_a, t) \,
[ \, {\cal B}_R (t), \chi \, ] \,
( \QQ \, \mP (t, s_b) ) \\
& =
\biggl\{ \, \Psi, \int_{s_a}^{s_b} dt \, \mP (s_a, t) \,
[ \, {\cal B}_R (t), \chi \, ] \, \mP (t, s_b) \, \biggr\}
- \int_{s_a}^{s_b} dt \, \mP (s_a, t) \,
\{ \, \Psi, [ \, {\cal B}_R (t), \chi \, ] \, \} \, \mP (t, s_b) \,.
\end{split}
\end{equation}
Since
\begin{equation}
\{ {\cal B}_R (t), [ \, \Psi, \chi \, ]  \}
+ \{ \, \Psi, [ \, {\cal B}_R (t), \chi \, ] \, \}
= [ \, \{ B_R (t), \Psi \, \} , \chi \, ] \,,
\end{equation}
we find
\begin{equation}
\begin{split}
\delta_\chi \mP (s_a, s_b)
& =  \QQ \int_{s_a}^{s_b} dt \, \mP (s_a, t) \,
[ \, {\cal B}_R (t), \chi \, ] \, \mP (t, s_b) \\
& \quad ~ + \biggl\{ \, \Psi, \int_{s_a}^{s_b} dt \, \mP (s_a, t) \,
[ \, {\cal B}_R (t), \chi \, ] \, \mP (t, s_b) \, \biggr\} \\
& \quad ~ - \int_{s_a}^{s_b} dt \, \mP (s_a, t) \,
[ \, {\cal L}_R (t) + \{ {\cal B}_R (t), \Psi \}, \chi \, ] \,
\mP (t, s_b) \,.
\end{split}
\end{equation}
Using~(\ref{dtmP}), we obtain the following final expression:
\begin{equation}
\begin{split}
\delta_\chi \mP (s_a, s_b)
& =  \QQ \int_{s_a}^{s_b} dt \, \mP (s_a, t) \,
[ \, {\cal B}_R (t), \chi \, ] \, \mP (t, s_b) \\\
& \quad ~ + \biggl\{ \, \Psi, \int_{s_a}^{s_b} dt \, \mP (s_a, t) \,
[ \, {\cal B}_R (t), \chi \, ] \, \mP (t, s_b) \, \biggr\}
- [ \, \chi, \mP (s_a, s_b) \, ] \,.
\end{split}
\end{equation}
It follows from this and~(\ref{Acycl}) that
\begin{equation}
\delta_\chi \oint_{s_b - s_a} \mP (s_a, s_b)
=  Q \oint_{s_b - s_a}
\int_{s_a}^{s_b} dt \, \mP (s_a, t) \,
[ \, {\cal B}_R (t), \chi \, ] \, \mP (t, s_b) \,.
\end{equation}
We thus have
\begin{equation}
\delta_\chi \ket{\mB (\Psi)}
=  Q \, \Bigl[ \,
e^{\frac{\pi^2}{s} ( L_0 + \tilde{L}_0 )} \,
\oint_s \int_0^s dt \, \mP (0, t) \,
[ \, {\cal B}_R (t), \chi \, ] \, \mP (t, s) \, \Bigr]
\end{equation}
and conclude that
 \begin{equation}\label{deltachi}
    \boxed{~~
    \delta_\chi \ket{\mB (\Psi)}\,=\, Q-\text{exact}\,.
    \phantom{\Bigl(}}
 \end{equation}

The inner product
$\bra{{\cal V}} \, ( c_0 - \tilde{c}_0 ) \ket{\mB (\Psi)}$
is then invariant under gauge transformations
of open string field theory
for any closed string state $\ket{\cal V}$ annihilated
by the BRST operator because
$\bra{{\cal V}} \, ( c_0 - \tilde{c}_0 ) \, Q \, \ket{\Omega} = 0$
for any closed string state $\ket{\Omega}$.
This can be shown in the following way
Any closed string state has to be annihilated
by $b_0 -\tilde{b}_0$.
Thus $\ket{\Omega}$ can be written as
\begin{equation}
\ket{\Omega} = ( b_0 -\tilde{b}_0 ) \, \ket{\widetilde{\Omega}} \,.
\end{equation}
Since $\ket{\cal V}$ is annihilated
by both $Q$ and $b_0 -\tilde{b}_0$, we have
\begin{equation}
\begin{split}
& \bra{{\cal V}} \, ( c_0 - \tilde{c}_0 ) \, Q \, \ket{\Omega}
= \bra{{\cal V}} \, ( c_0 - \tilde{c}_0 ) \, Q \,
( b_0 -\tilde{b}_0 ) \, \ket{\widetilde{\Omega}} \\
& = \bra{{\cal V}} \, \{ c_0 - \tilde{c}_0 ,\, Q \} \,
( b_0 -\tilde{b}_0 ) \, \ket{\widetilde{\Omega}}
= \bra{{\cal V}} \, [ \, \{ c_0 - \tilde{c}_0 ,\, Q \} ,\,
b_0 -\tilde{b}_0 \, ] \, \ket{\widetilde{\Omega}} \,.
\end{split}
\end{equation}
Using the Jacobi identity, we find
\begin{equation}
\begin{split}
[ \, \{ c_0 - \tilde{c}_0 ,\, Q \} ,\,
b_0 -\tilde{b}_0 \, ]
& = {}- [ \, \{ Q ,\, b_0 -\tilde{b}_0 \} ,\,
c_0 - \tilde{c}_0 \, ]
- [ \, \{ b_0 -\tilde{b}_0 ,\, c_0 - \tilde{c}_0 \} ,\,
Q \, ] \\
& = {}- [ \, L_0 -\tilde{L}_0 ,\, c_0 - \tilde{c}_0 \, ] = 0 \,.
\end{split}
\end{equation}
Therefore
$\bra{{\cal V}} \, ( c_0 - \tilde{c}_0 ) \, Q \, \ket{\Omega}$
vanishes for any closed string state $\ket{\Omega}$
and thus we conclude that
$\bra{{\cal V}} \, ( c_0 - \tilde{c}_0 ) \ket{\mB (\Psi)}$
is gauge-invariant
for any closed string state $\ket{{\cal V}}$
annihilated by the BRST operator.

\section{Dependence on the choice of the propagator strip}
\label{sec4}
\setcounter{equation}{0}

The closed string state $\ket{\mB (\Psi)}$
for a given solution $\Psi$
depends on the total strip length~$s$ and
the operator $\BB$ for the gauge-fixing condition.
In this section we study the dependence
of the state $\ket{\mB (\Psi)}$
on $s$ and ${\cal B}$.

\subsection{Variation of the propagator}
\label{secindprop}

Let us consider an infinitesimal change of
the gauge-fixing condition~(\ref{classgaugecond})
for the propagator.
The corresponding changes
of $\BB_R$ and $\LL_R$ are
\begin{equation}\label{deltagauge}
    \BB_R(t)~\to~\BB_R(t)+\delta\BB_R(t)\,,\qquad
    \LL_R(t)~\to~\LL_R(t)+\{\QQ_R(t),\delta\BB_R(t)\}\,.
\end{equation}
Thus the modified half-propagator strip $\mP (s_a, s_b)$
changes as follows:
\begin{equation}
\begin{split}
& \delta \mP (s_a, s_b)
= - \int_{s_a}^{s_b} dt \, \mP (s_a, t) \,
\delta [ \, {\cal L}_R (t)
+ \{ {\cal B}_R (t), \Psi \} \, ] \, \mP (t, s_b) \\
& = - \int_{s_a}^{s_b} dt \, \mP (s_a, t) \,
\{ \QQ_R(t), \delta {\cal B}_R (t) \} \, \mP (t, s_b)
- \int_{s_a}^{s_b} dt \, \mP (s_a, t) \,
\{ \delta {\cal B}_R (t), \Psi \} \, \mP (t, s_b) \\
& = {}- \QQ \int_{s_a}^{s_b} dt \, \mP (s_a, t) \,
\delta {\cal B}_R (t) \, \mP (t, s_b)
- \biggl\{ \Psi, \int_{s_a}^{s_b} dt \, \mP (s_a, t) \,
\delta {\cal B}_R (t) \, \mP (t, s_b) \biggr\} \,,
\end{split}
\end{equation}
where we used~(\ref{mQmP}) in the last step.
We therefore
find that
\begin{equation}
\delta \oint_{s_b - s_a} \mP (s_a, s_b)
= {}- Q \oint_{s_b - s_a}
\int_{s_a}^{s_b} dt \, \mP (s_a, t) \,
\delta {\cal B}_R (t) \, \mP (t, s_b) \,.
\end{equation}
We conclude that
the closed string state $\ket{\mB (\Psi)}$
changes by a BRST-exact term
under a variation~(\ref{deltagauge}) of
the gauge-fixing condition:
 \begin{equation}
    \boxed{~~
    \delta \ket{\mB (\Psi)}\,=\, Q-\text{exact}\,.
    \phantom{\Bigl(}}
 \end{equation}

\subsection{Change of the strip length and the action of $L_0+\tilde L_0$}
\label{sec4.2}

To understand the $s$ dependence of $\ket{\mB(\Psi)}$, let us begin by relating closed string states of the type $\oints{s_b-s_a}\Sigma(s_a,s_b)$ defined in~(\ref{Sigma})
with different values for $s_b$.
Consider an infinitesimal change in $s_b$.
A change in the
strip length $s_b-s_a$ affects the gluing of the strip to the closed string coordinate,
as can be seen from~(\ref{wsigma}). We thus need to reparameterize the closed string boundary. Infinitesimally, we account for this change by inserting a line integral of
the energy-momentum tensor
along the closed string boundary.
{}From~(\ref{wsigma})
it follows that the vector field $u$ which adjusts the parameterization of the closed string boundary to an infinitesimal change in $s_b$ is given by
\begin{equation}
    u(w)
    =\frac{w-\left(\gamma(\tfrac{\pi}{2})+s_a\right)}{s_b-s_a}\,.
\end{equation}
This vector field is tangential to the closed string boundary $\Im(w)=\frac{\pi}{2}$, vanishes at $w=\gamma(\tfrac{\pi}{2})+s_a$, and satisfies
\begin{equation}
    u\left(\gamma(\tfrac{\pi}{2})+s_b\right)=1\,.
\end{equation}
It thus follows that
the corresponding line integral of the energy-momentum tensor,
\begin{equation}\label{Lrepara}
    \LrepHB\,\,=\,
    \int_{\gamma(\frac{\pi}{2})+s_a}^{\gamma(\frac{\pi}{2})+s_b}
    \biggl[\frac{dw}{2\pi i}\,\frac{w-\gamma(\tfrac{\pi}{2})-s_a}{s_b-s_a}\,T(w)
    +\frac{d\bar w}{2\pi i}\,\frac{\bar w-\overline{\gamma(\tfrac{\pi}{2})}-s_a}{s_b-s_a}\,\widetilde T(\bar w)\biggr] \,,
\end{equation}
generates the desired linear stretching of the closed string boundary:
\begin{equation}\label{dsbC}
    \del_{s_b}\oints{s_b-s_a}\hskip-.2cm\Sigma(s_a,s_b)
    =\oints{s_b-s_a}\hskip-.2cm\del_{s_b}\Sigma(s_a,s_b)
    +\oints{s_b-s_a}\hskip-.2cm\LrepHB\,\Sigma(s_a,s_b)\,.
\end{equation}
Note that the constant part of  the vector field $u(w)$ has an imaginary contribution $\frac{-i\pi}{2(s_b-s_a)}$ arising from $\gamma(\frac{\pi}{2})$
and thus
we cannot immediately derive a useful identity
analogous to~(\ref{commLR}).
This contribution to
the operator $\LrepHB$ is proportional to $L_0+\tilde L_0$,
which can be written in the $w$ frame as
\begin{equation}
\begin{split}
  L_0+\tilde L_0
  &= \int_{|\zeta|=\exp(-\frac{\pi^2}{s_b-s_a})}
  \biggl[\frac{d\zeta}{2\pi i}\,\zeta\, T(\zeta)+\frac{d\bar \zeta}{2\pi i}\,\bar \zeta\,\widetilde T(\bar \zeta)\biggr]\\
  &= \frac{s_b-s_a}{2\pi i}\int_{\gamma(\frac{\pi}{2})+s_a}^{\gamma(\frac{\pi}{2})+s_b}
  \biggl[\frac{dw}{2\pi i}\, T(w)-\frac{d\bar w}{2\pi i}\,\widetilde T(\bar w)\biggr]\,.
\end{split}
\end{equation}
Therefore, if we define
\begin{equation}\label{Lrepara2}
    \LreppHB\,\,=\,
    \int_{\gamma(\frac{\pi}{2})+s_a}^{\gamma(\frac{\pi}{2})+s_b}
    \biggl[\frac{dw}{2\pi i}\,\frac{w-\Re\left(\gamma(\tfrac{\pi}{2})\right)-s_a}{s_b-s_a}\,T(w)
    +\frac{d\bar w}{2\pi i}\,\frac{\bar w-\Re\left(\gamma(\tfrac{\pi}{2})\right)-s_a}{s_b-s_a}\,\widetilde T(\bar w)\biggr]
\end{equation}
for $\LreppHB$ acting on $\Sigma(s_a,s_b)$, we have

\begin{equation}\label{dsb}
\begin{split}
    \del_{s_b}\oints{s_b-s_a}\hskip-.2cm\Sigma(s_a,s_b)
    &=\oints{s_b-s_a}\hskip-.2cm\del_{s_b}\Sigma(s_a,s_b)
    +\oints{s_b-s_a}\hskip-.2cm\LreppHB\Sigma(s_a,s_b)
    +\frac{\pi^2}{(s_b-s_a)^2}(L_0+\tilde L_0)\oints{s_b-s_a}\hskip-.2cm\Sigma(s_a,s_b)\,.
\end{split}
\end{equation}
We introduce
$\Lrepp_R(t)$ with the same integrand as $\LreppHB$ by
\begin{equation}\label{UR}
    \Lrepp_R(t)\,\,=\,
    \int_{t}^{\gamma(\frac{\pi}{2})+t}
    \biggl[\frac{dw}{2\pi i}\,\frac{w-\Re\left(\gamma(\tfrac{\pi}{2})\right)-s_a}{s_b-s_a}\,T(w)
    +\frac{d\bar w}{2\pi i}\,\frac{\bar w-\Re\left(\gamma(\tfrac{\pi}{2})\right)-s_a}{s_b-s_a}\,\widetilde T(\bar w)\biggr]\,,
\end{equation}
and we have the following identity analogous to~(\ref{commLR}):
\begin{equation}
\Lrepp_R(s_a) \, \PR (s_a,s_b)
-\PR (s_a,s_b) \, \Lrepp_R(s_b)
+\LreppHB \, \PR (s_a,s_b) =0 \,.
\end{equation}
Note that unlike the line integrals $\LL_R$, $\BB_R$ and $\QQ_R$, the integrand of $\Lrepp_R$ is not invariant under the identification $w\sim w+(s_b-s_a)$.
We instead have
\begin{equation}\label{mapUR}
    \oints{s_b-s_a}\hskip-.2cm\Lrepp_R(s_a)\Sigma(s_a,s_b)-\oints{s_b-s_a}\hskip-.2cm\Sigma(s_a,s_b)\Lrepp_R(s_b)
    +\oints{s_b-s_a}\hskip-.2cm\Sigma(s_a,s_b)\LL_R(s_b)=0\,.
\end{equation}
Adding the left-hand side of~(\ref{mapUR})
to~(\ref{dsb})  allows us to
localize the integration contour
around the slits where the open strings are inserted. We obtain
\begin{equation}\label{dsb2}
\begin{split}
    \del_{s_b}\oints{s_b-s_a}\hskip-.2cm\Sigma(s_a,s_b)
    =&\oints{s_b-s_a}\hskip-.2cm\bigl[\del_{s_b}\Sigma(s_a,s_b)+\Sigma(s_a,s_b)\LL_R(s_b)\bigr]
    +\frac{\pi^2}{(s_b-s_a)^2}(L_0+\tilde L_0)\oints{s_b-s_a}\hskip-.2cm\Sigma(s_a,s_b)\\
    &+\sum_{i=1}^k\oints{s_b-s_a}\hskip-.2cm
    \PR (s_a,s_1)\ast A_1 \,\cdots\,
    [\Lrepp_R(s_i), A_i]\,\cdots
    \,  A_k \ast \PR (s_{k},s_{b})\,.
\end{split}
\end{equation}
Let us apply this result to the expression
on the right-hand side of~(\ref{defB}) for the boundary state~$\ket{B}$.
The last term in~(\ref{dsb2}) vanishes for this case as
$\PR (0,s)$
does not contain any open string state insertions $A_i$. Furthermore, the first term in~(\ref{dsb2}) also vanishes
because of~(\ref{dsPR}).
We find
\begin{equation}\label{delsB}
\begin{split}
    \del_{s}\ket{B}=\del_{s} \biggl[e^{\frac{\pi^2}{s} ( L_0 + \tilde{L}_0 )} \, \oint_s \PR (0,s) \biggr]
    =-\frac{\pi^2}{s^2}(L_0+\tilde L_0)\ket{B}+ e^{\frac{\pi^2}{s}( L_0 + \tilde{L}_0 )} \,
    \del_{s}  \oint_s \PR (0,s)
    =0\,.
\end{split}
\end{equation}
We have thus confirmed~(\ref{dsB}),
and the right-hand side of~(\ref{defB})
reproduces the BCFT boundary state
of the original theory independent of $s$.

\subsection{Variation of $s$}

We now use the results
of the previous subsection to study the $s$ dependence
of the closed string state $\ket{\mB(\Psi)}$.
Recalling that
\begin{equation}
    \del_{s}\mP(0,s)=-\mP(0,s)\bigl(\LL_R(s)+\{\BB_R(s),\Psi\}\bigr)\,,
\end{equation}
we find
\begin{equation}\label{delsBmC}
\begin{split}
    \del_{s} \ket{\mB(\Psi)}
    &= -  \frac{\pi^2}{s^2}
    \, ( L_0 + \tilde{L}_0 ) \,
    e^{\frac{\pi^2}{s}( L_0 + \tilde{L}_0 )} \, \oint_s \mP (0,s)
    + e^{\frac{\pi^2}{s}( L_0 + \tilde{L}_0 )} \,
    \partial_s \oint_s \mP (0,s)\\
    &=e^{\frac{\pi^2}{s}( L_0 + \tilde{L}_0 )}
    \oint_s \biggl[\LreppHB\,\mP (0,s)-\mP (0,s)\bigl(\LL_R(s)+\{\BB_R(s),\Psi\}\bigr)\biggr]\,,
\end{split}
\end{equation}
where we have used~(\ref{dsb}).
We define the $b$-ghost line integral $\BreppHB$
as
$\LreppHB$ in~(\ref{Lrepara2})
with $T(z)$ and $\widetilde T(\bar z)$  replaced by $b(z)$ and $\tilde b(\bar z)$, respectively.  It follows that
$\{\QQ,\BreppHB\}=\LreppHB$.
Consider now  the first term on the above right-hand side.
We have
\begin{equation}
    \oint_s \LreppHB\,\mP (0,s)
= \oint_s \{\QQ,\BreppHB\}\mP (0,s)
=Q\,\oint_s\BreppHB\,\mP (0,s)+\oint_s\BreppHB[\Psi\,,\,\mP (0,s)]\,.
\end{equation}
The commutator term vanishes using the cyclicity property~(\ref{Acycl}),\footnote{The cyclicity property is unaffected by the presence of the line integral $\BreppHB$ along the closed string boundary.}
and we conclude that the first term in~(\ref{delsBmC}) is BRST exact:
\begin{equation}\label{UHBC}
    e^{\frac{\pi^2}{s}( L_0 + \tilde{L}_0 )} \oint_s \LreppHB\,\mP (0,s)
    =Q\,\biggl[e^{\frac{\pi^2}{s}( L_0 + \tilde{L}_0 )} \oint_s \BreppHB\,\mP (0,s)\biggr]\,.
\end{equation}
The second term in~(\ref{delsBmC}) is also BRST exact.
In fact, we again use the cyclicity property
to find
\begin{equation}
\begin{split}
    \oint_s \mP (0,s)\bigl(\LL_R(s)+\{\BB_R(s),\Psi\}\bigr)
    &=\oint_s \biggl(\mP (0,s)\{\QQ_R(s),\BB_R(s)\}-\bigl[\,\Psi\,,\,\mP (0,s)\,\bigr]\,\BB_R(s)\biggr)\\[.5ex]
    &=Q\oint_s \mP (0,s)\BB_R(s)\,.
\end{split}
\end{equation}
As both terms in~(\ref{delsBmC}) are BRST exact, we conclude that $\ket{\mB(\Psi)}$ changes by a BRST exact piece under a variation of $s$:
\begin{equation}\label{dsmBPsi}
\boxed{~~
    \del_s \ket{\mB(\Psi)} \,=\, Q-\text{exact} \,.
\phantom{\Bigl(}}
\end{equation}
Using the formula
\begin{equation}
\oint_s \Brepp_R (0) \, \mP (0,s)
-\oint_s \mP (0,s) \, \Brepp_R (s)
+\oint_s \mP (0,s) \, \BB_R(s)=0 \,,
\end{equation}
which can be derived in the same way as~(\ref{mapUR}),
the total BRST-exact term
 in~(\ref{dsmBPsi})
can be written as
\begin{equation}
  \del_s \ket{\mB(\Psi)} =
 Q \, \biggl[ \,
 e^{\frac{\pi^2}{s}( L_0 + \tilde{L}_0 )}
 \oint_s \, \Bigl( \, \BreppHB\,\mP (0,s)
 + \Brepp_R (0) \, \mP (0,s)
 - \mP (0,s) \, \Brepp_R (s) \, \Bigr)
 \, \biggr]
\end{equation}
so that the three $b$-ghost contours can be connected.

\subsection{The $s\to0$ limit}
\label{secs0}

Let us now consider the limit $s\to 0$ of the state $\ket{\mB(\Psi)}$.
The first term in
the expansion~(\ref{mBPsiexpansion})
is the original boundary state $\ket{B}$.
It is independent of $s$ and thus the limit $s \to 0$ is trivial.
The next term $\ket{\mB^{(1)} (\Psi)}$ can be written as
\begin{equation}\label{rots1-99}
\begin{split}
\ket{\mB^{(1)} (\Psi)}
&= {}- e^{\frac{\pi^2}{s} ( L_0 + \tilde{L}_0 )}\int_{0}^{s} \!ds_1 \oint_s\,
 \PR (0, s_1)\, \{{\cal B}_R (s_1), \Psi \}\PR (s_1, s)\\
&={}- e^{\frac{\pi^2}{s} ( L_0 + \tilde{L}_0 )} \, \frac{2\pi i}{s}\int_{0}^{s} \!ds_1\,(b_0-\tilde{b}_0) \, \oint_s\,
 \PR (0, s_1)\,\Psi \,\PR (s_1, s)  \,,
\end{split}
\end{equation}
where we used~(\ref{b0-}).
We can further transform it using
an integrated version of~(\ref{L0-})
as follows:
\begin{equation}\label{rots1}
\begin{split}
\ket{\mB^{(1)} (\Psi)}
&={}- e^{\frac{\pi^2}{s} ( L_0 + \tilde{L}_0 )}\,\frac{2\pi i}{s}\int_{0}^{s} \!ds_1\,(b_0-\tilde{b}_0) \,
e^{-s_1\,\frac{2\pi i}{s} ( L_0 - \tilde{L}_0 )} \, \oint_s\, \Psi\,\PR (0, s)\\
&= -i\,e^{\frac{\pi^2}{s} ( L_0 + \tilde{L}_0 )}\int_{0}^{2\pi} \! d\theta\,(b_0-\tilde{b}_0)\,
e^{-i\theta ( L_0 - \tilde{L}_0 )} \, \oint_s\, \Psi\,\PR (0, s)\,,
\end{split}
\end{equation}
where in the last step we defined $\theta = 2\pi s_1/s$.
One might have suspected that
the state
\begin{equation}
\lim_{s \to 0}
\int_{0}^{s} \!ds_1 \oint_s\,
 \PR (0, s_1)\, \{{\cal B}_R (s_1), \Psi \}\PR (s_1, s)
\end{equation}
vanishes because the integration region of $s_1$ collapses
in the limit $s\to0$.
We see from~(\ref{rots1}), however, that
the integration over $s_1$ effectively rotates the surface state
once around the closed string coordinate.
The vanishing integration region over $s_1$ was only a coordinate effect of our parameterization of the integral over the rotational modulus,
and
the final expression in~(\ref{rots1})
is clearly nonvanishing
in the limit $s\to0$ for generic $\Psi$.

Let us next consider
inner products
$\langle \, {\cal V} \, | \, ( c_0 - \tilde{c}_0 )
\ket{\mB^{(k)} (\Psi)}$ with $k \ge 2$,
where ${\cal V}$ is an arbitrary on-shell closed string state:
\begin{equation}
\label{more-than-one}
\begin{split}
& \langle \, {\cal V} \, | \, ( c_0 - \tilde{c}_0 )
\ket{\mB^{(k)} (\Psi)} \\
& = (-1)^k \,
\langle \, {\cal V} \, | \, ( c_0 - \tilde{c}_0 ) \, |
\oint_s \int_0^s ds_1 \int_{s_1}^s ds_2 \ldots
\int_{s_{k-1}}^s ds_k \,
\PR (0, s_1)
\, \{ {\cal B}_R (s_1), \Psi \} \, \PR (s_1, s_2) \ldots \\
& \qquad \qquad \qquad \qquad \qquad \qquad \qquad
\qquad \qquad \qquad \qquad \times
\PR (s_{k-1}, s_k) \,
\{ {\cal B}_R (s_k), \Psi \} \, \PR (s_k, s) \, \rangle \,.
\end{split}
\end{equation}
The factor $e^{\frac{\pi^2}{s} ( L_0 + \tilde{L}_0 )}$
did not contribute because ${\cal V}$ is a primary field
of weight $(0,0)$.
The limit $s \to 0$ of these inner products were essentially
discussed in section~4 of~\cite{Ellwood:2008jh},
where it was argued that the terms with $k \ge 2$ vanish
for a certain regular class of solutions.
Let us review the argument in~\cite{Ellwood:2008jh}.\footnote{ The analysis in~\cite{Ellwood:2008jh} was based on the Siegel propagator strip, but this choice does not enter the following argument
in an essential way.}

As we did in~(\ref{rots1-99}),
one
can extract a factor of
$b_0 - \tilde{b}_0$ from the expression in~(\ref{more-than-one}).
It is accompanied by a factor of $2 \pi i/s$,
thus the integrand in~(\ref{more-than-one})
is singular in the limit $s \to 0$.
However, the $k$ dimensional integral
should be transmuted into one integral
for the overall rotation and a $k-1$ dimensional integral,
and the Jacobian should cancel the singularity
of the factor $2 \pi i/s$ as in~(\ref{rots1}).
It was argued in~\cite{Ellwood:2008jh} that
the resulting integrand is finite in the limit $s \to 0$,
while the $k-1$ dimensional integration region vanishes.
Then the inner products in~(\ref{more-than-one}) vanish
in the limit $s \to 0$ for $k \ge 2$.

As was remarked in one of the footnotes of~\cite{Ellwood:2008jh},
however, it is difficult to identify
necessary regularity conditions on the solution
for the finiteness of the resulting integrand in the limit $s \to 0$
and thus difficult to prove
rigorously
that the inner products in~(\ref{more-than-one}) vanish
in the limit $s \to 0$ for $k \ge 2$.
For example,
the open string midpoint of the solution
approaches the closed string vertex operator
in the limit $s \to 0$, and we may find singular
operator products.
In fact, the analytic solutions
in Schnabl gauge
constructed
in~\cite{0511286, 0701248, 0701249}
contain $b$-ghost integrals
extending up to the open string midpoint,
and their operator products with ${\cal V}$
can potentially be singular.
Furthermore, we have to be careful
when we judge whether expressions are finite
using the $w$ frame
because conformal factors associated with the map
to a disk coordinate can potentially be singular
in the limit $s \to 0$.
In fact, we have seen that the singular factor in~(\ref{rots1-99})
arose from such a conformal factor.
Similarly, we have
\begin{equation}
\label{singular-example}
\begin{split}
& \oint_s \, \PR (0, s_1)\, \{{\cal B}_R (s_1), \Psi \} \,
\PR (s_1, s_2)\, \{{\cal B}_R (s_2), \Psi \} \, \PR (s_2, s) \\
& = \frac{2 \pi i}{s} \, ( b_0 - \tilde{b}_0 )
\oint_s \, \PR (0, s_1)\, \Psi \,
\PR (s_1, s_2)\, \{{\cal B}_R (s_2), \Psi \} \, \PR (s_2, s) \,.
\end{split}
\end{equation}
This is also singular in the limit $s \to 0$, but we expect that
\begin{equation}
\label{regular-example}
\oint_s \, \PR (0, s_1)\, \Psi \,
\PR (s_1, s_2)\, \{{\cal B}_R (s_2), \Psi \} \, \PR (s_2, s)
\end{equation}
is finite in the limit $s \to 0$ if the solution $\Psi$ is regular.
The difference in the behavior as $s \to 0$
between~(\ref{singular-example})
and~(\ref{regular-example})
is not so obvious.

We therefore do not make a general claim that
the inner products in~(\ref{more-than-one}) vanish
in the limit $s \to 0$ for $k \ge 2$.
On the other hand, an advantage of our approach
is that the state $\ket{\mB (\Psi)}$ is explicitly calculable
for solutions based on
wedge states
if we choose the propagator strip of Schnabl gauge,
as we demonstrate
in section~\ref{secwedge}.
We revisit the limit $s \to 0$
in \subs\ref{thestozerorevisited},
and we
examine
this suppression of higher-order terms
explicitly in section~\ref{secBCFT}
for various  known analytic solutions.
We indeed find that the states $\ket{\mB^{(k)} (\Psi)}$ with $k\geq2$
vanish in the limit $s\to0$ for all explicit examples that we consider.

 \smallskip

If the inner products
$\langle \, {\cal V} \, | \, ( c_0 - \tilde{c}_0 )
\ket{\mB^{(k)} (\Psi)}$ with $k \ge 2$
vanish
 in the limit $s\to0$
for a given regular solution $\Psi$, we find
\begin{equation}\label{contractV}
\begin{split}
    &\lim_{s\to0}\,\,\langle{\cal V}|(c_0-\tilde{c}_0)|\mB(\Psi)\rangle- \langle{\cal V}|(c_0-\tilde{c}_0)|B\rangle
= \lim_{s\to0}\,
\langle{\cal V}|(c_0-\tilde{c}_0)|\mB^{(1)} (\Psi)\rangle \\
    &= -i\,\lim_{s\to0}\int_{0}^{2\pi} \! d\theta\,\langle\,{\cal V}\,|\,(c_0-\tilde{c}_0)\,(b_0-\tilde{b}_0)\,
  |\, \oint_s\, \Psi\,\PR (0, s)\,\rangle\\[1ex]
    &=-4\pi i\,\lim_{s\to0} \,\langle\,{\cal V}\,|\oint_s\, \Psi\,\PR (0, s)\,\rangle
\,.
\end{split}
\end{equation}
Here we used that $\ket{\cal V}$
is annihilated by $L_0\pm\tilde{L}_0$ and $b_0-\tilde{b}_0$.
The parameter $s$ in~(\ref{contractV}) is simply a regularization of a contraction of $\Psi$ with the identity state.
In the limit $s\to0$, the closed string vertex
operator is inserted at the open string midpoint of $\Psi$,
and we
recover  the familiar string field theory observables
$W ({\cal V}, \Psi)$:
\begin{equation}\label{observablesSiegel}
    \lim_{s\to0}\,\<{\cal V}|(c_0-\tilde{c}_0)|\mB(\Psi)\>- \<{\cal V}|(c_0-\tilde{c}_0)|B\>\,
    =\,-4\pi i\,W ({\cal V}, \Psi)\,,
\end{equation}
where $W ({\cal V}, \phi)$ for a generic
 open string
state
 $\ket{\phi}$
in the Fock space is defined by
\begin{equation}
W ({\cal V}, \phi)
= \langle \, {\cal V}(i) \, f_I \circ \phi (0) \, \rangle_{\rm UHP}
\,.
\end{equation}
We denote
 by $f_I \circ \phi (0)$
the conformal transformation
of the operator $\phi (0)$ corresponding to the state $\ket{\phi}$
under the identity map
\begin{equation}
\label{identity-map}
f_I (\xi) = \frac{2 \, \xi}{1-\xi^2} \,.
\end{equation}

As is well known, the observables $W ({\cal V}, \Psi)$ are
gauge-invariant.
The vanishing of terms with two or more solution insertions in the limit $s\to0$ is thus consistent because
$\ket{\mB^{(0)} (\Psi)} + \ket{\mB^{(1)} (\Psi)}$
is gauge-invariant in this limit.
Furthermore, when we vary $s$,
the state $|\mB(\Psi)\>$ only changes by a BRST-exact term
which has a vanishing inner product
with $\bra{{\cal V}}(c_0-\tilde{c}_0)$, and thus
we conclude that~(\ref{observablesSiegel})
holds even for finite $s$:
\begin{equation}\label{observablesall}
   \<{\cal V}|(c_0-\tilde{c}_0)|\mB(\Psi)\>- \<{\cal V}|(c_0-\tilde{c}_0)|B\>\,
   =\,-4\pi i\,W ({\cal V}, \Psi)
\quad \text{for any $s$} \,.
\end{equation}
It was argued in~\cite{Ellwood:2008jh} that
the observables
$W ({\cal V}, \Psi)$
represent the difference
between the original boundary state $\ket{B}$
and the boundary state $\ket{B_\ast}$
contracted with
$\langle {\cal V} | ( c_0 - \tilde{c}_0 )$:\footnote{
The relation in the notation of~\cite{Ellwood:2008jh} is
$W ({\cal V}, \Psi)
={\cal A}_\Psi^{\rm disk}({\cal V})
-{\cal A}_0^{\rm disk}({\cal V})$,
where ${\cal A}_\Psi^{\rm disk}({\cal V})$
and ${\cal A}_0^{\rm disk}({\cal V})$ are related to
the inner products as
$\<{\cal V}|(c_0-\tilde{c}_0)|B_\ast\>
=-4\pi i{\cal A}^{\rm disk}({\cal V})$
and $\<{\cal V}|(c_0-\tilde{c}_0)|B\>
=-4\pi i{\cal A}_0^{\rm disk}({\cal V})$.
}
\begin{equation}\label{fromIan}
    \<{\cal V}|(c_0-\tilde{c}_0)|B_\ast\>-\<{\cal V}|(c_0-\tilde{c}_0)|B\>
    =-4\pi i\,W ({\cal V}, \Psi) \,.
\end{equation}
This implies that
\begin{equation}\label{VmBBCFT}
     \<{\cal V}|(c_0-\tilde{c}_0)|\mB(\Psi)\>\,=\,\<{\cal V}|(c_0-\tilde{c}_0)
     |B_\ast\>\,
\end{equation}
and thus
\begin{equation}\label{mB=BBCFT+Q}
    \ket{\mB(\Psi)}\,=\,\ket{B_\ast}\,
    +~( Q-\text{exact} ) \,.
\end{equation}
To summarize, if the inner products
$\langle \, {\cal V} \, | \, ( c_0 - \tilde{c}_0 )
\ket{\mB^{(k)} (\Psi)}$ with $k \ge 2$
vanish for a given regular solution $\Psi$,
the relation~(\ref{fromIan}) which was argued
in~\cite{Ellwood:2008jh} to hold in general
implies that
the closed string state $\ket{\mB (\Psi)}$
coincides with the BCFT boundary state $\ket{B_\ast}$
up to a possible BRST-exact term.
Again, instead of attempting to prove this relation in general,
we explicitly calculate $\ket{\mB (\Psi)}$
for various known analytic solutions in section~\ref{secBCFT}.
We find, surprisingly, that the possible BRST-exact term
\emph{vanishes} for arbitrary $s$,
and we precisely obtain the BCFT boundary state
for all the solutions we consider in section~\ref{secBCFT}.

\section{The boundary state and open-closed vertices}
\label{secocsft}
\setcounter{equation}{0}

In this section we explain
that $\ket{\mB(\Psi)}$ encodes a set of open-closed vertices.
They are generically ``complex'' open-closed vertices
and the reality condition
is satisfied only for certain choices of the propagator strip.
As we will review below, open-closed interactions
 feature prominently in the construction of
open-closed string field theory~\cite{Zwiebach:1990qj,Zwiebach:1997fe}.
Using a natural classical
 sector of this theory and
 \emph{assuming} its physical
 background
independence,
we will do the following:
\begin{itemize}

\item Give a brief proof  that
the observable $W({\cal V}, \Psi)$
encodes the change in one-point functions of on-shell closed
strings
on a disk
upon change of open string background.

\item Give a brief proof that the closed string state
$\ket{\mB(\Psi)}$ has the correct on-shell content,
{\em i.e.},
agrees on-shell with $\ket{B_*}$.

\end{itemize}

\subsection{Open-closed string field theory and background independence}

We consider the following open-closed
string field theory
action:
\be
\begin{split}
\label{ocsft-ocv}
S_{\rm{oc}} (\Psi, {\cal V})  &= S_{\rm{o}}(\Psi) + \Bigl( \bra{B} + \sum_{k=1}^\infty
\bra{B_k} \underbrace{\Psi\rangle  \ldots \ket{\Psi} }_{k-\rm{times}}
 \Bigr)
(c_0 - \tilde c_0) \ket{\mathcal{V}}\\
&~~ + \,\sum_{p=2}^\infty \sum_{k=0}^\infty \bra{B_{k,p}} \underbrace{\Psi\rangle  \ldots \ket{\Psi} }_{k-\rm{times}}
~\underbrace{ \ket{{\cal V}} \ldots \ket{{\cal V}} }_{p-\rm{times}} \,.
\end{split}
\ee
Here
$S_{\rm{o}}(\Psi)$ is the familiar classical open string field theory
action
of  Witten,
which is well known to reproduce correctly {\em all}
amplitudes that involve only external open string states.   The action
$S_{\rm{oc}} (\Psi, {\cal V})$
also includes, in the first line, a series of
terms that are linear in the closed string field $\ket{\mathcal{V}}$
but include
 any number of
open string fields.  These terms are particularly
important for us.
The state $\bra{B}$ is the boundary state of the
original
BCFT represented
by a vanishing open string field.
The state $\bra{B_1}$ arises from an open-closed vertex that couples
one open string to one closed string.  More generally, the state $\bra{B_k}$ couples $k$ open strings to
 one closed string.
In the second line
 in~(\ref{ocsft-ocv})
we include the terms that
couple, via a disk,
two or more closed strings to arbitrary numbers of open strings.
The above action gives the correct amplitudes for processes that
include any numbers of closed
and open strings scattering through a disk
and all states can be off-shell.
Note that the  action $S_{\rm{oc}} (\Psi, {\cal V})$
 includes neither
a closed string kinetic term
nor the classical, genus-zero, closed string interactions.
The closed string field should not be treated as dynamical.

As explained in~\cite{Zwiebach:1990qj,Zwiebach:1997fe}, one can view $S_{\rm{oc}} (\Psi, {\cal V})$   as an action for open
strings propagating in  a nontrivial closed string background.
Indeed, the open string gauge invariance of $S_{\rm{o}}(\Psi)$
can be extended to a gauge invariance of (\ref{ocsft-ocv}) when
the closed string configuration $\ket{\cal V}$
 satisfies the equation of motion
 of pure closed string field theory~\cite{Zwiebach:1992ie}.
It is
a natural generalization of the pure open string field theory,
and the action $S_{\rm{oc}} (\Psi, {\cal V})$ defines
a consistent classical theory.

The action (\ref{ocsft-ocv}) gives the correct amplitudes
mentioned above when the
open-closed vertices
satisfy sets of recursion relations.  In particular, for the vertices
$\bra{B_k}$  with $k=1, 2, \ldots \infty$
that couple to a
single closed string, the recursion
relations
express
the BRST action on $\bra{B_k}$   as
a series of terms in which $\bra{B_{k-1}}$ is glued to the three-open-string vertex:
\be
\label{rec-rel-oc-vert}
\begin{split}
\bra{B_k;\, 1,\ldots , k} \Bigl(Q_c + \sum_{p=1}^k Q^{(p)} \Bigr)
&\sim ~ \bra{V_3;  1,2,x} ~~\bra{B_{k-1}; x, 3, 4, \ldots , k} \, R;x,x'\rangle\\
& ~+ \bra{V_3; 2,3,x} ~~\bra{B_{k-1}; x, 4, \ldots , k,1} \, R;x,x'\rangle\\
&\qquad  \vdots \hskip120pt  \vdots \\
&~+ \bra{V_3; k,1,x}~ \bra{B_{k-1}; x,2,\ldots,
k\hskip-2pt-\hskip-2pt 1}   \,R;x,x'\rangle\,.
\end{split}
\ee
The right-hand side is a sum of $k$ terms in which
 $\bra{B_{k-1}}$ is glued to the three-open-string vertex $\bra{V_3}$
 by the reflector $\ket{R}$. The labels for the external open
 string states have been written out while the closed string label
 remains implicit.
 The various terms arise from the possible cyclic ordering of the
 external state labels.  Open-closed vertices are said to be
 consistent if they satisfy the above recursion relations.

We also note that in (\ref{ocsft-ocv}) the state that couples
linearly to the closed string field and couples to no open string
field is the boundary state $\bra{B}$.
 It  encodes the on-shell one-point
functions of closed string states on a disk.
Let  $\bar \Psi$ denote a {\em fixed} string field solution of $S_{\rm{o}}$
(or of $S_{\rm{oc}} (\Psi, {\cal V}=0)$).
We explore the new background by shifting the open string
field by the classical solution, namely, letting   $\Psi\to \bar \Psi + \Psi$
 in the action.  In the background defined
by $\bar \Psi$  the role of boundary
state is played by the terms in $S_{\rm{oc}}(\bar\Psi + \Psi, {\cal V})$
that couple to ${\cal V}$  and to no  open
string field $\Psi$.  This boundary state $\bra{B_*^{oc}(\bar \Psi)}$ is
therefore defined
by
\be
\label{ocstdkflk99}
\bra{B_*^{\rm{oc}}(\bar \Psi)}\equiv   \bra{B}
+ \sum_{k=1}^\infty
  \bra{B_k}\, (\ket{ \bar\Psi})^k\,.
\ee
Without possible confusion, we revert to our earlier notation where
$\Psi$ denotes a classical solution and thus simply write
\be
\label{ocstdkflk}
\bra{B_*^{\rm{oc}}( \Psi)}\equiv   \bra{B}
+ \sum_{k=1}^\infty
  \bra{B_k}\, (\ket{\Psi})^k\,.
\ee
 By construction, $\bra{B_*^{\rm{oc}}(\Psi)}$ gives
 one-point functions
of closed string states in the background defined
by $\Psi$. In fact, the role of the sum  in (\ref{ocstdkflk}) as a
``energy-momentum tensor" associated
with
the open string field was suggested long ago in~\cite{Zwiebach:1990qj}.
The statement of physical background independence
for the open-closed
string field theory (\ref{ocsft-ocv}) includes the
claim
 that the physical one-point
functions determined by $\bra{B_*^{\rm{oc}}(\Psi)}$ agree
with those of the BCFT  boundary state $\bra{B_*}$ associated with
the
new background:
\be
\label{ifbiholds}
\hbox{Background independence}
\quad \Longrightarrow \quad
\ket{B_*^{\rm{oc}}(\Psi)} =  \ket{B_*}
 \,+~ ( \, Q-\text{exact} \, )
\,.
\ee
The background independence of (\ref{ocsft-ocv}) has
not been proven, but it is motivated by the fact that (\ref{ocsft-ocv}) is a consistent, gauge-invariant
extension of the familiar open string field theory.\footnote{The existing proofs~\cite{Sen:1993mh} of background independence apply to
classical open
string field theory
and to quantum closed string field theory
and have been established for
infinitesimal marginal deformations only. }
In a nutshell,
the claim in (\ref{ifbiholds}) is that the state built
as in (\ref{ocstdkflk}) using  {\em any} consistent set of open-closed
vertices,
{\em i.e.},
vertices
 satisfying~(\ref{rec-rel-oc-vert}),
agrees
on-shell with the boundary state.

\smallskip
A few consistency checks can be readily performed.
The recursion relations (\ref{rec-rel-oc-vert})
 guarantee that
$\bra{B_*^{\rm{oc}}(\Psi)}$ is BRST
closed whenever $\Psi$ is an open string field
theory
solution.  Moreover, under a gauge transformation of the solution
$\Psi$ the state $\bra{B_*^{\rm{oc}}(\Psi)}$ changes
by a BRST-exact term.
Under geometric changes of the open-closed vertices,
the state
$\bra{B_*^{\rm{oc}}(\Psi)}$ also changes
by a BRST-exact term.
For  regular open-closed vertices the local
coordinate for the closed string insertion is non-singular and the inner
product $\bra{B_*^{\rm{oc}}(\Psi)} (c_0 - \tilde c_0) \ket{\phi_c}$ can be evaluated
for
any off-shell closed string state
$\ket{\phi_c}$.

\smallskip
While $\ket{B_*^{\rm{oc}}(\Psi)}$ provides a solution to the problem of
constructing a boundary state associated with the background
represented by the solution $\Psi$,
it is difficult to evaluate it
explicitly for any previously
known set of
open-closed string vertices.
There was, in addition, no evidence that
 there is a choice of open-closed
vertices and
a representative of the solution
for which the resulting
 $\ket{B_*^{\rm{oc}}(\Psi)}$
coincides with
$\ket{B_*}$ without any
BRST-exact term.

We will show that,
 up to the action of $e^{\frac{\pi^2}{s} ( L_0 + \tilde{L}_0 )}$,
the state $\ket{B_*(\Psi)}$ can be viewed as the state
$\ket{B_*^{\rm{oc}}(\Psi)}$ associated with
a set of
 ``complex''
open-closed
vertices.
For the complex
vertices derived from the Schnabl propagator strip
the state
becomes calculable.

\subsection{Recovering the gauge-invariant observables $W(\Psi, {\cal V})$}

The open-closed vertices that define
$\bra{B_*^{\rm{oc}}(\Psi)}$ are not unique.
A one-parameter
 family of vertices was introduced and discussed
  in~\cite{Zwiebach:1992bw}.
The vertices in~\cite{Zwiebach:1992bw}
arise from the minimal area metrics subject to the
constraint that all nontrivial open curves be longer than or equal to $\pi$
while all nontrivial closed curves be longer than or
equal to a parameter $\ell_c$.  For a vertex
 which couples
one open string to one closed string, for example,
one considers a disk
with one closed string puncture
 (in the interior)
and one open string puncture
(on the boundary) and searches for the appropriate minimal area metric.  In this metric the neighborhood of the open string puncture is isometric to a semi-infinite flat strip of width
$\pi$ and the neighborhood of the closed string puncture is isometric to a semi-infinite flat cylinder of circumference $\ell_c$. The coordinate
curves that define the vertex are the natural boundaries of the strip
and the cylinder.
For any value of $\ell_c>0$, one finds a set of
consistent open-closed vertices
 $\bra{B_k}$ and $\bra{B_{k,p}}$.
This family of
open-closed vertices is  parameterized by~$\ell_c$.

\smallskip
It was noted in~\cite{Zwiebach:1992bw} that the construction
simplifies dramatically  as $\ell_c \to 0$.  In this limit, the
open-closed vertex $\bra{B_1}$ alone
constructs, through the Feynman rules, a full cover of the moduli
space of all disks that involve any number of  closed strings and any number
of open strings.  No higher vertices are thus needed and
we can
set them to zero.
The open-closed string field theory (\ref{ocsft-ocv}) thus becomes
\be
\label{ocsft-ocv99}
S_{\rm{oc}}(\Psi, {\cal V}) = S_{\rm{o}}(\Psi) +
\Bigl( \bra{B}+ \bra{B_1} \Psi\rangle \Bigr)
(c_0 - \tilde c_0) \ket{\mathcal{V}} \,  \qquad
 \text{for }~\ell_c \to 0
\,.
\ee
The open-closed vertex $\bra{B_1}$ obtained
for  $\ell_c\to 0$  is  the
vertex used to define $W({\cal V} , \Psi)$. Indeed,  in this
limit the
closed string cylinder disappears
and the open string semi-infinite
strip terminates by having its left and right
 half-string edges
identified,
an operation implemented by the
 open string identity field.
The closed string local coordinate
is singular in this vertex.  As a result the vertex can only be used
for on-shell closed string states.

Not only is the simplification dramatic, but the action (\ref{ocsft-ocv99})
has one important advantage over (\ref{ocsft-ocv}). While it can only
be used for on-shell closed string states,  it reproduces
{\em all} amplitudes involving any number of external open string
and closed string states!  It does so for surfaces of any topology
that have at least one boundary.  In other words,
 it produces all
amplitudes except those of pure closed string
theory~\cite{Zwiebach:1992bw}.  The fact that
(\ref{ocsft-ocv99})
correctly reproduces all on-shell amplitudes
shows that
the {\em on-shell}
contributions
of all the higher open-closed vertices vanish as $\ell_c \to 0$.

For the action (\ref{ocsft-ocv99})  the
associated boundary state $\bra{ B_*^{\rm{oc}}(\Psi)}$
in (\ref{ocstdkflk}) becomes
\be
\bra{ B_*^{\rm{oc}}(\Psi)} =  \bra{B} + \bra{B_1} \Psi\rangle\, \qquad
 \text{for }~\ell_c \to 0
\,.
\ee
It is a boundary state for the
background
described by $\Psi$.  As we noted above, the open-closed
vertex $\bra{B_1}$ defines
the gauge-invariant observable $W({\cal V}, \Psi)$.
Therefore,
\be
 \bra{B_1} \Psi\rangle
(c_0 - \tilde c_0) \ket{\mathcal{V}}
=  -4\pi i \,W ({\cal V} , \Psi)
\ee
for on-shell ${\cal V}$.
It follows from the last two equations that
\be
\bra{ B_*^{\rm{oc}}(\Psi)}
(c_0 - \tilde c_0) \ket{\mathcal{V}}
-\bra{ B}
(c_0 - \tilde c_0) \ket{\mathcal{V}}
=  -4\pi i \,W ({\cal V} , \Psi)\,.
\ee
Using the background independence statement (\ref{ifbiholds}) and the
on-shell property of ${\cal V}$,
we can replace
$\bra{ B_*^{\rm{oc}}(\Psi)}$ by $\bra{ B_*}$ in the above expression.
The result is
\be
\bra{ B_*}
(c_0 - \tilde c_0) \ket{\mathcal{V}}
-\bra{ B}
(c_0 - \tilde c_0) \ket{\mathcal{V}}
=  -4\pi i \,W ({\cal V} , \Psi)\,.
\ee
 This is the claim~(\ref{ellwood-claim}) that was
  discussed
 in~\cite{Ellwood:2008jh}.

\subsection{Generalized open-closed vertices from
$\ket{\mB(\Psi)}$}\label{genopen-closed}

By expanding the closed string state $\ket{\mB(\Psi)}$
in powers of $\Psi$, we can
deduce
a series of couplings of open string states to a single
closed string state.
Indeed, its expansion
(\ref{mBPsiexpansion})
looks like the
set of terms (\ref{ocsft-ocv}) in the open-closed string field theory.
To elucidate this statement and examine its limitations we
will first consider the coupling
encoded in
$\ket{\mB^{(1)}(\Psi)}$,
a coupling of one open string to one closed string.

The closed string local coordinate in $\ket{\mB(\Psi)}$ is obtained
by acting with the scaling operator
 $e^{{\pi^2\over s} (L_0 + \tilde L_0)}$
on the local coordinate defined by
$\oint_s {\cal P}_*(0,s)$,
as can be seen in the definition~(\ref{defmB}).
Since the {\em scale} of the
closed string local
 coordinate
patch will not be relevant to our discussion, we
 focus on  $\oint_s {\cal P}_*(0,s)$, for which the closed string coordinate
 curve is the one traced by the open string midpoint.\footnote{After the
 action of
$\exp \, [ \, {\pi^2\over s} (L_0 + \tilde L_0) \, ]$
 one cannot generally define the closed string coordinate curve.
Naively, it seems to be at the boundary of the disk, but this only
holds if there are no open string insertions,
as in the case of $\ket{B}$.
In the presence of the open string slits
a closed string local coordinate that extends to the boundary would
fail to be continuous at the slits. The expanded local coordinate, however,
can be safely used to insert
closed string states in the Fock space.}

\begin{figure}[tb]
\centerline{
\hbox{\epsfig{figure=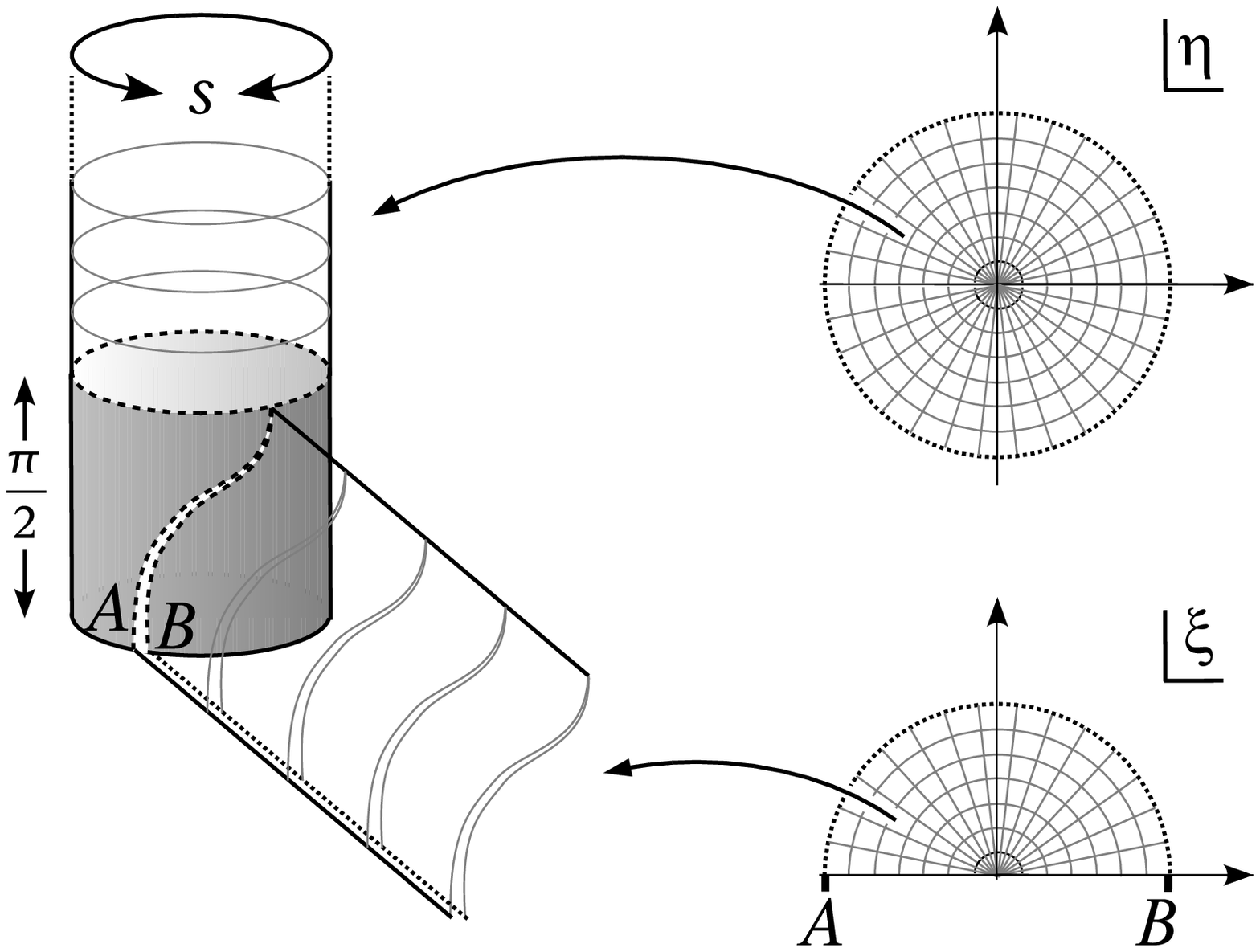, height=6.5cm}}
\hskip 3cm
\hbox{\epsfig{figure=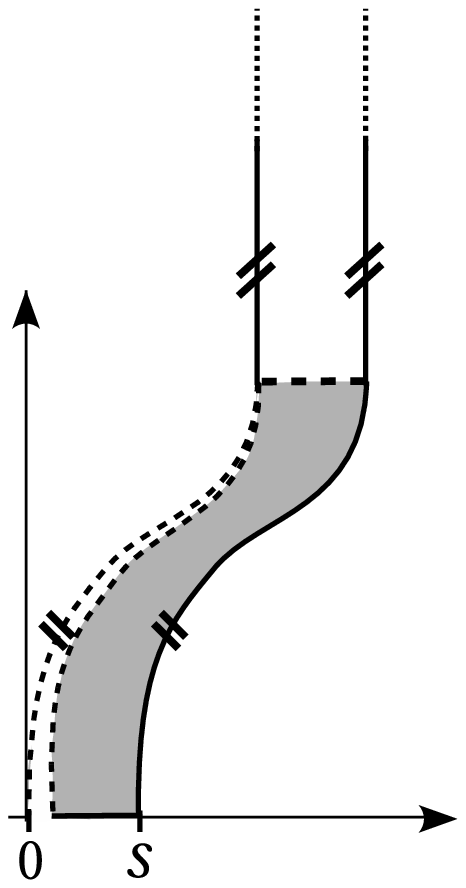, height=6.5cm}}
}
\smallskip
\centerline{\hskip 2.5cm (a) \hskip 8cm (b)}
\caption{
(a)
The open-closed vertex
constructed from
$\oint\Psi\,\PP(0,s)$.
The closed string local coordinate
is mapped from the unit disk of $\eta$
to the semi-infinite cylinder
 of circumference $s$
 bounded by
the geodesic at a distance $\pi/2$ from the
 open string boundary.
The open string local coordinate
is mapped from the half disk of $\xi$
to the strip that is attached at the slit on
the cylinder.
 ~\,(b) The $s\to 0$ limit of the open-closed vertex
constructed from
$\oint\Psi\,\PP(0,s)$.
The two sides of the slit
are identified in the limit
 and the semi-infinite closed string cylinder disappears.
}\label{ocv-fig1}
\end{figure}

 For the term in $\oint_s {\cal P}_*(0,s)$ with one open string field,
let us look at  (\ref{rots1}).
Up to  rotation of the closed
string local coordinate and the 
$b$-ghost integral insertion,
the interaction of
one open string with one closed string
is encapsulated by
\be
\label{flroiev}
\oint_s \Psi  \,  {\cal P} (0,s) \,.
\ee
The picture of the open-closed
 interaction described by (\ref{flroiev})  is shown in Figure~\ref{ocv-fig1}(a).
We can deduce from  Figure~\ref{ocv-fig1}  the limit of this interaction
as $s\to 0$.
 When the circumference of the cylinder becomes
smaller than the horizontal span of the slit it is convenient to redraw the
vertex as in
Figure~\ref{ocv-fig1}(b).
It is then clear that as $s\to 0$ the upper part of
the cylinder disappears and the two sides of the slit are identified.
The closed string insertion is now at the open string midpoint.
Since points at equal heights are identified, this implies the
identification $\xi \to -1/\xi$ on the boundary
$|\xi|=1$ with $\Im(\xi) \geq 0$
of the open string local coordinate frame.  This is precisely
the identification that defines the contraction of an open string field
with the identity.  We have thus recovered the singular ($\ell_c\to 0$)
open-closed vertex that defines the
gauge-invariant observables $W(\Psi, {\cal V})$:
\be
\lim_{s\to 0} \, \bra{{\cal V}\,}
\oint_s \Psi  \,  {\cal P} (0,s)
= W(\Psi, {\cal V}) \,.
\ee
This
relation
holds for an arbitrary choice
of the shape of the propagator strip.
In section~\ref{thestozerorevisited}
we will use the example of
the Schnabl propagator strip
to calculate the open string local coordinate
of the open-closed vertex for finite $s$.
This can be carried out explicitly and one
manifestly recovers the geometry of the open string identity field.

 For a general number of solution insertions, the construction of $\ket{\mB(\Psi)}$ suggests
the following definition of vertices $\bra{B_k}$:
 \begin{equation}\label{mBocv}
 \begin{split}
    (\bra{\Psi})^k\ket{B_k}
    &  =(-1)^k\oints{s}
    \int_{0}^{s} ds_1 \ldots \int_{s_{i-1}}^{s}\hskip-4pt ds_i \ldots
    \int_{s_{k-1}}^{s} \hskip-4pt ds_k \,
    \PR (0, s_1)
    \, \{ {\cal B}_R (s_1), \Psi \} \, \PR (s_1, s_2) \ldots \\
    &  \times \ldots
    \PR (s_{i-1}, s_i) \,
    \{ {\cal B}_R (s_i), \Psi \} \, \PR (s_i, s_{i+1}) \ldots
    \PR (s_{k-1}, s_k) \,
    \{ {\cal B}_R (s_k), \Psi \} \, \PR (s_k, s) \,.
 \end{split}
 \end{equation}
It is important to note that
 these vertices
satisfy the recursion
relations~(\ref{rec-rel-oc-vert}). This is geometrically clear.
Consider $\ket{B_k}$ in~(\ref{mBocv}) for any $k$.
The boundaries in the integration region
consist of configurations where $s_i=s_{i+1}$
for some~$i$.
These configurations represent
the collision of two slits, or more precisely,
two insertions of $\Psi$,  which then couple
through the three-string vertex.  The number of slits is effectively
reduced by one, and they are still integrated over the cylinder as in
 $\bra{B_{k-1}}$.
This is clearly what we see on the right-hand side
of~(\ref{rec-rel-oc-vert}).
The property that the vertices
satisfy
the recursion relations
provides an alternative
way to understand why $\ket{\mB(\Psi)}$ is BRST closed and why
it changes by a BRST-exact term under a gauge transformation
of the open string field.

\bigskip

There is an interesting complication
 that must be addressed.  The open-closed
 vertices~(\ref{mBocv}) typically fail to satisfy
a reality condition.  For each vertex the
reality condition guarantees that the associated
term in the action is real
when the open and closed string fields are real (as they must be).
For the vertex that couples one open string and one closed
string, reality requires an antiholomorphic
automorphism of the surface and the local coordinates that define
the vertex.  For vertices coupling more than one open string
the vertex includes a sum over surfaces. Reality then requires
the overall invariance of the set of surfaces under the action of an antiholomorphic automorphism on each surface.

For the vertex that couples one open string to a closed string,
the automorphism can be described easily
after a map that takes the disk to
the upper-half plane of $u$,
with the open string
puncture at $u=0$ and the closed string puncture at $u=i$.
Since the real axis
of the open string local
frame $\xi$ is mapped to the real
 axis of the upper-half plane, we have
$u(\bar\xi)= \overline{u(\xi)}$.
Reality requires that, in addition,  $u(-\xi) = - u(\xi)$.  As a result
one has
$u (-\bar\xi) = \overline{u(\xi)}$.
This implies that the surface
and the coordinate
curve is invariant under $u \to -\bar u$.
The same condition must
hold for the closed string coordinate curve.\footnote{
In the $\zeta$ frame this condition is invariance under
a reflection about the real axis.  Our closed string coordinate curve is
a circle in the $\zeta$ frame so the condition is clearly satisfied.}
 This is the antiholomorphic
automorphism required by reality.
This reality condition is not satisfied
for open-closed vertices based on non-BPZ-even gauge choices.
We will explicitly demonstrate it
in section~\ref{thestozerorevisited}
for the open-closed
vertex $\ket{B_1}$
associated with
the Schnabl propagator strip.

\begin{figure}[tb]
\centerline{\hbox{\epsfig{figure=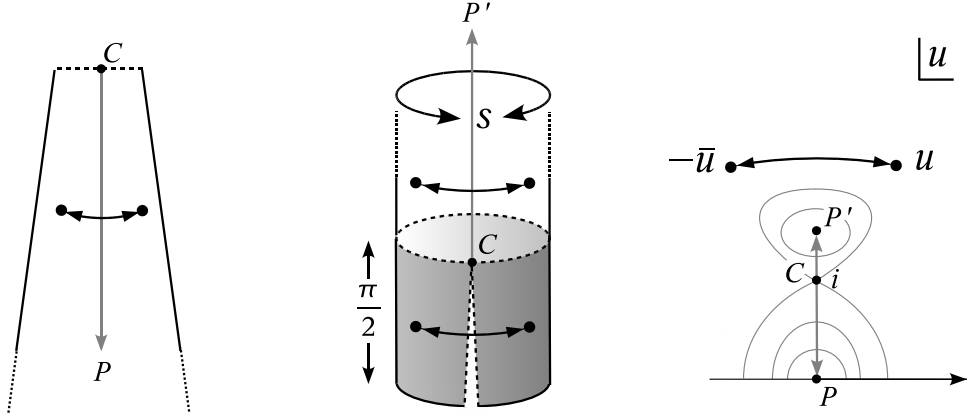, height=5.7cm}}}
\caption{
The vertex
associated with
$\ket{\mB^{(1)}(\Psi)}$
constructed from the propagator strip of a BPZ-even gauge condition.
The vertical slit must be attached to the boundary at the upper end of the
semi-infinite open strip. The antiholomorphic automorphism associated
with reality   is the reflection indicated by
 arrows. On the right side of the figure, the vertex is illustrated in
the
upper-half plane of $u$,
where the automorphism becomes $u \to -\bar u$.
}\label{ocv-fig3}
\end{figure}

On the other hand, we claim
that for propagators that arise from BPZ-even
gauge-fixing conditions, the vertices~(\ref{mBocv}) are
not only consistent
but also real.
 These vertices can thus be used in the action~(\ref{ocsft-ocv}) and
 the corresponding open-closed boundary state $\ket{B_*^{\rm oc}(\Psi)}$
 coincides with $\ket{\mB(\Psi)}$
up to the action of
$\exp \, [ \, {\pi^2\over s} (L_0 + \tilde L_0) \, ]$.
It is simplest to see the reality of
the vertex coupling one closed string to one open string.
A BPZ-even gauge condition implies
that the curve $\gamma(\theta)$
is vertical~\cite{Kiermaier:2007jg},
{\em i.e.},
$\Re(\gamma(\theta))=0$.
Thus the slit on the
 half-propagator strip
is vertical, as shown in Figure~\ref{ocv-fig3}.  Imagine now
 gluing a semi-infinite open string strip to the slit.
The surface has an automorphism
under
a combination of
reflection across the
 line traced by the open string midpoint on
the glued strip
and reflection
across
the slit and its extension up the cylinder. If the surface is mapped to the upper-half plane of $u$
with the open string puncture at $u=0$
  and the closed string puncture at $u=i$,
this automorphism can be presented as the transformation
  $u\to - \bar u$
 that leaves the open string
coordinate curve invariant.  This is the
 automorphism
that guarantees reality.

\begin{figure}[tb]
\centerline{\hbox{\epsfig{figure=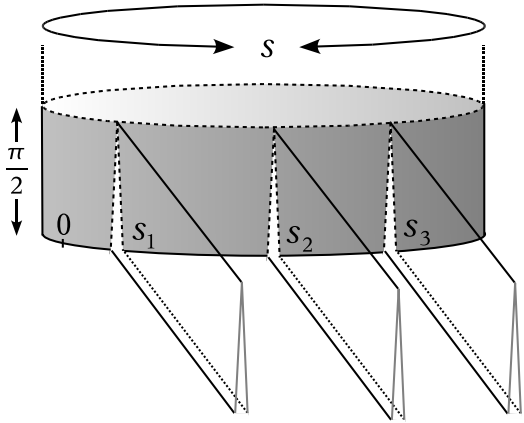, height=5.7cm}}}
\caption{
An open-closed vertex defined by  $\ket{\mB(\Psi)}$ when
the propagator strip is that of Siegel gauge.
}\label{ocv-fig4}
\end{figure}

If we choose Siegel gauge, which is a BPZ-even gauge choice,
the associated open-closed vertices~(\ref{mBocv})
take a simple form, shown in Figure~\ref{ocv-fig4}.  One can visualize
each vertex
as a semi-infinite cylinder of circumference $s$ that has
vertical slits of height $\pi/2$ cut from the bottom edge.\footnote{
This Siegel-gauge geometry was discussed
in the context of the factorization
of the closed string two-point function in~\cite{Ellwood:2008jh}.
}
Semi-infinite
strips of width $\pi$ can be folded and glued isometrically
to
the slits.   The separation between the slits is integrated
over.  It should be noted that
the vertex coupling one open string to one closed string
in this construction is not the same as
that of
the open-closed string field theory built
with $\ell_c=2\pi$. (See Figure~1 in~\cite{Zwiebach:1990qj}.)

\medskip
We have seen that for  propagator strips associated with BPZ-even gauges,
the vertices~(\ref{mBocv}) are consistent and real.
Up to scaling of the closed string coordinate
(which does not change the on-shell content),
the state $\ket{\mB(\Psi)}$ can be thought of as
the state $\ket{B_*^{\rm{oc}}(\Psi)}$ built
with the open-closed vertices (\ref{mBocv}).
These vertices can
be supplemented with vertices coupling
multiple closed strings to construct
a complete action
in the form of~(\ref{ocsft-ocv}).
If we then assume
background independence, as stated in (\ref{ifbiholds}),
we conclude that  $\ket{\mB(\Psi)}$ has the on-shell content of $\ket{B_*}$
for BPZ-even gauges.  Since
$\ket{\mB(\Psi)}$
changes by a BRST-exact term
when the choice of propagator strip
is modified,
this implies that
$\ket{\mB(\Psi)}$ always has the correct on-shell content.

\medskip
In summary,  propagators
corresponding to
non-BPZ-even
gauge conditions result
in states $\ket{\mB(\Psi)}$ that do not
give
real open-closed vertices.
In those cases $\ket{\mB(\Psi)}$ {\em cannot} be viewed
as a boundary state $\ket{B_*^{\rm{oc}}(\Psi)}$ of open-closed string field theory.
Indeed,
we use the propagator strip of Schnabl gauge,
which is a non-BPZ-even gauge choice,
for the explicit calculations of $\ket{\mB(\Psi)}$
in the following sections.
In this sense our proposal of $\ket{\mB(\Psi)}$
goes beyond the framework
of open-closed string field theory.
Since it is based on complex open-closed
vertices, one may wonder if
the state $\ket{\mB(\Psi)}$ based on Schnabl gauge
is real. In fact, our arguments
only guarantee that
its contraction
with on-shell closed string
states is real, but
the contraction
with off-shell closed string states could be complex.
 Of course, it is also possible
 that a state $\ket{\mB(\Psi)}$  that is not real for
arbitrary real open string states
turns out to be real for
open string states satisfying
the equation of motion.
We will find that
the state $\ket{\mB(\Psi)}$ based on Schnabl gauge
is indeed real for the solutions we consider
in section~\ref{secBCFT}.

\section{Regular and calculable boundary states}
\label{secwedge}
\setcounter{equation}{0}

\subsection{Simplifications for the Schnabl propagator and wedge-based solutions}

For any choice of parameter $s$, propagator
gauge-fixing condition $\BB$,
 and classical solution $\Psi$,
we can construct the closed string state $\ket{\mB(\Psi)}$.
In general,
however, it is difficult to calculate $\ket{\mB(\Psi)}$ explicitly
because
the gluing of insertions of classical solutions to the slits
in the $w$ frame
generically
requires calculations of correlation functions
on a complicated Riemann surface.
A drastic simplification occurs if we choose Schnabl's gauge condition $\BB=B$ with\footnote{
The simplification also occurs for
$\BB\propto (B+\alpha B^\star)$ with $\alpha\neq1$.}
\begin{equation}
 B = \oint \frac{d \xi}{2 \pi i} \,
 v_{\cal S}(\xi)\, b(\xi) \,,
\end{equation}
where
\begin{equation}\label{Schnablf}
 v_{\cal S}(\xi)=\frac{f(\xi)}{f'(\xi)}\,,\qquad
f(\xi) = \frac{2}{\pi} \, \arctan \xi \,.
\end{equation}
We use the doubling trick in this section and in the next section.
An explicit
mode expansion
of $B$ is
\begin{equation}
    B=b_0+2\sum_{j=1}^\infty\frac{(-1)^{j+1}}{4j^2-1}b_{2j}\,.
\end{equation}
If we choose $\BB=B$ and
a classical solution $\Psi$ based on wedge states,
we can use the results of~\cite{Kiermaier:2008jy} to map the resulting surface to an
annulus
and calculate
the state $\ket{\mB(\Psi)}$ explicitly.
In this section we assemble
the main ingredients necessary for this calculation.

We define the wedge region ${\cal W}_\alpha$ by
the semi-infinite strip on the upper-half plane
of~$z$
between the vertical lines
$\Re(z)=-\frac{1}{2}$ and $\Re(z)=\frac{1}{2}+\alpha$
with these lines identified by translation.
The wedge state $W_\alpha$ is defined by
\begin{equation}
\label{wedge}
\langle \, \phi, W_\alpha \, \rangle
= \langle \ f \circ \phi (0) \, \rangle_{{\cal W}_\alpha}
\,.
\end{equation}
Here and in what follows we denote a generic open string state
in the Fock space by $\ket{\phi}$
and its corresponding operator by $\phi (0)$.
When a solution is made of wedge states with operator insertions,
we call it a wedge-based solution.

\begin{figure}[tb]
\centerline{\hbox{\epsfig{figure=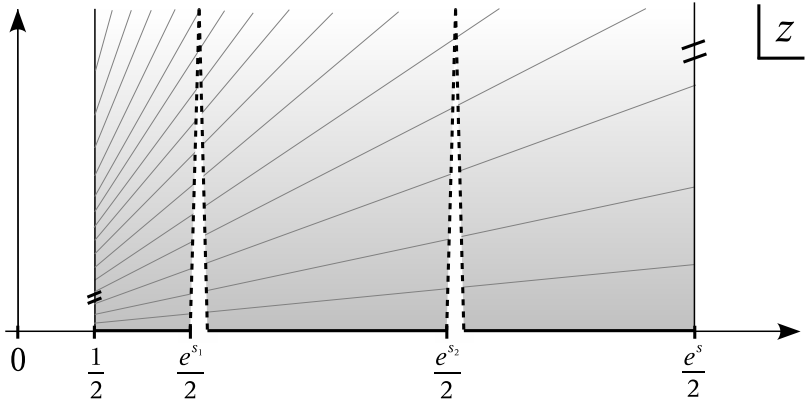, height=5.7cm}}}
\caption{
Illustration of~(\ref{mBm}) for $k=2$
before inserting the states $A_{\alpha_1}$ and $A_{\alpha_2}$.
 The two slits are depicted as dashed lines in the figure.
The grey lines illustrate the rescaling of the parameterization from the left boundary to the right boundary. The double lines indicate the identification between these two boundaries.
}\label{figSchnablslits}
\end{figure}

We now map the $w$-frame geometry with its parameterized slits to a  frame which is convenient for the propagator choice $\BB=B$.
It is related to the $w$ frame via
\begin{equation}
    z=\frac{1}{2}\,e^{w}\,.
\end{equation}
The $z$ frame is closely related to the familiar sliver frame. In fact, the image of the curve $\gamma(\theta)$ in the $z$ frame coincides with the sliver-frame coordinate line $f(e^{i\theta})$. More generally,
$s_i+\gamma(\theta)$, which parameterizes
a slit in a half-propagator strip or its boundary,
becomes vertical in this frame and is located at $\Re(z)=\frac{1}{2}e^{s_i}$. The parameterization in the $z$ frame is given by
\begin{equation}\label{gammazpara}
    s_i+\gamma(\theta)\,\to\,z=e^{s_i}f(e^{i\theta})=\frac{1}{2}\,e^{s_i}
    +i \, e^{s_i}\biggl[ \, \frac{2}{\pi}
    \, {\rm arctanh} \,
    \Bigl( \, \tan\frac{\theta}{2} \, \Bigr) \, \biggr] \,.
\end{equation}
Since the slits are infinite,
 the closed string boundary is
\emph{hidden} at $z\to i\infty$.
This property can be traced back to the fact that Schnabl gauge is not a \emph{regular} linear $b$ gauge. In fact, the vector field $v_{\cal S}(\xi)$ vanishes at $\xi=i$ and thus violates the condition~(\ref{regularv}) at the open string midpoint.
This simplifies the analysis in Schnabl gauge, but all manipulations of surfaces must be justified by regularizing the propagator
and taking the Schnabl limit. Fortunately,
all manipulations that our analysis requires were already justified in~\cite{Kiermaier:2008jy}, so we can simply apply the prescriptions developed there.
In particular, it is important to understand the contour
of the integrals $\BB_R (t)$ and $\LL_R (t)$ in the $z$ frame
as a limit of a regulated curve.
We denote the contour after using the doubling trick by $C(t)$:
\begin{equation}\label{SchnablBRLR}
\BB_R (t) \to \int_{C(t)} \frac{dz}{2 \pi i} \,
 z \, 
b(z) \,, \qquad
\LL_R (t) \to \int_{C(t)} \frac{dz}{2 \pi i} \,
 z \, T(z) \,.
\end{equation}
 In the regularized analysis, the closed string boundary
 is a finite segment on the imaginary $z$ axis, and $t$ parameterizes the endpoint of the contour $C(t)$ on that line segment.\footnote{
 In the notation of~\cite{Kiermaier:2008jy}, the endpoints of the contour $C(t)$ are $\pm  ie^t\Lambda$. In the Schnabl limit we have $\Lambda\to\infty$.}
 In the Schnabl limit, the location of the closed string boundary diverges to $i\infty$.
The contour $C(t)$
 in this limit
naively runs from $-i \, \infty$ to $i \, \infty$
along
 the vertical line $\Re(z)=\frac{e^t}{2}$
and
 the $t$ dependence of its endpoints on the imaginary axis is hidden.
As we discussed in section~\ref{sec2}, however, it
 \emph{does}
depend on $t$ even in the limit, namely,
 \begin{equation}
  \int_{C(t)} \frac{dz}{2 \pi i} \,
   z \, 
  b(z) \,
  -\int_{C(t')} \frac{dz}{2 \pi i} \,
   z \, 
  b(z) \,
  \ne 0
\end{equation}
for $t' > t$ even when there are no operator insertions
between the two contours.
 This is the $z$-frame representation of
 \begin{equation}
      \BB_R(t)\,\PR (t,t')-\PR (t,t')\,\BB_R(t')\neq 0
 \end{equation}
 which follows from the inequality~(\ref{nocommutator}).
Let us now consider
a calculation of
\begin{equation}\label{mBm}
(-1)^k
\oint_s\,
\prod_{i=1}^k
\Biggl[\int_{s_{i-1}}^{s} \!\!ds_i\, \PR (s_{i-1}, s_i)
\, \{ {\cal B}_R (s_i), A_{\alpha_i} \}\Biggr] \PR (s_k, s)\,,
\end{equation}
where $A_{\alpha_i}$ is
 a Grassmann-odd state
made of the wedge state $W_{\alpha_i}$ with operator
insertions,
and
$s_0=0$.
Before gluing the states $A_{\alpha_i}$
to the parameterized slits, the total surface is located in the region
\begin{equation}
    \frac{1}{2}\leq\Re(z)\leq \frac{1}{2}\,e^{s}\,,
\end{equation}
and the parameterizations of its vertical boundaries are
given by
$f(e^{i\theta})$ for the left boundary
and $e^s f(e^{i\theta})$ for the right boundary.
These boundaries are glued by the operation $\oints s$ which forms a closed string state from the surface, and this gluing is
compatible with the identification $z\sim e^s z$. See Figure~\ref{figSchnablslits}.
The map to the annulus frame $\zeta$ is
\begin{equation}\label{zetaz}
    \zeta=\exp\Bigl(\frac{2\pi i}{s}\,\ln 2z\Bigr)\,,
\end{equation}
which is compatible with the identification $z\sim e^s z$.

\begin{figure}[tb]
\centerline{\hbox{\epsfig{figure=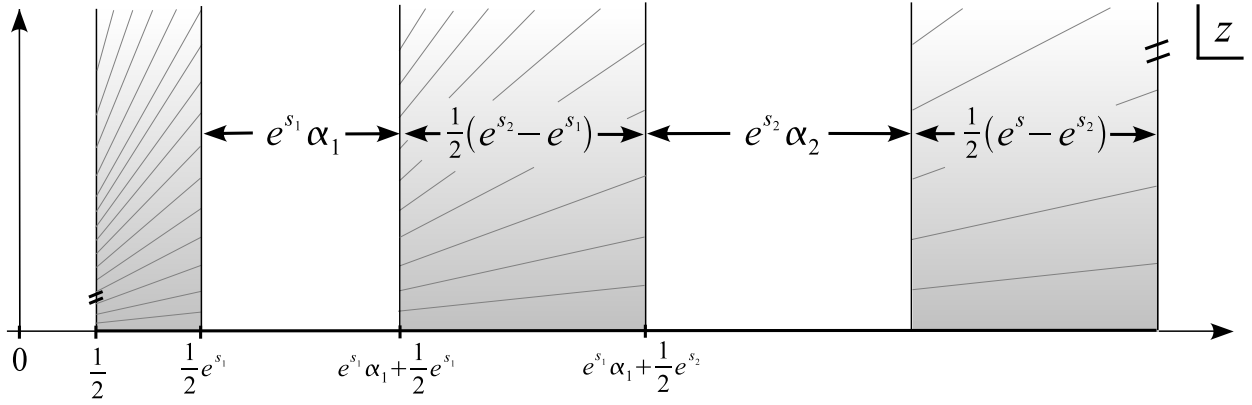, height=5.7cm}}}
\caption{
Illustration of~(\ref{mBm}) for $k=2$
with gaps of width $e^{s_1}\alpha_1$ and of width $e^{s_2}\alpha_2$
inserted at the two slits.
Compare this with  Figure~\ref{figSchnablslits}.
The states $A_{\alpha_1}$ and $A_{\alpha_2}$
are then mapped into these gaps.
}
\label{figSchnablB}
\end{figure}

{}From the parameterization~(\ref{gammazpara})
it is obvious how to
insert the wedge-based state $A_{\alpha_1}$
to the slit at
$\Re(z)=\frac{e^{s_1}}{2}$: simply translate
the remaining surface
in the region
$\Re(z)>\frac{e^{s_1}}{2}$ horizontally to the right
by $e^{s_1} \alpha_1$
and map
$A_{\alpha_1}$ from
its sliver frame $z^{(1)}$ used in~(\ref{wedge})
to the resulting gap in the $z$ frame via
\begin{equation}
    z=e^{s_1}z^{(1)}\,.
\end{equation}
The next slit is now located at
$\Re(z)=e^{s_1}\alpha_1+\frac{1}{2}e^{s_2}$.
We translate the remaining surface
in the region $\Re(z)>e^{s_1}\alpha_1+\frac{1}{2}e^{s_2}$
by $e^{s_2} \alpha_2$,
and map the state $A_{\alpha_2}$
into the resulting gap via
\begin{equation}
    z=e^{s_1}\alpha_1+e^{s_2}z^{(2)}\,.
\end{equation}
See Figure~\ref{figSchnablB}.
The construction iterates.
For the insertion of $A_{\alpha_i}$
after having inserted the previous $i-1$ states,
the slit is located at
\begin{equation}
    \Re(z)=\sum_{j=1}^{i-1}e^{s_j}\alpha_j+\frac{1}{2}\,e^{s_i}\,,
\end{equation}
and we translate the remaining surface to the right of the slit by
$e^{s_i}\alpha_i$.
Then we map
the state $A_{\alpha_i}$
into the resulting gap via
\begin{equation}\label{zSPsini}
    z=\sum_{j=1}^{i-1}e^{s_j}\alpha_j+e^{s_i}
z^{(i)}\,.
\end{equation}
At the end of the process, the whole surface corresponding to~(\ref{mBm}), which we denote by $\Sigma$, is located in the $z$ frame in the region
\begin{equation}
    \frac{1}{2}\leq\Re(z)\leq
    \sum_{j=1}^{k}e^{s_j}\alpha_j+\frac{1}{2}\,e^{s}\,.
\end{equation}
The parameterization of the identified left and right boundaries of this resulting surface $\Sigma$ are not related by scaling $z\sim e^s z$\,, so we cannot map $\Sigma$ directly to the annulus frame via~(\ref{zetaz}).
To restore this relation, we use the prescription
given in section 6.1 of~\cite{Kiermaier:2008jy}.
We shift the entire surface $\Sigma$ horizontally by
\begin{equation}\label{a0}
    a_0=\frac{1}{e^s-1}\sum_{j=1}^{k}
    e^{s_j}\alpha_j\,.
\end{equation}
With this value of the shift,
$\Sigma$ is then located
in the region
\begin{equation}
    \frac{1}{2}+a_0\leq\Re(z)\leq e^s\Bigl(\frac{1}{2}+a_0\Bigr) \,,
\end{equation}
and the gluing of the left to the right boundary of $\Sigma$ is now compatible with the identification $z\sim e^s z$.
This translated frame is called {\it the natural $z$ frame.}
The total
map from
the coordinate $z^{(i)}$ of the wedge surface
on which $A_{\alpha_i}$ is defined to the natural $z$ frame is
a combination of the map~(\ref{zSPsini})
and a horizontal translation by $a_0$.
It is thus given by
 \begin{equation}\label{znatPsini}
    \boxed{\phantom{\biggl(}
    z=\ell_i+e^{s_i} z^{(i)}\,,
    ~~}
 \end{equation}
 where
 \begin{equation}
 \label{ell-definition}
 \ell_i = \sum_{j=1}^{i-1} \alpha_j \, e^{s_j} + a_0 \,, \quad
 \ell_1 = a_0 \,.
 \end{equation}
 It is consistent with the identification $z\sim e^s z$ in the natural $z$ frame to
map $\Sigma$ to the annulus frame $\zeta$ via~(\ref{zetaz}).
The gluing to the closed string coordinate patch is unaffected by our manipulations in the $z$ frame, as was shown in~\cite{Kiermaier:2008jy}. One may worry that the horizontal translations of the surfaces may have resulted in a relative rotation in the $\zeta$ frame. This is not the case, as can be easily seen from the regularized analysis of~\cite{Kiermaier:2008jy}.
 We can thus analytically map the surface that defines the closed string state~(\ref{mBm}) to an annulus that has
exactly the same
modulus as the annulus that defines $\oint_s\PP(0,s)$.
 In particular,
all closed string
surface states contributing to $\ket{\mB(\Psi)}$
for wedge-based solutions are
represented
on exactly the {\em same  Riemann surface}.
This remarkable property of the Schnabl propagator strip will be crucial
for our explicit calculations in
section~\ref{secBCFT}.

\medskip

The $b$-ghost line integrals $\BB_R(s_i)$ in~(\ref{mBm}) have a simple representation in the natural $z$ frame. Indeed,
 mapping the line integrals $\BB_R(s_i)$ from their initial $z$-frame representation~(\ref{SchnablBRLR}) in the presence of slits to their final location in the natural $z$ frame, we find
\begin{equation}\label{anticommBA}
{}- \{ {\cal B}_R (s_i), A_{\alpha_i} \}
    ~\to~
-\int_{C(s_i)}
\frac{dz}{2\pi i}\,(z-\ell_i)b(z)\,[\ldots]\,
- \,[\ldots]\,
\int_{C(s_i)}
\frac{dz}{2\pi i}\,(z-\ell_{i+1})b(z)\,.
\end{equation}
The dots $[\ldots]$ in~(\ref{anticommBA}) represent
the operator insertions
for $A_{\alpha_i}$.
Note that the difference
in the endpoints of the contour $C(s_i)$
generated by translation of the contour
in the direction of the real axis vanishes
in the Schnabl limit as discussed in~\cite{Kiermaier:2008jy}.
Since both contours in~(\ref{anticommBA}) have the same endpoint
on the closed string boundary,
the contours can be connected.
{}From the relation
$\ell_{i+1} = \ell_i +e^{s_i} \alpha_i $ we find
\begin{equation}\label{anticommBAconnect}
\boxed{\phantom{\Biggl(}
 {}-\{ {\cal B}_R (s_i), A_{\alpha_i} \}~\to~
    \oint\frac{dz}{2\pi i}\,(z-\ell_i)b(z)\,[\ldots]\,
+ \,e^{s_i}\alpha_i\,[\ldots]\,\BB_R^+\,, ~}
\end{equation}
where
the contour encircles the operators $[\ldots]$ counterclockwise and
\begin{equation}\label{defBRplus}
    \BB_R^+=
\int_{-i \infty}^{i\infty}
\frac{dz}{2\pi i}\,b(z) \,.
\end{equation}
 We do not write the $t$ dependence on the endpoints of the
  integration contour of $\BB_R^+$
 because the integral does not depend on the choice of $t$.
 To see this, note that
 \begin{equation}
    \LL_R^+\equiv\{Q,\BB_R^+\}
\end{equation}
 generates horizontal translations in the $z$ frame. It was shown
 in~\cite{Kiermaier:2008jy} that the closed string boundary is unaffected by such translations in the Schnabl limit. It follows that the integrand in~(\ref{defBRplus}) vanishes along the closed string boundary.\footnote{This can also be explicitly confirmed by mapping~(\ref{defBRplus}) to the annulus frame $\zeta$ and evaluating the integrand at
 $|\zeta|=e^{-\frac{\pi^2}{s}}$.}
Thus the operator $\BB_R^+$ does not
 depend on the choice of $t$.
The position of this insertion  is given implicitly by the
operator ordering in the correlator.

\medskip

We have assembled all the ingredients required to explicitly calculate  $\ket{\mB(\Psi)}$ with $\BB=B$ for wedge-based solutions. In fact,
the map from the wedge surfaces on which the solutions are defined to
the natural $z$ frame
are explicitly given
in~(\ref{znatPsini})
and the positions and conformal factors of operator insertions on the surface $\Sigma$ can be explicitly calculated.
The $b$-ghost line integrals $\BB_R(s_i)$ have the simple representation~(\ref{anticommBAconnect}) in the natural $z$ frame.
Finally, the map from the natural $z$
frame
to the annulus frame is explicitly given in~(\ref{zetaz}).
We conclude that the boundary state $\ket{\mB(\Psi)}$ is explicitly calculable for
wedge-based
solutions if we use
the half-propagator strips
associated with Schnabl's $B$. We will now use this
knowledge to illustrate
the vanishing of terms with more than one solution insertion in the limit $s\to 0$.

\subsection{The $s\to0$ limit revisited}\label{thestozerorevisited}

In this subsection we first examine the open-closed vertex
encoded in
$\ket{\mB^{(1)} (\Psi)}$
constructed using the Schnabl propagator strip.
The open string local coordinate
can be calculated explicitly and one can confirm that this open-closed
vertex is not real
for finite $s$.
We can also confirm
that
the reality condition
is recovered in the $s \to 0$
limit,  where  the vertex becomes the familiar singular one
used in the definition of $W({\cal V}, \Psi)$
discussed in \subs\ref{genopen-closed}.
We then consider $\ket{\mB^{(k)} (\Psi)}$ with $k \ge 2$
constructed using the Schnabl propagator strip
and argue that they
vanish
in the $s\to 0$ limit.

\medskip
As explained earlier
in~(\ref{flroiev}),
the geometrical configuration
of $\ket{\mB^{(1)} (\Psi)}$
can be reduced to that of
$\oint_s \Psi  \,  {\cal P} (0,s)$.
To examine the reality of this vertex, it is sufficient to consider
the case where the open string field is a generic state $\ket{\phi}$
in the Fock space.
Since a state in the Fock space is
a wedge state of unit width with a local operator insertion,
the surface associated with $\phi \,  {\cal P} (0,s)$
has total width ${1\over 2} (e^s+1)$ in
the $z$ frame.
The necessary shift
$a_0$ for the natural $z$ frame from the formula~(\ref{a0})
with $k=1$, $\alpha_1=1$, and $s_1=0$ is
 \be
 \label{twodf}
 a_0
=  {1\over e^s-1}\,.
 \ee
 For small $s$, it
can be expanded as
 \begin{equation}
    a_0   \simeq  {1\over s} -{1\over 2} + \mathcal{O}(s)\,.
 \end{equation}
The position $z_p$ of
 the operator insertion associated with $\ket{\phi}$
in the natural $z$ frame is
\be
z_p= 1+ a_0   = \, {e^s \over e^s-1} \,,
\ee
and the local coordinate for the open string
is given by
\be
z(\xi) = z_p + f(\xi)
\ee
 with $f(\xi)$ defined in~(\ref{Schnablf}).
Since the natural $z$ frame with the identification $z \sim e^s z$
is singular in the limit $s \to 0$,  it is
convenient to map this
to the annulus $\zeta$ frame.
We then map it further to the upper-half plane
to compare it with $W({\cal V}, \Psi)$.
The local coordinate for the open string
in the $\zeta$ frame is
\be
\zeta (\xi)
= \exp \biggl[ \, {2\pi i\over s} \,
( \, \ln 2 z(\xi)- \ln 2 z_p \, ) \, \biggr]
= \exp \biggl( \, {2\pi i\over s}
\ln \Bigl[ \, 1+ {f(\xi)\over 1+ a_0} \,
\Bigr] \, \biggr)\,,
\ee
where we rotated the $\zeta$ frame
to satisfy $\zeta (0) = 1$ for convenience.
The surface associated with the state
$\oint_s \phi  \,  {\cal P} (0,s)$
in the $\zeta$ frame
is an annulus with inner radius $r= e^{-{\pi^2\over s}}$.
We further map this to the upper-half plane of $u$
by the following conformal transformation:
\be
u=  i  \,\, {1-\zeta \over 1+ \zeta} \,.
\ee
The puncture is located at $u=0$
and the outer boundary of unit radius in the $\zeta$ frame
is mapped to the real axis.
For small $s$,
the inner boundary of the annulus is
mapped to
a small, almost circular closed curve around $u=i$
in this frame.
The local coordinate now takes the form
\be
\label{zofxi-coor}
u(\xi) =  \tan \Bigl( {\pi \over s}  \ln \Bigl[ 1+ {f(\xi)\over 1+ a_0}
\Bigr]  \Bigr)  \,.
\ee
When $s$ is small and as $\theta \to \pi/2$,
$u(e^{i\theta})$ is a curve
that rotates
along
a small circle around the point $u= i$.
As discussed
in \subs\ref{genopen-closed}, the reality
of the open-closed vertex requires  $u(-\xi) = - u(\xi)$.
While  $f(-\xi) =
- f(\xi)$, the coordinate $u(\xi)$ in (\ref{zofxi-coor})
does not satisfy
$u(-\xi) = - u(\xi)$.
Thus the
open-closed vertex is not real for finite $s$.

Let us now consider the limit $s \to 0$.
We can make  use of the expansion
\be
\ln \Bigl[ 1+ {f\over 1+ a_0}
\Bigr]  \simeq  s f - {1\over 2} s^2 (f+ f^2)  + {\cal O}(s^3)\,,
\ee
which is valid for any
 $|\xi | \leq1$ satisfying $|\xi-i|>\epsilon$ with fixed
$\epsilon>0$
because $|f(\xi)|$ is bounded in this region.
We then find
\be
\label{zofxi-coor-99}
\begin{split}
u(\xi) &=  \tan \Bigl(\pi f -{\pi\over 2} s (f+ f^2) + {\cal O}(s^3) \Bigr)
  \,, \\
  &=  \tan (\pi f) -{\pi\over 2} s (f+ f^2) \sec^2 (\pi f)
  + {\cal O}(s^3)
  \,.
  \end{split}
\ee
The limit $s\to 0$ is perfectly well defined and we find
\be
\label{oienvg}
\lim_{s\to 0}  u(\xi) = \tan (\pi f) = \tan ( 2
\arctan
(\xi)) = {2\xi\over 1-\xi^2}\,.
\ee
This result is valid for
the region $|\xi-i|>\epsilon$ on the unit disk, as stated above.
We can see
that (\ref{oienvg})  is, in fact,  the local coordinate
map~(\ref{identity-map})
for the identity state. The closed string coordinate curve becomes
a vanishingly small curve around $u=i$ in this limit, and
we have
recovered, as expected, the open-closed vertex used to define
the observable $W({\cal V}, \Psi)$.
In this limit, of course, the open-closed vertex is real.

\medskip

Let us now examine the limit $s\to 0$ of
the state $\ket{\mB^{(k)} (\Psi)}$ in~(\ref{defmBexp})
for arbitrary $k$.
Naively, one might expect that only the leading term
$k=0$,
the boundary state $\ket{B}$,  survives in this limit because
the integration region for every Schwinger
parameter
collapses as $s\to0$. We have seen in
section~\ref{secs0} that this reasoning fails
for $\ket{\mB^{(1)} (\Psi)}$.
This is the open-closed vertex examined above.
In this case, the integral over $s_1$
covers
the entire range of the rotational modulus
of the associated surface
even in the limit $s\to0$.
See~(\ref{rots1}).
For general choices of propagator strips,
the Schwinger parameters $s_i$ represent
moduli of an annulus
with insertions of the classical solution.
The $b$-ghost line integrals
$\BB_R(s_i)$
provide the measure for this integration over moduli.
For the
Schnabl propagator strip and wedge-based
solutions, the modulus
of the annulus depends only on $s$,
 and the Schwinger parameters $s_i$
represent position moduli for the insertions.
It is instructive to  calculate explicitly the position moduli
by examining the maps of solution insertions to the annulus. We will show that in the limit $s\to0$, the integration over all Schwinger parameters
 $s_1,\ldots,s_k$
only covers a one-dimensional subspace of the
position moduli.
 We will argue that this is
consistent with the vanishing of
$\ket{\mB^{(k)} (\Psi)}$ with $k \ge 2$
in the limit $s\to0$ for regular solutions.

Consider any point $t^{(i)}$
on the boundary of
the wedge surface of width $\alpha_i$
for the state $A_{\alpha_i}$ in~(\ref{mBm}).
We have
 $\half\leq t^{(i)} \leq\half+\alpha_i$.
By~(\ref{zetaz})
 and~(\ref{znatPsini}),
this term is mapped on the annulus frame to the point
\begin{equation}
    \zeta_i=e^{i\theta_i}\qquad\text{ with }\qquad
    \theta_i=\frac{2\pi}{s}\,\ln 2\Bigl[a_0+
     \sum_{j=1}^{i-1}e^{s_j}\alpha_j
    +e^{s_i}t^{(i)}\Bigr]\,.
\end{equation}
 It will be useful to consider the angular separation between a point
  $t^{(p)}$ on $A_{\alpha_p}$ and a point $t^{(q)}$ on $A_{\alpha_q}$
 in the limit $s\to0$. We choose  $q>p$
 for definiteness. Recalling $a_0=\OO(s^{-1})$, we obtain
\begin{equation}\label{thetaki}
\begin{split}
     \theta_q-\theta_p
     &=\frac{2\pi}{s}\,\ln
     \Biggl[\frac{a_0+\sum_{j=1}^{q-1}e^{s_j}\alpha_j + e^{s_q}t^{(q)}}
     {a_0+\sum_{j=1}^{p-1}e^{s_j}\alpha_j+e^{s_p}t^{(p)}}\Biggr]\\
     &=\frac{2\pi}{s}\,\ln \Biggl[1+\frac{1}{a_0}\Bigl(t^{(q)}-t^{(p)}+\sum_{j=p}^{q-1}\alpha_j\Bigr)+\OO(s^{2})\Biggr]\\
     &=2\pi\,\frac{t^{(q)}-t^{(p)}+\sum_{j=p}^{q-1}\alpha_j}{\sum_{j=1}^{k}\alpha_j}+\OO(s)\,,
\end{split}
\end{equation}
where we used the explicit expression for $a_0$ from~(\ref{a0}) in the last step and,
since
 $0\leq s_i \leq s$ for all $i$, we simply treated all ${\cal O}(s_i)$
terms as ${\cal O}(s)$ terms.
We also assumed that the sum of
 $\alpha_i$'s
in the denominator above
is greater than zero.  This is clearly satisfied if
 $\alpha_i>0$ for all $i$.
We conclude from~(\ref{thetaki}) that the angular separation between
 $t^{(p)}$ and $t^{(q)}$
on the unit circle  $|\zeta|=1$
freezes up
in the limit $s\to0$:
it becomes independent of all Schwinger parameters $s_i$. Thus the relative positions
of all points mapped to the open string boundary of the annulus
are fixed in this limit.  As the closed string coordinate patch in the annulus frame is
also independent of all $s_i$
and is only a function of $s$,
the only remaining modulus that can vary as
we integrate over the $s_i$ is the relative angle of all operator insertions with respect to the closed string coordinate patch. We conclude that
 a  $k$ dimensional integral over the Schwinger parameters $s_i$ is confined to a one-dimensional subspace of moduli in the limit $s\to0$. As the
  line integrals $\BB_R(s_i)$
 supply the correct measure for the
 $k$-dimensional integration and the integration region becomes degenerate,
 this is consistent with the vanishing of the integrals with $k\geq 2$ in the limit $s\to0$.

 We can apply this analysis to the state
 $\ket{\mB^{(k)}(\Psi)}$
with
$k\geq2$ for solutions $\Psi$ which are defined on wedge
  surfaces  of nonvanishing width.
   Again, as in \subs\ref{secs0},
   we assume that the solution $\Psi$ satisfies some regularity condition that ensures
a nonsingular behavior
in the limit $s\to0$. Then we conclude from the above argument that all $\ket{\mB^{(k)}(\Psi)}$ with $k\geq2$ vanish in this limit.
This explicit analysis of the $s_i$ dependence of the maps to the annulus
for the
Schnabl gauge propagator thus illustrates the more general
argument
of section~\ref{secs0}.

\section{The BCFT boundary state from analytic solutions}
\label{secBCFT}
\setcounter{equation}{0}

In this section we
explicitly calculate $\ket{\mB(\Psi)}$
constructed using the Schnabl propagator strips
for various known wedge-based
solutions.
We analyze Schnabl's tachyon vacuum solution
and the two known analytic solutions for marginal deformations
with regular operator products.
We find that
\begin{equation}
\ket{\mB(\Psi)}\,=\,0\quad\text{ for all }s
\end{equation}
in the case of the tachyon vacuum solution and
\begin{equation}\label{mB=BBCFT}
    \ket{\mB(\Psi)}\,=\,\ket{\mB}\quad\text{ for all }s
\end{equation}
with no additional BRST-exact term
in the case of marginal deformations.

\subsection{Schnabl's solution for tachyon condensation}
\label{sec7.1}

Schnabl's solution $\Psi_{\cal S}$
for tachyon condensation is given by~\cite{0511286}
\begin{equation}\label{definePsiS}
\Psi_{\cal S} = \lim_{N \to \infty} \biggl[ \,
\sum_{n=0}^N
\psi'_n - \psi_N \, \biggr] \,, \qquad
\psi'_n \equiv \frac{d}{d n}
 \,
\psi_n  \,,
\end{equation}
where
\begin{equation}
\langle \phi, \psi_{n-1} \rangle
= - \langle \, f \circ \phi (0) \, c(1) \, {\cal B}^+_R \, c(n) \,
\rangle_{{\cal W}_n}
\quad\text{ for }~n>1
\end{equation}
and
\begin{equation}
\label{psi_0}
\langle \, \phi, \psi_0 \rangle
\,\equiv\,\lim_{n\to1}\langle \, \phi, \psi_{n-1} \rangle
\,= \,\langle \, f \circ \phi (0) \, c(1) \, \rangle_{{\cal W}_1}\,.
\end{equation}
The goal of this subsection is the calculation
of $\ket{\mB(\Psi_{\cal S})}$.
As a warm-up exercise, it is instructive to calculate
\begin{equation}\label{warmup}
\ket{\mB^{(1)} (\psi_0)} =
{}- e^{\frac{\pi^2}{s} ( L_0 + \tilde{L}_0 )} \,\oint_s \int_0^s ds_1 \,
\PR (0, s_1) \, \{ {\cal B}_R (s_1), \psi_0  \} \,
\PR (s_1, s) \,.
\end{equation}
The required shift $a_0$ from~(\ref{a0})
is given by
\begin{equation}
a_0 = \frac{e^{s_1}}{e^s-1} \,,
\end{equation}
and the operator $c(1)$ in~(\ref{psi_0}) is mapped
to $e^{-s_1} c(e^{s_1} + a_0)$.
The operator insertions in the natural $z$ frame are
\begin{equation}\label{tachvacopins}
\begin{split}
& - e^{-s_1}
\int_{C(s_1)}
\frac{dz}{2 \pi i} \,
(z-a_0) \, b(z) \, c(e^{s_1} + a_0)
- e^{-s_1}c (e^{s_1} + a_0)
\int_{C(s_1)}
\frac{dz}{2 \pi i} \,
(z-a_0-e^{s_1}) \, b(z) \\
& = e^{-s_1}\oint \frac{dz}{2 \pi i} \,
(z-a_0) \, b(z) \, c(e^{s_1} + a_0)
+  c(e^{s_1} + a_0) \,\BB_R^+
\\& =
 1+c(e^{s_1} + a_0) \, {\cal B}^+_R
= {}- {\cal B}^+_R \, c (e^{s_1} + a_0) \,,
\end{split}
\end{equation}
where
the first line corresponds to~(\ref{anticommBA}),
the second line corresponds to~(\ref{anticommBAconnect}),
and we used
the anticommutation relation
$\{ \, {\cal B}^+_R, c(t) \, \} = -1 \,$
in the last step.
{}From the identification $z_+ = e^s \, z_-$
in the natural $z$ frame, we find
\begin{equation}
\label{B^+_R-transformation}
\int \frac{dz_+}{2 \pi i} \, b(z_+)
= e^{-s} \int \frac{dz_-}{2 \pi i} \, b(z_-)
\end{equation}
and therefore
\begin{equation}
\int_{C(t+s)}
\frac{dz_+}{2 \pi i} \, b(z_+)
- \int_{C(t)}
\frac{dz_-}{2 \pi i} \, b(z_-)
= (e^{-s}-1)
\int_{C(t)}
\frac{dz_-}{2 \pi i} \, b(z_-) \,.
\end{equation}
We thus obtain the following formula:\footnote{One might worry that the ${\cal B}^+_R$ integral cannot be closed without taking into account a contribution from the hidden boundary similar to the contribution $\BHB$ in~(\ref{commBRQ}) for $\BB_R$. Fortunately, this is not the case because the integrand of ${\cal B}^+_R$ \emph{vanishes} along the
 closed string
boundary in the Schnabl limit,
 as we discussed in section~\ref{secwedge}.
}
\begin{equation}\label{closeBRplus99}
{\cal B}^+_R \,
 \,[\ldots]\,
= - \frac{e^s}{e^s-1}
\oint \frac{dz}{2 \pi i} \, b(z) \,
 [\ldots]
\,,
\end{equation}
where
 the dots $[\ldots]$ represent arbitrary insertions of local operators and
the contour encircles all
 these operators
counterclockwise.
Using this formula, the operator insertions
 (\ref{tachvacopins})
in the natural $z$ frame
can be calculated as
\begin{equation}\label{tachghostsector}
{}- {\cal B}^+_R \, c (e^{s_1} + a_0)
= \frac{e^s}{e^s-1}
\oint \frac{dz}{2 \pi i} \, b(z) \, c(e^{s_1} + a_0)
 = \frac{e^s}{e^s-1}\,.
\end{equation}
For~(\ref{warmup}) we thus obtain
\begin{equation}
\begin{split}
\ket{\mB^{(1)} (\psi_0)} & =
{}- e^{\frac{\pi^2}{s} ( L_0 + \tilde{L}_0 )} \,\oint_s \int_0^s ds_1 \,
\PR (0, s_1) \, \{ {\cal B}_R (s_1), \psi_0 \} \,
\PR (s_1, s) \\
& = \int_0^s ds_1 \frac{e^s}{e^s-1}\, \, \ket{B}
= \frac{s\,e^s}{e^s-1} \, \ket{B} \,.
\end{split}
\end{equation}

Let us now
generalize this calculation to
closed string states of the following form:
\begin{equation}\label{Phin1nk}
\begin{split}
& \ket{\Phi \, (n_1, \ldots , n_k)} \\ \equiv & {}~
(-1)^k \, e^{\frac{\pi^2}{s} ( L_0 + \tilde{L}_0 )} \,
\oint_s \PR (0,s_1) \, \{ {\cal B}_R (s_1), \psi_{n_1-1} \} \,
\PR (s_1,s_2) \, \{ {\cal B}_R (s_2), \psi_{n_2-1} \} \\
& \qquad \qquad \qquad \qquad ~ \times
\PR (s_2,s_3) \, \{ {\cal B}_R (s_3), \psi_{n_3-1} \} \, \ldots
\PR (s_{k-1},s_k) \, \{ {\cal B}_R (s_k), \psi_{n_k-1} \} \,
\PR (s_k,s) \,,
\end{split}
\end{equation}
where $n_i\geq1$.  Note that $\ket{\mB^{(k)}(\Psi_{\cal S})}$ with $\Psi_S$ given in~(\ref{definePsiS}) can be expressed in terms of
 states of the form $\ket{\Phi \, (n_1, \ldots , n_k)}$. For example,
 \begin{equation}
 \begin{split}
    \ket{\mB^{(2)}(\Psi_{\cal S})}=\lim_{N_1,N_2\to\infty}\int_0^s ds_1 \int_{s_1}^s ds_2
    &\Biggl[\,\sum_{n_1=1}^{N_1}\sum_{n_2=1}^{N_2}\frac{\del}{\del n_1}\frac{\del}{\del n_2}\,\ket{\Phi \, (n_1 , n_2)}
    ~-\,\sum_{n_1=1}^{N_1}\frac{\del}{\del n_1}\,\ket{\Phi \, (n_1 , N_2)}\\
    &~-\,\sum_{n_2=1}^{N_2}\frac{\del}{\del n_2}\,\ket{\Phi \, (N_1,n_2)}
    ~+\,\ket{\Phi \, (N_1,N_2)}~\Biggl]\,.
 \end{split}
 \end{equation}
In the calculation of $\ket{\Phi \, (n_1, \ldots , n_k)}$,
the operators we insert for $\psi_{n_i-1}$ in the natural $z$ frame
are
\begin{equation}
{}- e^{-s_i} \, c(e^{s_i} + \ell_i) \, {\cal B}^+_R \,
c(n_i \, e^{s_i} + \ell_i) \,,
\end{equation}
where
\begin{equation}
\ell_i = \sum_{j=1}^{i-1} n_j \, e^{s_j} + a_0 \,, \quad
\ell_1 = a_0 \,.
\end{equation}
All the operators for the state
$\ket{\Phi \, (n_1, \ldots , n_k)}$ can be written
using the formula~(\ref{anticommBAconnect}) as
\begin{equation}
\begin{split}
&
\prod_{i=1}^k \,
\biggl[ \, {}- e^{-s_i} \oint \frac{dz}{2 \pi i} \,
(z-\ell_i) \, b(z) \,
c(e^{s_i} + \ell_i) \, {\cal B}^+_R \,
c(n_i \, e^{s_i} + \ell_i)
- n_i \, c(e^{s_i} + \ell_i) \, {\cal B}^+_R \,
c(n_i \, e^{s_i} + \ell_i) \, {\cal B}^+_R \,
\biggr] \\
& = \prod_{i=1}^k \,
\biggl[ \, {}-
{\cal B}^+_R \, c(n_i \, e^{s_i} + \ell_i)
- n_i \, c(e^{s_i} + \ell_i) \, {\cal B}^+_R \,
+ n_i \, c(e^{s_i} + \ell_i) \, {\cal B}^+_R \,
\biggr] \\
& = \prod_{i=1}^k \, \Bigl[ \,
{}- {\cal B}^+_R \, c(n_i \, e^{s_i} + \ell_i) \, \Bigr] \,.
\end{split}
\end{equation}
Using the anticommutation relation
$\{ \, {\cal B}^+_R, c(t) \, \} = -1$
and $( \, {\cal B}^+_R \, )^2 = 0$ repeatedly, we find
\begin{equation}\label{BcBcBc}
{\cal B}^+_R \, c(t_1) \,
{\cal B}^+_R \, c(t_2) \, \ldots
{\cal B}^+_R \, c(t_k)
= (-1)^{k-1} \, {\cal B}^+_R \, c(t_k)
= \frac{(-1)^{k} \, e^s}{e^s-1} \,,
\end{equation}
 where we used~(\ref{closeBRplus99}) in the last step.
Therefore we have
\begin{equation}\label{tachconst}
\prod_{i=1}^k \, \Bigl[ \,
{}- {\cal B}^+_R \, c(n_i \, e^{s_i} + \ell_i) \, \Bigr]
= \frac{e^s}{e^s-1}
\end{equation}
and thus
\begin{equation}
\label{final-Phi}
\ket{\Phi \, (n_1, \ldots , n_k)}
= \frac{e^s}{e^s-1} \, \ket{B} \,.
\end{equation}

Note that this is independent of $n_i$.
This means that the $\psi'_n$ piece of the solution
does not contribute
to $\ket{\mB(\Psi_{\cal S})}$ because all derivatives
of $\ket{\Phi \, (n_1, \ldots , n_k)}$
with respect to $n_i$ vanish.
 In particular,
mixed terms that involve $\psi'_n$ and $\psi_N$
do not contribute to $\ket{\mB(\Psi_{\cal S})}$.
Therefore the whole contribution
to $\ket{\mB(\Psi_{\cal S})}$ comes entirely from
the `phantom' term $-\psi_N$ of the solution,
namely,
 \begin{equation}\label{onlyPsiN}
 \ket{\mB^{(k)} (\Psi_{\cal S})} =(-1)^k
 \lim_{N_1,\ldots,N_k\to\infty}\,\int_0^s ds_1 \int_{s_1}^s ds_2 \ldots \int_{s_{k-1}}^s ds_k \,
 \ket{\Phi \, (N_1, N_2,\ldots , N_k)}.
 \end{equation}
 Since $\ket{\Phi \, (N_1, \ldots , N_k)}$ is independent of $N_i$,
 the limit $N_1, \ldots, N_k \to \infty$ is trivial.\footnote{
 In particular, the limit $N_1, \ldots, N_k \to \infty$
 is independent of the order in which we take
$N_i\to\infty$.}
The result~(\ref{final-Phi})
is also independent of $s_i$.
Thus the integrals over $s_i$
 in~(\ref{onlyPsiN})
simply gives the following factor:
\begin{equation}
\int_0^s ds_1 \int_{s_1}^s ds_2 \int_{s_2}^s ds_3 \ldots
\int_{s_{k-1}}^s ds_k = \frac{s^k}{k!} \,.
\end{equation}
We therefore conclude that
\begin{equation}\label{mBPsiS}
\ket{\mB (\Psi_{\cal S})}
= \biggl[ \, 1 + \sum_{k=1}^\infty \frac{s^k}{k!} \,
(-1)^k \, \frac{e^s}{e^s-1} \, \biggr]
\ket{B}
= \biggl[ \, 1 + ( e^{-s} - 1 ) \, \frac{e^s}{e^s-1} \, \biggr]
\ket{B} = 0\,.
\end{equation}
 In~\cite{Ellwood:2008jh,Kawano:2008ry}
 it was shown that the gauge-invariant
observables $W ({\cal V}, \Psi)$ vanish for the tachyon vacuum solution $\Psi_{\cal S}$.
Their result can be reproduced by calculating the
on-shell part of the $k=1$ term $\ket{\mB^{(1)}(\Psi_{\cal S})}$
and taking the limit $s \to 0$.
The result~(\ref{mBPsiS}) can thus be viewed as
the generalization of the calculation in~\cite{Ellwood:2008jh}
to the off-shell part and to finite $s$.
Indeed, the terms with $k\geq2$ in~(\ref{mBPsiS}) are suppressed for small $s$, consistent with our analysis in \subs\ref{thestozerorevisited}. In the limit $s\to0$ the $k=1$ term  by itself cancels the original
boundary state $\ket{B}$.
In summary, we conclude that $\ket{\mB (\Psi_{\cal S})}$
for Schnabl's tachyon vacuum solution $\Psi_{\cal S}$ vanishes
for any finite $s$:
\begin{equation}
\boxed{\phantom{\Biggl\{}
\ket{\mB (\Psi_{\cal S})} \,= \, 0
\qquad\text{for $\BB=B$ and any finite $s$.}
~}
\end{equation}
This is consistent with Sen's conjecture that
the D-brane disappears at the tachyon vacuum.

\subsection{Factorization of $\ket{\mB (\Psi)}$ into matter and ghost sectors}
\label{sectrivialghost}

We have seen
that
$\ket{\mB^{(k)} (\Psi_{\cal S})}$ is proportional to $\ket{B}$ for any $k$.
In particular, this means
 that the ghost sector of $\ket{\mB^{(k)} (\Psi_{\cal S})}$ is
the same as the ghost sector
of $\ket{B}$, namely,
the boundary state $\ket{B^{(bc)}}$ of the $bc$ CFT.
This boundary state satisfies the relations
\begin{equation}
\label{bc-conditions}
( \, b_n - \tilde{b}_{-n} \, ) \, \ket{B^{(bc)}} = 0 \,,
\qquad
( \, c_n + \tilde{c}_{-n} \, ) \, \ket{B^{(bc)}} = 0
\end{equation}
 for all $n\in\mathbb{Z}$.
If the state $\ket{\mB (\Psi)}$ factorizes into
matter and ghost sectors as
\begin{equation}
\label{factorized}
\ket{\mB (\Psi)} = \ket{\mB^{({\rm matter})} (\Psi)}
\otimes \ket{B^{(bc)}} \,,
\end{equation}
it follows from $Q \, \ket{\mB (\Psi)} = 0$
and (\ref{bc-conditions}) that
the matter part $\ket{\mB^{({\rm matter})} (\Psi)}$
satisfies the
relation for conformal boundary conditions
\begin{equation}
( \, L^{({\rm matter})}_n
- \tilde{L}^{({\rm matter})}_{-n} \, ) \,
\ket{\mB^{({\rm matter})} (\Psi)} = 0 \,.
\end{equation}
While our claim is that
the state $\ket{\mB (\Psi)}$ coincides
with the BCFT boundary state $\ket{B_\ast}$
up to a possible BRST-exact term,
the state $\ket{\mB (\Psi)}$
factorized as~(\ref{factorized})
can be a consistent BCFT boundary state
without any BRST-exact term.
It is  therefore
important to examine
for what solutions
the ghost part of $\ket{\mB (\Psi)}$
becomes the boundary state $\ket{B^{(bc)}}$ of the $bc$ CFT.

We now claim that
the ghost part of a closed string state
of the form
\begin{equation}
\label{trivial-ghost}
\begin{split}
& e^{\frac{\pi^2}{s} ( L_0 + \tilde{L}_0 )} \,
\oint_s \, \PR (0, s_1)
\, \{ {\cal B}_R (s_1), A_1 \} \, \PR (s_1, s_2) \,
\, \{ {\cal B}_R (s_2), A_2 \} \, \PR (s_2, s_3) \,
\, \{ {\cal B}_R (s_3), A_3 \} \, \ldots \\
& \qquad \qquad \qquad \times
\PR (s_{k-1}, s_k) \,
\{ {\cal B}_R (s_k), A_k \} \, \PR (s_k, s)
\end{split}
\end{equation}
coincides with the boundary state of the $bc$ CFT
if open string fields $A_1, A_2, \ldots, A_k$
of ghost number one are made of wedge states
with
\begin{itemize}
\item local operator insertions of the $c$ ghost and its derivatives,
\item $b$-ghost line integrals $\BB_R^+$,
\item arbitrary insertions of matter operators, and
\item  line integrals $\LL_R^+=\{Q,\BB_R^+\}$ of the energy-momentum tensor.
We demand that there are no other operators
on the contour of each $\LL_R^+$.
\end{itemize}
This can be shown in the following way.
First consider the case where there are no line integrals of $\LL^+_R$.
The $b$-ghost integral $\BB_R (s_i)$ in
$\{ \BB_R (s_i),A_i \}$
can be written in the form~(\ref{anticommBAconnect}):
\begin{equation}
\label{ghost-integrals}
 {}\{ {\cal B}_R (s_i), A_{\alpha_i} \}~\to~
 {}-
\oint\frac{dz}{2\pi i} \, (z-\ell_i) \, b(z) \, [\ldots] \,
 {}-
e^{s_i}\alpha_i\,[\ldots]\,\BB_R^+\,,
\end{equation}
where $\alpha_i$ is the length of the wedge state
associated with $A_i$
and $\ell_i$ is defined in~(\ref{ell-definition}).
Nonvanishing contributions to the first term
come only from local operator insertions of the $c$ ghost
and its derivatives in $[\ldots]$.
Therefore, after performing the integrals
of the form~(\ref{ghost-integrals}),
remaining operator insertions in the ghost sector
are insertions of $\BB_R^+$
and insertions of the $c$ ghost and its derivatives.
It then follows from $(\BB^+_R)^2 = 0$
and the transformation property~(\ref{B^+_R-transformation})
that there must be at least one insertion
of the $c$ ghost or its derivatives
between two insertions of $\BB^+_R$
for the result to be nonvanishing.
However, since the total ghost number vanishes,
there must be only one insertion
of the $c$ ghost or its derivatives
between two insertions of $\BB^+_R$.
Then such contributions can be calculated
using the formula~(\ref{BcBcBc}).
In fact, terms which contain derivatives of the $c$ ghost
vanish because the right-hand side of~(\ref{BcBcBc})
is independent of $t_i$'s.
Nonvanishing contributions are of the form~(\ref{BcBcBc})
and no ghost operators remain in the end.
We have thus shown that the ghost part of~(\ref{trivial-ghost})
coincides with
$\ket{B^{(bc)}}$.

Let us next consider the case where there is
one line integral of $\LL^+_R$.
The insertion of $\LL_R^+$ in the definition of a state $A_i$ appears
in the form
\begin{equation}
    \langle \phi\,,\,A_i\rangle
    = \langle \, f \circ \phi (0) \,\, [\ldots]_1\,\LL_R^+\,[\ldots]_2\rangle_{{\cal W}_{\alpha_i}}\,,
 \end{equation}
where we denoted all the operator insertions
to the left of the $\LL^+_R$ line integral by $[\ldots]_1$
and those to the right by $[\ldots]_2$.
Such separation is possible because of our assumption that
there are no operators on the contour of $\LL^+_R$.
The state $A_i$ can be written as
\begin{equation}\label{ridofLR+}
A_i= -\partial_t A_i (t) \Bigr|_{t=0} \,,
\end{equation}
where the state $A_i (t)$ for $t > 0$ is defined by\footnote{
The definition of $A_i (t)$ can be extended
to $t < 0$ until operator insertions in $[\ldots]_1$
and $g \circ [\ldots]_2$ collide.
The derivative of $A_i (t)$ at $t=0$ is therefore well defined.
}
\begin{equation}
\begin{split}
\langle \, \phi \,,\,A_i (t) \, \rangle
= \langle \, f \circ \phi (0) \,\, [\ldots]_1 \,\,
g \circ [\ldots]_2\,\rangle_{{\cal W}_{\alpha_i+t}}
\quad \text{ with }\quad g(z) = z+t \,.
\end{split}
\end{equation}
Since $A_i (t)$ belongs to the class of states
considered before,
the ghost part of~(\ref{trivial-ghost})
with $A_i$ replaced by $A_i (t)$ for any positive $t$
coincides with
$\ket{B^{(bc)}}$.
Therefore, the ghost part of~(\ref{trivial-ghost})
with $A_i$ replaced by $\partial_t A_i (t)$
also coincides with
$\ket{B^{(bc)}}$.
It thus follows that
the ghost part of~(\ref{trivial-ghost})
with $A_i$ given by~(\ref{ridofLR+})
also coincides with
$\ket{B^{(bc)}}$.
It is straightforward to generalize the proof
to the case with an arbitrary number of insertions of $\LL^+_R$.
We conclude that the state $\ket{\mB (\Psi)}$
takes the form~(\ref{factorized})
if the solution $\Psi$ consists of wedge states
with local operator insertions of the $c$ ghost
and its derivatives, line integrals $\BB_R^+$
and $\LL_R^+$,
and arbitrary insertions of matter operators.\footnote{
In general, if we have a one-parameter family
of closed string states of the form~(\ref{trivial-ghost})
with their ghost sectors being $\ket{B^{(bc)}}$,
ghost sectors of closed string states
obtained by taking derivatives
with respect to the parameter are also given by $\ket{B^{(bc)}}$.
We considered the case with line integrals of $\LL^+_R$
as a particular example of this generalization
because $\psi'_n$ in the tachyon vacuum solution
contains a line integral of $\LL^+_R$,
but various other generalizations will be possible.
Derivatives of the $c$ ghost can also be treated
in this way.}

This condition on the solution
for which $\ket{\mB (\Psi)}$ satisfies~(\ref{factorized})
is a sufficient condition and is not a necessary condition.
However, this class of states covers
all known wedge-based analytic solutions
such as Schnabl's tachyon vacuum solution~\cite{0511286}
as analyzed
in \subs\ref{sec7.1}
and the solutions associated with marginal deformations
for both regular and singular operator products constructed
in~\cite{0701248,0701249,0704.2222,0707.4472}.

In the case of marginal deformations
with regular operator products,
the solutions in Schnabl gauge
constructed in~\cite{0701248,0701249}
and those in~\cite{0707.4472}
are expected to be gauge-equivalent.
We just argued that these different solutions
give the same state $\ket{\mB (\Psi)}$
in the form~(\ref{factorized}).
We thus expect that the form~(\ref{factorized})
is preserved for a certain class of gauge transformations.
We have shown in \subs\ref{secgaugeinv} that
the state $\ket{\mB (\Psi)}$ changes
under a gauge transformation
$\delta_\chi \Psi = Q \chi + [ \, \Psi, \chi \, ]$
by the following BRST-exact term:
\begin{equation}
\delta_\chi \ket{\mB (\Psi)}
=  Q \, \Bigl[ \,
e^{\frac{\pi^2}{s} ( L_0 + \tilde{L}_0 )} \,
\oint_s \int_0^s dt \, \mP (0, t) \,
[ \, {\cal B}_R (t), \chi \, ] \, \mP (t, s) \, \Bigr] \,.
\end{equation}
Consider a closed string state of the form
\begin{equation}
\label{trivial-transformation}
\begin{split}
& e^{\frac{\pi^2}{s} ( L_0 + \tilde{L}_0 )} \,
\oint_s \, \PR (0, s_1)
\, \{ {\cal B}_R (s_1), A_1 \} \, \PR (s_1, s_2) \,
\, \{ {\cal B}_R (s_2), A_2 \} \, \PR (s_2, s_3) \,
\, \{ {\cal B}_R (s_3), A_3 \} \, \ldots \\
& \qquad \qquad \qquad \times
\PR (s_{i-1}, s_i) \,
\, [ \, {\cal B}_R (s_i), A_i \, ] \, \PR (s_i, s_{i+1}) \,
\ldots
\PR (s_{k-1}, s_k) \,
\{ {\cal B}_R (s_k), A_k \} \, \PR (s_k, s) \,,
\end{split}
\end{equation}
where $A_i$ carries ghost number zero and all other states
$A_1, A_2, \ldots, A_k$ carry ghost number one.
They are made of wedge states
with local operator insertions of the $c$ ghost and its derivatives,
 line integrals $\BB_R^+$ and $\LL_R^+$,
and arbitrary insertions of matter operators.
 The $b$-ghost integral $\BB_R (s_i)$ in $[ \, \BB_R (s_i),A_i \, ]$
 for the Grassmann-even state $A_i$
 can be written as
\begin{equation}
\label{ghost-integral-2}
 [ {\cal B}_R (s_i), A_{\alpha_i} ]~\to~
 -
\oint\frac{dz}{2\pi i} \, (z-\ell_i) \, b(z) \, [\ldots] \,
{}+ e^{s_i}\alpha_i\,[\ldots]\,\BB_R^+\,,
\end{equation}
which follows from the same manipulations that led to~(\ref{anticommBAconnect}) for Grassmann-odd states.
Line integrals of $\LL^+_R$ can be treated
in the same way as before so that it is sufficient to consider
the case when they are absent.
After performing the integrals
of the form~(\ref{ghost-integrals}) and~(\ref{ghost-integral-2}),
remaining operator insertions in the ghost sector
are again insertions of $\BB_R^+$
and insertions of the $c$ ghost and its derivatives.
In this case, however, the total ghost number is $-1$
and thus we have one more insertions of $\BB_R^+$
than insertions of the $c$ ghost and its derivatives.
Any term with more than one insertion of $\BB_R^+$ immediately
vanishes because of $(\BB^+_R)^2 = 0$
and the transformation property~(\ref{B^+_R-transformation}).
In the case of one insertion of $\BB_R^+$,
we do not have any other insertions of ghost operators,
and it follows from the
 formula~(\ref{closeBRplus99}) that the contribution vanishes.
We thus conclude that
$\delta_\chi \ket{\mB (\Psi)}$ vanishes
if the solution $\Psi$ and the gauge parameter $\chi$
consist of wedge states
with local operator insertions of the $c$ ghost
and its derivatives, line integrals $\BB_R^+$
 and $\LL_R^+$,
and arbitrary insertions of matter operators.
In particular, gauge transformations generated
by gauge parameters made of wedge states
with only matter operator insertions do not change
the state $\ket{\mB (\Psi)}$.

\subsection{Marginal deformations with regular operator products}

Deformations of BCFT generated by a matter primary field $V$
of weight one
are exactly marginal
when operator products
$V(t_1) \, V(t_2) \, \ldots \, V(t_n)$  are regular.
In this case we expect to have a one-parameter family of solutions
to the equation of motion of open string field theory.
Analytic solutions for such marginal deformations
with regular operator products were constructed
in~\cite{0701248,0701249,0707.4472}.
 In~\cite{Ellwood:2008jh,Kishimoto:2008zj},
the gauge-invariant
observables $W ({\cal V}, \Psi)$ were calculated
for these solutions, and the results confirmed the relation~(\ref{ellwood-claim}). These calculations essentially extracted the on-shell part of $\ket{\mB^{(1)}(\Psi)}$ in the limit $s\to0$.
In this section we calculate the
full state $\ket{\mB (\Psi)}$
constructed using the Schnabl propagator strips
for these analytic solutions
to see if it coincides with the BCFT boundary state
\begin{equation}\label{wanttoshow}
\ket{\mB} =
\exp \,
\biggl[ \, \lambda \int_0^{2 \pi} d \theta \,
V(\theta) \, \biggr] \, \ket{B} \,,
\end{equation}
where $\theta$ with $0 \le \theta \le 2 \pi$
parameterizes the boundary of the disk.

\subsubsection{The leading term}

The leading term $\Psi^{(1)}$ defined by
\begin{equation}
\langle \, \phi, \Psi^{(1)} \, \rangle
= \langle \, f \circ \phi (0) \, cV (1) \, \rangle_{{\cal W}_1}
\end{equation}
is identical for all solutions~\cite{0701248,0701249,0707.4472}
associated with
a
marginal operator $V$.
Let us calculate
\begin{equation}
\ket{\mB^{(1)} (\Psi^{(1)})} =
{}- e^{\frac{\pi^2}{s} ( L_0 + \tilde{L}_0 )} \,\oint_s \int_0^s ds_1 \,
\PR (0, s_1) \, \{ {\cal B}_R (s_1), \Psi^{(1)} \} \,
\PR (s_1, s) \,.
\end{equation}
The required shift $a_0$ from~(\ref{a0})
is given by
\begin{equation}
a_0 = \frac{e^{s_1}}{e^s-1} \,.
\end{equation}
 Using~(\ref{anticommBAconnect}), the operator insertions in the natural $z$ frame are
 \begin{equation}
 \begin{split}
  &\oint
  \frac{dz}{2 \pi i} \,
  (z-a_0) \, b(z) \, cV (e^{s_1} + a_0)
  +e^{s_1}\,cV (e^{s_1} + a_0)\,\BB_R^+ \\[.3ex]
  &= e^{s_1}\,V (e^{s_1} + a_0)+e^{s_1}\,cV (e^{s_1} + a_0)\,\BB_R^+\\
  &=-e^{s_1}\,\BB_R^+\,cV (e^{s_1} + a_0)\,.
 \end{split}
 \end{equation}
 The ghost sector of these operator insertions is identical
 to the one calculated in~(\ref{tachghostsector}) for
 the tachyon vacuum
solution. We obtain
 \begin{equation}
  -e^{s_1}\,\BB_R^+\,cV (e^{s_1} + a_0)
   = \frac{e^s \, e^{s_1}}{e^s-1} \, V (e^{s_1} + a_0) \,.
 \end{equation}
If we define
\begin{equation}
u_1 = e^{s_1} + a_0
  = \frac{e^s \, e^{s_1}}{e^s-1} \,,
\end{equation}
we observe that
\begin{equation}
\int_0^s ds_1 \, \frac{e^s \, e^{s_1}}{e^s-1} \, V (e^{s_1} + a_0)
 = \int_{\frac{e^{s}}{e^s-1}}^{\frac{e^{2s}}{e^s-1}}
 du_1 \, V(u_1) \,.
\end{equation}
Namely,
the measure factor
\begin{equation}
\frac{\partial u_1}{\partial s_1}
 =\frac{e^s \, e^{s_1}}{e^s-1} \end{equation}
has been correctly provided through the calculation.
Since the point
 $\frac{e^{s}}{e^s-1}$ and the point $\frac{e^{2s}}{e^s-1}$
are identified
in the natural $z$ frame, the operator corresponds to
\begin{equation}
\int_0^{2 \pi} d \theta \, V(\theta)
\end{equation}
in the $\zeta$ frame,
where $\theta$ parameterizes the boundary
$| \zeta | = 1$ as $\zeta = e^{i \theta}$.
Therefore we find
\begin{equation}
\ket{\mB^{(1)} (\Psi^{(1)})} =
{}- e^{\frac{\pi^2}{s} ( L_0 + \tilde{L}_0 )} \,\oint_s \int_0^s ds_1 \,
\PR (0, s_1) \, \{ {\cal B}_R (s_1), \Psi^{(1)} \} \,
\PR (s_1, s)
= \int_0^{2 \pi} d \theta \, V(\theta) \, \ket{B} \,.
\end{equation}
 This is indeed the $\ord\lambda$ term in the path-ordered
exponential~(\ref{wanttoshow}) that we expected.

\subsubsection{Regular marginal deformations in Schnabl gauge}

Let us next calculate $\ket{\mB (\Psi)}$
with $\Psi$ being the Schnabl-gauge
solutions for marginal deformations
constructed in \cite{0701248,0701249}.
The solution $\Psi$ is given by
\begin{equation}
\Psi = \sum_{n=1}^\infty \lambda^n \, \Psi^{(n)} \,,
\end{equation}
where\footnote{
Note that the operator $\BB$ in~\cite{0701249}
corresponds to $-\BB^+_R$ in this paper.
This expression can be derived from~(3.3) of \cite{0701249}
by replacing $\BB$ with $-\BB^+_R$
and by using $\{ \, {\cal B}^+_R, c(t) \, \} = -1$
and $( \, {\cal B}^+_R \, )^2 = 0$ recursively.}
\begin{equation}
\label{KORZ-solution}
\begin{split}
\langle \, \phi, \Psi^{(n)} \, \rangle
&  = - \int_0^1 \hskip-3pt dt_1 \int_0^1 \hskip-3pt dt_2 \ldots
\int_0^1 \hskip-3pt dt_{n-1} \,
\langle \, f \circ \phi (0) \, c V (1) \\
& \qquad \quad {}\times
V (1+t_1) \, V (1+t_1+t_2) \, \ldots
V (1+t_1+t_2+ \ldots +t_{n-2}) \\
& \qquad \quad {}\times
{\cal B}^+_R \, c V (1+t_1+t_2+ \ldots +t_{n-1}) \,
\rangle_{{\cal W}_{1+t_1+t_2+ \ldots +t_{n-1}}} \,.
\end{split}
\end{equation}
The calculation of the ghost sector has been done
in \subs\ref{sec7.1} so that
we only need to calculate the matter sector.\footnote{
The separation of matter and ghost sectors depends on a frame.
We separate them in the sliver frame used in~(\ref{KORZ-solution}).
For example, the operator $cV$ in the sliver frame $z^{(i)}$
is divided into $e^{-s_i} c$ and $e^{s_i} V$ in the natural $z$ frame
via the conformal transformation~(\ref{znatPsini}).}

There are two terms which contribute to $\ket{\mB (\Psi)}$
at ${\cal O} (\lambda^2)$.
The first one is
\begin{equation}
\begin{split}
& \ket{\mB^{(2)} (\Psi^{(1)})}
= e^{\frac{\pi^2}{s} ( L_0 + \tilde{L}_0 )} \,\oint_s \int_0^s ds_1 \int_{s_1}^s ds_2 \,
\PR (0, s_1) \, \{ {\cal B}_R (s_1), \Psi^{(1)} \} \\
& \qquad \qquad \qquad \qquad \qquad \qquad \qquad \qquad \qquad
\times
\PR (s_1, s_2) \, \{ {\cal B}_R (s_2), \Psi^{(1)} \} \,
\PR (s_1, s) \,.
\end{split}
\end{equation}
The operator insertion in the natural $z$ frame is given by
\begin{equation}\label{KORZterm1}
\frac{e^s}{e^s-1} \, \int_0^s ds_1 \int_{s_1}^s ds_2 \,
e^{s_1} \,
V \Bigl( \, \frac{e^s e^{s_1} + e^{s_2}}{e^s-1} \, \Bigr) \,
e^{s_2} \,
V \Bigl( \, \frac{e^s e^{s_1} + e^s e^{s_2}}{e^s-1} \, \Bigr) \,,
\end{equation}
 where we used~(\ref{tachconst}) to calculate the ghost sector.
The second one is
\begin{equation}
\ket{\mB^{(1)} (\Psi^{(2)})} =
{}- e^{\frac{\pi^2}{s} ( L_0 + \tilde{L}_0 )} \, \oint_s \int_0^s ds_1 \,
\PR (0, s_1) \, \{ {\cal B}_R (s_1), \Psi^{(2)} \} \,
\PR (s_1, s) \,.
\end{equation}
The operator insertion in the natural $z$ frame is given by
\begin{equation}
\frac{e^s}{e^s-1} \, \int_0^s ds_1 \int_0^1 dt_1 \,
e^{s_1} \,
V \Bigl( \, \frac{e^s e^{s_1} + t_1 \, e^{s_1}}{e^s-1} \, \Bigr) \,
e^{s_1} \,
V \Bigl( \, \frac{e^s e^{s_1} + e^s \, t_1 \, e^{s_1}}{e^s-1} \,
\Bigr) \,.
\label{Psi^(2)-intermediate}
\end{equation}
If we define
\begin{equation}
s'_1 = s_1 \,, \quad s'_2 = s_1 + \ln t_1 \,,
\end{equation}
this can be written as follows:
\begin{equation}\label{KORZterm2}
\frac{e^s}{e^s-1} \, \int_0^s ds'_1 \int_{-\infty}^{s'_1} ds'_2 \,
e^{s'_1} \,
V \Bigl( \, \frac{e^s e^{s'_1} + e^{s'_2}}{e^s-1} \, \Bigr) \,
e^{s'_2} \,
V \Bigl( \, \frac{e^s e^{s'_1} + e^s e^{s'_2}}{e^s-1} \, \Bigr) \,.
\end{equation}
Note that the second factor of $e^{s_1}$
in~(\ref{Psi^(2)-intermediate})
has been changed to $e^{s'_2}$ because of the Jacobian
\begin{equation}
\frac{\partial (s'_1, s'_2)}{\partial (s_1, t_1)}
= \frac{1}{t_1} \,.
\end{equation}
The two expressions
 (\ref{KORZterm1}) and~(\ref{KORZterm2})
are combined to give
\begin{equation}
\frac{e^s}{e^s-1} \, \int_0^s ds_1 \int_{-\infty}^s ds_2 \,
e^{s_1} \,
V \Bigl( \, \frac{e^s e^{s_1} + e^{s_2}}{e^s-1} \, \Bigr) \,
e^{s_2} \,
V \Bigl( \, \frac{e^s e^{s_1} + e^s e^{s_2}}{e^s-1} \, \Bigr) \,.
\end{equation}
If we define
\begin{equation}
u_1 = \frac{e^s e^{s_1} + e^{s_2}}{e^s-1} \,, \quad
u_2 = \frac{e^s e^{s_1} + e^s e^{s_2}}{e^s-1} \,,
\end{equation}
this expression can be written in the following form:
\begin{equation}
\int_{\Gamma^{(2)'}} du_1 du_2 \,
V (u_1) \, V (u_2) \,,
\end{equation}
where we have used
\begin{equation}
\frac{\partial (u_1, u_2)}{\partial (s_1, s_2)}
= \frac{e^s}{e^s-1} \, e^{s_1} \, e^{s_2} \,.
\end{equation}
Since
\begin{equation}
e^{s_1} = u_1 - e^{-s} u_2 \,, \quad
e^{s_2} = u_2 - u_1 \,,
\end{equation}
the integration region $\Gamma^{(2)'}$
can be characterized by
\begin{equation}
1 \le u_1 - e^{-s} u_2 \le e^s \,, \quad
0 \le u_2 - u_1 \le e^s \,.
\end{equation}
Using the identification $z \sim e^s z$
in the natural $z$ frame, this integral can also be written as
\begin{equation}
\int_{\Gamma^{(2)}} du_1 du_2 \,
V (u_1) \, V (u_2)
\end{equation}
with $\Gamma^{(2)}$ given by
\begin{equation}
1 \le e^s u_1 - u_2 \le e^s \,, \quad
0 \le u_2 - u_1 \le 1 \,.
\end{equation}

It is straightforward to generalize the calculations
to the case of ${\cal O} (\lambda^n)$.
The details
are presented in appendix~\ref{appKORZ1}.
The matter sector of the boundary state $\ket{\mB (\Psi)}$
can be written in the $z$ frame
with the identification $z \sim e^s z$ as follows:
\begin{equation}
\sum_{n=0}^\infty \lambda^n
\int_{\Gamma^{(n)}} du_1 du_2 \ldots du_n \,
V(u_1) \, V(u_2) \, \ldots V(u_n) \,,
\label{KORZ-matter}
\end{equation}
where the region $\Gamma^{(n)}$ is given by
\begin{equation}
0 \le u_2 - u_1 \le 1 \,, \quad
0 \le u_3 - u_2 \le 1 \,, \quad \ldots \quad
0 \le u_n - u_{n-1} \le 1 \,, \quad
1 \le e^s u_1 - u_n \le e^s \,.
\end{equation}
In appendix~\ref{appKORZ2}
we use the identification $z\sim e^sz$ in the $z$ frame repeatedly
to show that
 \begin{equation}
 \int_{\Gamma^{(n)}}\!\! du_1 du_2 \,.\,.\, du_n \,
 V(u_1) \, V(u_2) \,.\,.\, V(u_n)
 = \int^{e^s}_1\!\!\! du_1\int_{u_1}^{e^s}\!\!\! du_2\,.\,.\,\int_{u_{n-1}}^{e^s}\!\!\! du_n\,V(u_1) \, V(u_2) \,.\,.\, V(u_n)\,.
 \end{equation}
 It  follows that
\begin{equation}
\sum_{n=0}^\infty \lambda^n
\int_{\Gamma^{(n)}} du_1 du_2 \ldots du_n \,
V(u_1) \, V(u_2) \, \ldots V(u_n)
= \exp \,
\biggl[ \, \lambda \int_1^{e^s} du \, V(u) \, \biggr] \,.
\end{equation}
Therefore we conclude that
\begin{equation}
\boxed{\phantom{\Biggl\{}
\ket{\mB (\Psi)} =
\exp \,
\biggl[ \, \lambda \int_0^{2 \pi} d \theta \,
V(\theta) \, \biggr] \, \ket{B} \,.
~}
\end{equation}

\subsubsection{Other solutions for regular marginal deformations}

We have so far considered solutions in Schnabl gauge,
and thus the choice of the propagator strip $\BB=B$
and the gauge condition on the solution $B \Psi = 0$
are correlated.
However, neither the explicit calculability of $\ket{\mB (\Psi)}$
nor the factorizability~(\ref{factorized})
depends on the Schnabl-gauge condition $B \Psi = 0$.
We next consider the analytic solutions
for regular marginal deformations constructed in~\cite{0707.4472},
which do not satisfy $B \Psi = 0$.
While there is an obstruction
in the construction of solutions in Schnabl gauge
for marginal deformations with singular operator products,
the solutions in~\cite{0707.4472}
can be generalized to such singular cases
and they still belong to the class of solutions
discussed in \subs\ref{sectrivialghost}.

The solution $\Psi_L$ constructed in~\cite{0707.4472} is given by
\begin{equation}
    \Psi_L=\sum_{n=1}^\infty \Psi_L^{(n)}\,,
\end{equation}
where
\begin{equation}\label{regPsiLn}
    \langle \, \phi \,, \Psi_L^{(n)} \, \rangle =
    \lambda^n\int_1^{2} dt_2 \int_{t_2}^{3}dt_3 \int_{t_3}^{4}dt_4
     \, \ldots \, \int_{t_{n-1}}^{n}dt_{n}\,
     \langle \, f \circ \phi (0) \,  \, c V(1) \, V(t_2) \, V(t_3) \, \ldots \, V(t_{n})\,\rangle_{{\cal W}_n} \,.
\end{equation}
While $\Psi_L$ solves the equation of motion, it does not satisfy
the reality condition
\begin{equation}\label{reality}
   \Psi={\rm hc}^{-1}(\Psi^\star)
\end{equation}
on the string field, where
${\rm hc}$ denotes hermitian conjugation.
A real solution $\Psi_{\rm real}$ that satisfies~(\ref{reality}) was constructed from $\Psi_L$ using a gauge transformation~\cite{0707.4472}.
As the gauge parameter $\chi$ of the required gauge transformation
 is of the type described in section~\ref{sectrivialghost},\footnote{
The
infinitesimal
gauge parameter
$\chi$ for this transformation is
proportional to $\ln \sqrt{U}$ with
the state $U$ defined in~\cite{0707.4472}.
As $U$ is a wedge-based state with only matter operator insertions
and
$U=1+\ord\lambda$, we conclude that $\chi$ is well
defined perturbatively in $\lambda$ and of the
form described in \subs\ref{sectrivialghost}.}
we conclude from
 the discussion in that section
that
\begin{equation}\label{L=real}
    \ket{\mB(\Psi_L)}=\ket{\mB(\Psi_{\rm real})}\,.
\end{equation}
It is thus sufficient to calculate
$\ket{\mB(\Psi_L)}$.

Consider the term with $k$
insertions $\Psi_L^{(n_1)},\ldots,\Psi_L^{(n_k)}$
in $\ket{\mB(\Psi_L)}$:
\begin{equation}\label{mBorderk}
\begin{split}
& (-1)^k \, e^{\frac{\pi^2}{s} ( L_0 + \tilde{L}_0 )} \, \oint_s
\int_0^s ds_1 \int_{s_1}^s ds_2 \ldots \int_{s_{k-1}}^s ds_k \,
\PR (0,s_1) \, \{ {\cal B}_R (s_1), \Psi_L^{(n_1)} \} \\
& \quad \times
\PR (s_1,s_2) \, \{ {\cal B}_R (s_2), \Psi_L^{(n_2)} \} \,
\PR (s_2,s_3) \, \{ {\cal B}_R (s_3), \Psi_L^{(n_3)} \} \, \ldots \\
& \quad \times
\PR (s_{k-1},s_k) \, \{ {\cal B}_R (s_k), \Psi_L^{(n_k)} \} \,
\PR (s_k,s) \,.
\end{split}
\end{equation}
At
$\ord{\lambda^n}$,
all partitions $\vec n=\{n_1,\dots,n_k\}$ with
\begin{equation}
    \sum_{i=1}^kn_i=n
\end{equation}
contribute.
The solution $\Psi_L$ has the
structure discussed in \subs\ref{sectrivialghost}
and
 we can thus eliminate all ghost-sector operators
in the natural $z$ frame as described in that section.
 We show in appendix~\ref{KOghost} that the  numerical factor which remains after this manipulation is given by
 \begin{equation}\label{expectedfactor}
 \Delta_k\,\prod_{i=1}^k \, e^{s_i}
\quad\text{ with }\quad\Delta_k = 1 + \frac{1}{e^s-1}
 - \frac{1}{e^s-1} \prod_{i=1}^k \, (1 - n_i) \,.
 \end{equation}

Let us denote
the marginal operators from
the insertion
$\Psi^{(n_i)}_L$ in the natural $z$-frame picture
as $V(t^{(i)}_1), V(t^{(i)}_2), \ldots, V(t^{(i)}_{n_i})$
with $t^{(i)}_1 \le t^{(i)}_2 \le \ldots \le t^{(i)}_{n_i}$.
Note that
$cV(t^{(i)}_1)$ is the only unintegrated vertex operator from
$\Psi_L^{(n_i)}$ and the others
$V(t^{(i)}_2), \ldots, V(t^{(i)}_{n_i})$
are integrated vertex operators.
After
the calculation of the ghost sector
in the natural $z$ frame,
we expect the
factor~(\ref{expectedfactor})
to provide the correct measure
 for the integration over positions of
the operators $V(t^{(i)}_1)$
with $i=1\ldots k$. This is indeed the case:
\begin{equation}
\frac{\partial (t_1^{(1)}, t_1^{(2)}, \ldots , t_1^{(k)})}
{\partial (s_1, s_2, \ldots , s_k)} =\Delta_k\,
 \prod_{i=1}^k \, e^{s_i}
\,,
\end{equation}
as we show in appendix~\ref{KOmeasure}.

The operator insertions in the natural $z$ frame of any term~(\ref{mBorderk}) with partition $\vec n$ thus
 take
the form
\begin{equation}
    \int_{\Gamma(\vec n)}dt^{(1)}_1dt^{(1)}_2\dots dt^{(k)}_{n_k-1}dt^{(k)}_{n_k}~V(t^{(1)}_1)V(t^{(1)}_2)\ldots V(t^{(k)}_{n_k-1})V(t^{(k)}_{n_k})
\end{equation}
for some integration region $\Gamma(\vec n)$ associated with the particular partition $\vec n$. The integration regions  are complicated, and we were not able to explicitly combine
the regions $\Gamma(\vec n)$ of all partitions into
the expected form.
We instead choose
a different approach to show that the BCFT boundary state is indeed recovered. Consider any point in the integration region $\Gamma(\vec n)$ of any partition $\vec n$ which contributes to $\ket{\mB(\Psi)}$ at $\ord{\lambda^n}$. The associated positions
$\{t^{(1)}_1,\ldots,t^{(k)}_{n_k}\}$
 in the $z$ frame
are mapped
to a set of angles
 $\{\theta_1,\ldots,\theta_n\}$
on the unit circle in the $\zeta$ frame:
\begin{equation}
    \{t^{(1)}_1,\ldots,t^{(k)}_{n_k}\}~\to~ \{e^{i\theta_1},\ldots,e^{i\theta_n}\}
    \qquad\text{with }\quad 0\leq\theta_1\leq \dots\leq\theta_q\leq\dots\leq\theta_n\leq2\pi\,.
\end{equation}
Note that this is a map between sets, and the order of insertion points $t^{(i)}_j$ in the $z$-frame will in general be a cyclic permutation of the ordering of the angles $\theta_q$. In particular, we do not expect that $t_1^{(1)}$ is necessarily mapped to $\theta_1$.

We pick any one of the angles in the set $\{\theta_1,\ldots,\theta_n\}$, and denote its index by $\tilde q$. We now vary this angle, keeping all other angles fixed to their original values. In appendix~\ref{appKO3} we  show that for any $\theta_{\tilde q-1}\leq \tilde\theta\leq \theta_{\tilde q+1}$, we can find a partition $\tilde{\vec n}$ such that some point in its integration region $\Gamma(\tilde{\vec n})$ maps to the positions
\begin{equation}\label{zkalphamap}
     \{e^{i\theta_1},\dots,e^{i\theta_{\tilde q-1}},e^{i\tilde\theta},e^{i\theta_{\tilde q+1}},\dots,e^{i\theta_n}\}\,
\end{equation}
in the $\zeta$ frame.
As $\tilde\theta$ is varied, one generically reaches the boundary of the integration region
 $\Gamma(\vec n)$
of the current partition $\vec n$.
 We show that such points can always be smoothly matched to the boundary of
 the integration region $\Gamma(\tilde{\vec n})$ of
a different partition $\tilde{\vec n}$.
 The
variation of $\tilde\theta$ can thus
continue until it either coincides with $\theta_{\tilde q-1}$ or $\theta_{\tilde q+1}$.\footnote{
For the special case of $\tilde{q}=1$, the lower boundary
of the variation is $\theta_1=0$.
Similarly, the upper boundary of the variation for $\tilde{q}=n$
is $\theta_n=2\pi$.}
 Denoting the image of the region $\Gamma(\vec n)$ in the $\zeta$ frame by $\zeta\bigl(\Gamma(\vec n)\bigr)$, we conclude that
\begin{equation}\label{toproveinappendix}
    \{e^{i\theta_1},\dots,e^{i\theta_{\tilde q-1}},e^{i\tilde\theta},e^{i\theta_{\tilde q+1}},\dots,e^{i\theta_n}\}
    \,\subset\, \bigcup_{\vec n}\,\zeta\bigl(\Gamma(\vec n)\bigr)
    \quad\text{ for all }\theta_{\tilde q-1}\leq\tilde\theta\leq\theta_{\tilde q+1}\,.
\end{equation}
Note that the angles $\theta_q$
with $q\neq\tilde q$ are not arbitrary
because they are determined by the point in the integration  $\vec n$ that we picked originally as the starting point for the argument.
It is obvious, however, that
we can use the above argument iteratively for all $1\leq\tilde q\leq n$ and complete the integration
 region to all $\{\theta_1,\ldots,\theta_n\}$ satisfying
 \begin{equation}\label{expintreg}
    0\leq\theta_1\leq\ldots\leq \theta_n\leq2\pi\,.
 \end{equation}
This is the integration region
expected from~(\ref{wanttoshow})
at $\ord{\lambda^n}$.
As the above argument requires a
choice of a starting point, it
does not rule out multiple covering and we conclude that
\begin{equation}
\ket{\mB(\Psi_L)}
= \sum_{n=0}^\infty \, \frac{C_n}{n!}
\biggl[ \, \lambda \int_0^{2 \pi} d \theta \,
V(\theta) \, \biggr]^n \, \ket{B}
\quad\text{ for some }C_n\in\mathbb{N}\,.
\end{equation}

It now remains to show that  $C_n=1$ for all $n$.
Consider any partition $\vec n=\{n_1,\dots,n_k\}$ with $k\geq2$.
In the natural  $z$ frame,
the last operator insertion of  $\Psi^{(n_1)}$ and the first operator insertion of $\Psi^{(n_2)}$ are located at
\begin{equation}
    t_{n_1}^{(1)}
    \leq e^{s_1}n_1+a_0 \,,\qquad
    t_{1}^{(2)}
    =  \, e^{s_1}n_1 + e^{s_2} + a_0
\end{equation}
with
\begin{equation}
    a_0 = \frac{1}{e^s-1} \sum_{i=1}^k n_i \, e^{s_i}\leq\frac{e^sn}{e^s-1}\,.
\end{equation}
In the $\zeta$ frame, their
angular separation $\Delta\theta$ is
thus bounded from below:
\begin{equation}
    \Delta\theta\geq \Delta\theta_{\rm min}\quad\text{ with }\quad
    \Delta\theta_{\rm min}=\frac{2\pi}{s}\log\biggl[\frac{e^sn+(e^s-1)(n_1+1)}{e^sn+(e^s-1)n_1}\biggr]\,.
\end{equation}
The lower bound $\Delta\theta_{\rm min}>0$ is independent of the Schwinger parameters $s_i$.
Using cyclicity, we similarly conclude that
\begin{equation}
    \Delta\theta\leq 2\pi-\Delta\theta_{\rm min}\,.
\end{equation}
Now consider the subset of the integration
region~(\ref{expintreg}) where
all operator insertions are separated by less than the angle $\Delta\theta_{\rm min}$,
namely,
\begin{equation}\label{asmallball}
    |\theta_{i}-\theta_{j}|<\Delta\theta_{\rm min}\quad\text{ for all }1\leq i,j\leq n\,.
\end{equation}
This is a finite
region for finite $s$
which can only be covered by $\zeta\bigl(\Gamma(\vec n)\bigr)$
with the partition $\vec n=\{n\}$ of $k=1$.
On the other hand, we have shown in~(\ref{rots1})
that the region from $\ket{\mB^{(1)} (\Psi_L^{(n)})}$
covers the rotational modulus exactly once,
so the subset~(\ref{asmallball}) of the integration region is covered precisely once.
Therefore, there cannot be multiple coverings
of the integration region~(\ref{expintreg})
and we
obtain
\begin{equation}
    C_n=1\quad\text{ for all }~n\,.
\end{equation}
Recalling the relation~(\ref{L=real}),
we conclude that
\begin{equation}
\boxed{\phantom{\Biggl\{}
\ket{\mB (\Psi_L)} =\ket{\mB (\Psi_{\rm real})} =
\exp \,
\biggl[ \, \lambda \int_0^{2 \pi} d \theta \,
V(\theta) \, \biggr] \, \ket{B} \,.
~}
\end{equation}

\section{Discussion}
\label{secdisc}
\setcounter{equation}{0}

In this paper we have constructed a class of BRST-invariant closed string states $\ket{\mB(\Psi)}$ for any open string field solution $\Psi$.
The construction depends on a choice of a propagator strip.
Modifying
the propagator strip
or performing
a gauge transformation
on the classical solution
generically
changes
$\ket{\mB(\Psi)}$
by a BRST-exact term.
We calculated $\ket{\mB(\Psi)}$
for various known analytic solutions
choosing the Schnabl propagator strip
and found that
$\ket{\mB(\Psi)}$ precisely coincides
with the BCFT boundary state
$\ket{B_*}$ of the background that the solutions are expected to describe.
This is the first construction of the
full BCFT boundary state from solutions of
open string field theory.

 While we claim that the state $\ket{\mB (\Psi)}$ in general
 coincides with the BCFT boundary state $\ket{\mB}$
 up to a possible  BRST-exact term,
  such a term can be absent if 
 the state $\ket{\mB (\Psi)}$
 factorizes into the matter and ghost sectors as~(\ref{factorized}).
 We presented a sufficient condition on the solution $\Psi$
 such that $\ket{\mB (\Psi)}$, constructed with the choice of the Schnabl propagator strip,
 factorizes in this way.
It would be
useful to understand better
when the factorization~(\ref{factorized}) holds.
Our analysis indicates that
the remarkable simplifications associated with the choice
of Schnabl propagator strip
play an important role for the factorization~(\ref{factorized}).
It follows from the analysis in~\cite{Kiermaier:2008jy} that
all Riemann surfaces associated with
closed string states
of the form $\oint_s \Sigma (0,s)$
coincide in the Schnabl limit 
when $A_i$ in the definition~(\ref{Sigma}) of $\Sigma (0,s)$
are wedge-based states.
 In fact, the resulting Riemann surface is an annulus whose modulus only depends on $s$.
This outstanding feature was crucially important in this paper.
We expect from the analysis in~\cite{0611110}
that the simplifications
associated with
Schnabl's propagator
carry over
into
other projector-based propagators.

Because the boundary state is a basic
object in  BCFT, we believe that
our construction of $\ket{\mB}$ from a solution $\Psi$
provides an important step towards establishing
the map from classical solutions of open string field theory
to BCFT's.  Partial success in
the reverse map from BCFT's to classical solutions
was achieved
 in~\cite{0704.2222,0707.4472, 0708.3394}
for backgrounds connected by arbitrary marginal deformations.
A systematic procedure to construct solutions from
the BCFT operator that implements a change of boundary conditions
along a segment on the boundary was presented
for the bosonic string in~\cite{0707.4472}
and for the superstring in~\cite{0708.3394}.
Our ambitious goal is a complete understanding
of the relation between BCFT's
and classical solutions of open sting field theory.

The construction of the closed string state $\ket{\mB (\Psi)}$
is based on the representation~(\ref{defB})
of the original BCFT boundary state $\ket{B}$
in terms of the half-propagator strip $\PR (0,s)$.
The state $\ket{\mB (\Psi)}$ is obtained by
replacing $\PR (0,s)$ in~(\ref{defB})
with the half-propagator strip $\mP (0,s)$
associated with the background~$\Psi$.
This construction of $\ket{\mB (\Psi)}$
is the primary reason for our claim
that the state $\ket{\mB (\Psi)}$ coincides
with the BCFT boundary state $\ket{\mB}$
up to a possible BRST-exact term,
but we have not provided a proof for this claim.
We have in addition examined
two lines of argumentation.
The first one is based on the limit $s \to 0$
discussed in \subs\ref{secs0}.
Since $\ket{\mB(\Psi)}$ changes only by a BRST-exact term
as we vary the parameter $s$,
our claim follows if the relation~(\ref{observablesSiegel}) holds
and if the claim~(\ref{ellwood-claim}) discussed
in~\cite{Ellwood:2008jh} is proven.
The second argument is that our claim is
a consequence of the expected,
but not yet proven, background independence
of a version of open-closed string field theory.
We may obtain new insights from
further efforts towards a rigorous proof for the claim.

It would be interesting
to study the generalization of
our construction
to open superstring field theory
in the WZW formulation
by Berkovits~\cite{Berkovits:1998bt,Berkovits:2000fe}.
We can construct BRST-invariant closed string states
in the superstring by replacing $\Psi$ in $\ket{\mB (\Psi)}$
with $e^{-\Phi} Q \, e^\Phi$,
where $\Phi$ is the open superstring field
and $Q$ is the BRST operator in the superstring.
We expect that the resulting states
$\ket{\mB (e^{-\Phi} Q \, e^\Phi)}$ are related
to the BCFT boundary state.
It would be interesting to study such closed string states
extending the discussion in~\cite{Ellwood:2008jh}
on gauge-invariant
observables~\cite{Michishita:2004rx}
for the superstring.
The construction of the BCFT boundary state in the superstring from
open superstring fields can then be used
for consistency checks on solutions of open superstring field theory.
For example, a solution for the tachyon vacuum in this theory has been
proposed in~\cite{Fuchs:2008zx},
and it would be useful to examine if the state
$\ket{\mB (e^{-\Phi} Q \, e^\Phi)}$
vanishes for this solution.

We hope that our construction of $\ket{\mB (\Psi)}$
will pave the way to
the study of closed string physics within string field theory.
It may lead us to a novel formulation of open-closed string field theory,
and the set of open-closed vertices
encoded in
$\ket{\mB(\Psi)}$ is the
first step in this direction.
If we choose a propagator strip
associated with
a non-BPZ-even gauge condition,
we obtain
complex open-closed vertices. To obtain real vertices suitable for
open-closed string field theory,
it may be useful
to examine if our construction can be generalized
to the full propagator surface of regular linear $b$-gauges~\cite{Kiermaier:2007jg}.

Since gravity is contained in the closed string sector,
a consistent coupling of open strings
to an off-shell closed string in the framework of string field theory
can be thought of as a string theory generalization of
the energy-momentum tensor.
Apart from the reality issue we mentioned earlier,
the state $\ket{\mB (\Psi)}$ can thus be
regarded as giving such a generalized energy-momentum tensor.
Its expression
in terms of the path-ordered exponential~(\ref{defhalf-prop-strip})
is reminiscent of the energy-momentum tensor
of noncommutative gauge theory
in terms of open Wilson lines derived in~\cite{Okawa:2000sh}.
While the on-shell part of $\ket{\mB (\Psi)}$ is gauge-invariant,
the off-shell part is not.
We believe that information contained in its off-shell part,
especially when $\ket{\mB (\Psi)}$ coincides
with the BCFT boundary state $\ket{\mB}$,
is useful in understanding the map from solutions to BCFT's,
but it is an important open problem
whether physical observables
are contained in the off-shell part
in the context of string theory
or of string field theory.
For example, off-shell information in the BCFT boundary state
was used in the study of the rolling tachyon
by Sen~\cite{Sen:2002nu}.
Finally, the coupling of open string fields to closed string modes
plays an important role in the AdS/CFT correspondence.
The study of open string field theory
with such open-closed vertices
reviewed in section~\ref{secocsft}
indicates that a large amount of
the closed string physics
can in principle be reproduced,
and the results in this paper further provide
a prospect that they might actually be calculable.
We hope that exciting developments await us
in this direction.

%And we hope to derive the AdS/CFT correspondence
%using string field theory within three years.

\section*{Acknowledgements}

We would like to thank
Kentaro Hori, Taichiro Kugo, and Hiroshi Kunitomo for useful discussions.
We also
thank Ian Ellwood, Leonardo Rastelli, Martin Schnabl, and Ashoke Sen
for discussions and for their valuable comments
on an earlier version of the manuscript.
M.K.
would like to thank the members of
the University of Tokyo at the IPMU and at Komaba
for stimulating discussions and hospitality during this work.
Y.O. acknowledges the kind hospitality
of the IPMU at the University of Tokyo
during part of this work.
 The work of M.K. was  supported in part by the World Premier International
 Research Center Initiative of MEXT, Japan.
The work of M.K.
and B.Z.
is supported by the U.S. DOE grant DE-FG02-05ER41360.
The work of Y.O. is supported in part
by Grant-in-Aid for Scientific Research [No.~20340048 (B)] from
the Ministry of  Education, Science and Culture of Japan.

\bigskip

\appendix

\section{Marginal deformations in Schnabl gauge}
\setcounter{equation}{0}

\subsection{The matter sector to all orders}\label{appKORZ1}

In this appendix we derive~(\ref{KORZ-matter}).
We generalize the calculations in the subsection
to the following case:
\begin{equation}
\begin{split}
& (-1)^k e^{\frac{\pi^2}{s} ( L_0 + \tilde{L}_0 )} \,\oint_s
\int_0^s ds_1 \int_{s_1}^s ds_2 \ldots \int_{s_{k-1}}^s ds_k \,
\PR (0,s_1) \, \{ {\cal B}_R (s_1), \Psi^{(n_1)} \} \\
& \quad \times
\PR (s_1,s_2) \, \{ {\cal B}_R (s_2), \Psi^{(n_2)} \} \,
\PR (s_2,s_3) \, \{ {\cal B}_R (s_3), \Psi^{(n_3)} \} \, \ldots \\
& \quad \times
\PR (s_{k-1},s_k) \, \{ {\cal B}_R (s_k), \Psi^{(n_k)} \} \,
\PR (s_k,s) \,,
\end{split}
\end{equation}
where
\begin{equation}
\sum_{i=1}^k n_i = n \,.
\end{equation}
The operator insertion in the natural $z$ frame is given by
\begin{equation}
\begin{split}
& \frac{e^s}{e^s-1} \,
\int_0^s ds_1 \int_{s_1}^s ds_2 \ldots \int_{s_{k-1}}^s ds_k \,
\prod_{i=1}^k \, \biggl[ \,
\int_0^1 dt^{(i)}_1 \int_0^1 dt^{(i)}_2 \ldots
\int_0^1 dt^{(i)}_{n_i-1} \, ( e^{s_i} )^{n_i} \, \biggr] \\
& \quad ~ \times \prod_{a=1}^n
V \Bigl( \, \frac{e^s e^{s'_1} + \ldots + e^s e^{s'_a}
+ e^{s'_{a+1}} + \ldots + e^{s'_n}}{e^s-1} \, \Bigr) \,,
\end{split}
\end{equation}
where
\begin{equation}
s'_a = s_i + \ln t^{(i)}_{j-1} \quad
\text{with} \quad t^{(i)}_0 = 1
\end{equation}
for
\begin{equation}
a = n_1 + n_2 + \ldots + n_{i-1} + j \,, \quad
\text{with} \quad
j = 1, 2, \ldots , n_i \,.
\end{equation}
The integral can be written in terms of $s'_a$ as follows:
\begin{equation}
\frac{e^s}{e^s-1} \,
\int_{\Gamma (n_1, n_2, \ldots, n_k)}
ds'_1 ds'_2 \ldots ds'_n \, \prod_{a=1}^n
e^{s'_a} \, V \Bigl( \, \frac{e^s e^{s'_1} + \ldots + e^s e^{s'_a}
+ e^{s'_{a+1}} + \ldots + e^{s'_n}}{e^s-1} \, \Bigr) \,,
\end{equation}
where
 \begin{equation}
 \int_{\Gamma (n_1, n_2, \ldots, n_k)}
 ds'_1 ds'_2 \ldots ds'_n
 = \prod_{i=1}^k \int_{s'_{d_{i-1}+1}}^s
 ds'_{d_i+1}
 \int_{-\infty}^{s'_{d_i+1}}
 ds'_{d_i+2}
 \int_{-\infty}^{s'_{d_i+1}}
 ds'_{d_i+3} \ldots
 \int_{-\infty}^{s'_{d_i+1}}
 ds'_{d_i+n_i}
 \end{equation}
 with
 \begin{equation}
  d_1 = 0 \,, \quad
  d_i = \sum_{j=1}^{i-1} n_j \quad \text{for} \quad i > 1 \,, \quad
   s'_{d_0 + 1} = 0 \,.
 \end{equation}
For example, when $(n_1, n_2, n_3) = (3, 2, 1)$, we have
\begin{equation}
\int_{\Gamma (3, 2, 1)}
ds'_1 ds'_2 \ldots ds'_6
= \int_0^s ds'_1
\int_{-\infty}^{s'_1} ds'_2
\int_{-\infty}^{s'_1} ds'_3
\int_{s'_1}^{s} ds'_4
\int_{-\infty}^{s'_4} ds'_5
\int_{s'_4}^{s} ds'_6 \,.
\end{equation}
Note that $\Gamma (n_1, n_2, \ldots, n_k)$ satisfies
the following relation:
\begin{equation}
\begin{split}
& \int_{\Gamma (n_1, n_2, \ldots, n_k)}
ds'_1 ds'_2 \ldots ds'_n
\int_{-\infty}^s ds'_{n+1} \\
& \int_{\Gamma (n_1, n_2, \ldots, n_k)}
ds'_1 ds'_2 \ldots ds'_n
\int_{-\infty}^{s'_{b_k + 1}} ds'_{n+1}
+ \int_{\Gamma (n_1, n_2, \ldots, n_k)}
ds'_1 ds'_2 \ldots ds'_n
\int_{s'_{b_k + 1}}^s ds'_{n+1} \\
& = \int_{\Gamma (n_1, n_2, \ldots, n_{k-1}, n_k + 1)}
ds'_1 ds'_2 \ldots ds'_{n+1}
+ \int_{\Gamma (n_1, n_2, \ldots, n_k, 1)}
ds'_1 ds'_2 \ldots ds'_{n+1} \,.
\end{split}
\end{equation}
Using this relation recursively,
the integration regions from all partitions are combined to give
\begin{equation}
\sum_{n_1 + n_2 + \ldots + n_k = n}
\int_{\Gamma (n_1, n_2, \ldots, n_k)}
ds'_1 ds'_2 \ldots ds'_n
= \int_0^s ds'_1 \int_{-\infty}^s ds'_2
\int_{-\infty}^s ds'_3 \ldots \int_{-\infty}^s ds'_n \,.
\end{equation}
If we define
\begin{equation}
u_a = \frac{e^s e^{s'_1} + \ldots + e^s e^{s'_a}
+ e^{s'_{a+1}} + \ldots + e^{s'_n}}{e^s-1} \quad
\text{for} \quad 1 \le a \le n \,,
\end{equation}
the contributions from all partitions can be written
in the following form:
\begin{equation}
\begin{split}
& \frac{e^s}{e^s-1} \,
\int_0^s ds'_1 \int_{-\infty}^s ds'_2
\int_{-\infty}^s ds'_3 \ldots \int_{-\infty}^s ds'_n \,
\prod_{a=1}^n
e^{s'_a} \, V \Bigl( \, \frac{e^s e^{s'_1} + \ldots + e^s e^{s'_a}
+ e^{s'_{a+1}} + \ldots + e^{s'_n}}{e^s-1} \, \Bigr) \\
& = \int_{\Gamma^{(n)'}} du_1 du_2 \ldots du_n \,
V(u_1) \, V(u_2) \, \ldots \, V(u_n) \,.
\end{split}
\end{equation}
To see this, note that
\begin{equation}
\begin{split}
\frac{\partial (u_1, u_2, \ldots, u_n)}
{\partial (s'_1, s'_2, \ldots, s'_n)}
& = \frac{1}{(e^s-1)^n} \, \prod_{a=1}^n e^{s'_a} \,
\left|
\begin{array}{cccccc}
e^s & 1 & 1 & \ldots & 1 & 1 \\
e^s & e^s & 1 & \ldots & 1 & 1 \\
e^s & e^s & e^s  & \ldots & 1 & 1 \\
\vdots & \vdots & \vdots & \ddots & \vdots & \vdots \\
e^s & e^s & e^s  & \ldots & e^s & 1 \\
e^s & e^s & e^s  & \ldots & e^s & e^s \\
\end{array}
\right| \\
& = \frac{e^s (e^s-1)^{n-1}}{(e^s-1)^n} \,
\prod_{a=1}^n e^{s'_a}
= \frac{e^s}{e^s-1} \, \prod_{a=1}^n e^{s'_a} \,.
\end{split}
\end{equation}
Since
\begin{equation}
e^{s'_1} = u_1 - e^{-s} u_n \,, \quad
e^{s'_2} = u_2 - u_1 \,, \quad
e^{s'_3} = u_3 - u_2 \,, \quad \ldots \quad
e^{s'_n} = u_n - u_{n-1} \,,
\end{equation}
the integration region $\Gamma^{(n)'}$
can be characterized by
\begin{equation}
1 \le u_1 - e^{-s} u_n \le e^s \,, \quad
0 \le u_2 - u_1 \le e^s \,, \quad
0 \le u_3 - u_2 \le e^s \,, \quad \ldots \quad
0 \le u_n - u_{n-1} \le e^s \,.
\end{equation}
Using the identification $z \sim e^s z$
in the natural $z$ frame, this integral can also be written as
\begin{equation}
\int_{\Gamma^{(n)}} du_1 du_2 \ldots du_n \,
V (u_1) \, V (u_2) \, \ldots \, V(u_n)
\end{equation}
with $\Gamma^{(n)}$ given by
\begin{equation}\label{GammanA1}
1 \le e^s u_1 - u_n \le e^s \,, \quad
0 \le u_2 - u_1 \le 1 \,, \quad
0 \le u_3 - u_2 \le 1 \,, \quad \ldots \quad
0 \le u_n - u_{n-1} \le 1 \,.
\end{equation}
This completes the derivation of~(\ref{KORZ-matter}).

\subsection{Recovering the BCFT boundary state}\label{appKORZ2}
The matter sector of the boundary state $\ket{\mB (\Psi)}$
with $\Psi$ being solutions for marginal deformations
can be written in the $z$ frame
with the identification $z \sim e^s z$ as follows:
\begin{equation}
\sum_{n=0}^\infty \lambda^n
\int_{\Gamma^{(n)}} du_1 du_2 \ldots du_n \,
V(u_1) \, V(u_2) \, \ldots V(u_n) \,,
\end{equation}
 with $\Gamma^{(n)}$ given in~(\ref{GammanA1}).
In this appendix we show that
 \begin{equation}\label{Gammanclaim}
 \int_{\Gamma^{(n)}}\!\! du_1 du_2 \,.\,.\, du_n \,
 V(u_1) \, V(u_2) \,.\,.\, V(u_n)
 = \int^{e^s}_1\!\!\! du_1\int_{u_1}^{e^s}\!\!\! du_2\,.\,.\,\int_{u_{n-1}}^{e^s}\!\!\! du_n\,V(u_1) \, V(u_2) \,.\,.\, V(u_n)\,.
 \end{equation}
Let us first consider the
 case with $n=2$.
We define a region $U^{(2)}$ in the
$(u_1,u_2)$ plane by
\begin{equation}
U^{(2)}\,\equiv\,\bigl\{\,(u_1,u_2) ~\bigl|~ u_1 \le u_2 \,,~ u_2 \le e^s u_1\,\bigr\}\,.
\end{equation}
The region $\Gamma^{(2)}$ is the subset of $U^{(2)}$ given by
\begin{equation}
\Gamma^{(2)}\,\equiv\,\bigl\{\,(u_1,u_2) ~\bigl|~ 0 \le u_2 - u_1 \le 1 \,,~ 1 \le e^s u_1 - u_2 \le e^s\,\bigr\}\,.
\end{equation}
If we define $\Gamma (a_1, a_2)\subset U^{(2)}$ by
\begin{equation}
\Gamma (a_1, a_2)\,\equiv\,\bigl\{\,(u_1,u_2) ~\bigl|~ 0 \le u_2 - u_1 \le a_1 \,,~0 \le e^s u_1 - u_2 \le a_2 \,\bigr\}\,,
\end{equation}
the region $\Gamma^{(2)}$ can be written as
\begin{equation}
\Gamma^{(2)} = \Gamma (1,e^s) - \Gamma (1,1) \,.
\end{equation}
 Consider the map $g:~U^{(2)}\,\to\,U^{(2)}$ given by
 \begin{equation}
  g\bigl((u_1,u_2)\bigr)=(u_2,e^s u_1) \,.
 \end{equation}
 Note that the set of angles $\{\theta_1,\theta_2\}$ in the $\zeta$ frame that the points $(u_1,u_2)$ are mapped to via~(\ref{zetaz}) is invariant under $g$.
 The region $\Gamma (a_1, a_2)$ is mapped under $g$ as
\begin{equation}
 \Gamma (a_1, a_2) ~ \overset{g}{\longrightarrow} ~ \Gamma (a_2, e^s a_1) \,.
\end{equation}
Thus the region $\Gamma^{(2)}$ is mapped
by a sequence of maps $g$ as follows:
\begin{equation}
\begin{split}
\Gamma^{(2)} \,=\,\,\,& \Gamma (1,e^s) - \Gamma (1,1)
 ~ \overset{g}{\longrightarrow} ~
\Gamma (e^s,e^s) - \Gamma (1,e^s) \\
 ~ \overset{g}{\longrightarrow}& ~
\Gamma (e^s,e^{2s}) - \Gamma (e^s,e^s)
 ~ \overset{g}{\longrightarrow} ~
\Gamma (e^{2s},e^{2s}) - \Gamma (e^s,e^{2s})
   ~ \overset{g}{\longrightarrow} ~\ldots ~.
\end{split}
\end{equation}
 The map  $g$  is invertible and its inverse is given by
\begin{equation}
 g^{-1}\bigl((u_1,u_2)\bigr)=(e^{-s}u_2, u_1) \,.
\end{equation}
Under $g^{-1}$, the region $\Gamma (a_1, a_2)$ is mapped as
\begin{equation}
\Gamma (a_1, a_2)
 ~ \overset{~g^{-1}}{\longrightarrow} ~
 \Gamma (e^{-s} a_2, a_1) \,.
\end{equation}
Thus the region $\Gamma^{(2)}$ is mapped
by a sequence of maps $g^{-1}$ as follows:
\begin{equation}
\begin{split}
\Gamma^{(2)} \, =\,\,\,& \Gamma (1,e^s) - \Gamma (1,1)
 ~ \overset{~g^{-1}}{\longrightarrow} ~
\Gamma (1,1) - \Gamma (e^{-s},1) \\
 ~ \overset{~g^{-1}}{\longrightarrow}& ~
\Gamma (e^{-s},1) - \Gamma (e^{-s},e^{-s})
 ~ \overset{~g^{-1}}{\longrightarrow} ~
\Gamma (e^{-s},e^{-s}) - \Gamma (e^{-2s},e^{-s})
  ~ \overset{~g^{-1}}{\longrightarrow} ~ \ldots ~.
\end{split}
\end{equation}
Therefore, any point $(u_1, u_2)\in U^{(2)}$
can be mapped to $\Gamma^{(2)}$
either by a sequence of the map $g$ or by a sequence of the map $g^{-1}$. Let us denote this map by $G$. This map
\begin{equation}
    G:~U^{(2)}~\to~\Gamma^{(2)}
\end{equation}
is uniquely defined
because the images of $\Gamma^{(2)}$ under different sequences of either $g$ or $g^{-1}$ do not intersect.
Furthermore, $G$ is onto because it is the identity map when restricted to $(u_1,u_2)\in\Gamma^{(2)}$.

The region $\widetilde{\Gamma}^{(2)}$
for the path-ordered exponential
in~(\ref{Gammanclaim})
at ${\cal O} (\lambda^2)$
is the subset of $U^{(2)}$ given by
\begin{equation}
\widetilde{\Gamma}^{(2)}\,\equiv\,\bigl\{\,(u_1,u_2) ~\bigl|~ 1 \le u_1 \le u_2 \le e^s\,\bigr\}\,.
\end{equation}
If we define $\Gamma (u_i \le a)\subset U^{(2)}$ by
\begin{equation}
\Gamma (u_i \le a)\,\equiv\,\bigl\{\,(u_1,u_2) ~\bigl|~
u_1 \le u_2 \,,~
u_2 \le e^s u_1 \,,~
u_i \le a
\,\bigr\}\,,
\end{equation}
the region $\widetilde{\Gamma}^{(2)}$ can be written as
\begin{equation}
\widetilde{\Gamma}^{(2)}
= \Gamma (u_2 \le e^s) - \Gamma (u_1 \le 1) \,.
\end{equation}
Under the map $g$,
the region $\Gamma (u_i \le a)$ is mapped as
\begin{equation}
\Gamma (u_1 \le a)
 ~ \overset{g}{\longrightarrow} ~
 \Gamma (u_2 \le e^s a) \,, \quad
\Gamma (u_2 \le a)
 ~ \overset{g}{\longrightarrow} ~
 \Gamma (u_1 \le a) \,.
\end{equation}
Thus the region $\widetilde{\Gamma}^{(2)}$ is mapped
by a sequence of maps $g$ as follows:
\begin{equation}
\begin{split}
\widetilde{\Gamma}^{(2)}
\, =\,\,\,& \Gamma (u_2 \le e^s) - \Gamma (u_1 \le 1)
 ~ \overset{g}{\longrightarrow} ~  \Gamma (u_1 \le e^s) - \Gamma (u_2 \le e^s) \\
  ~ \overset{g}{\longrightarrow}& ~
 \Gamma (u_2 \le e^{2s}) - \Gamma (u_1 \le e^s)
 ~ \overset{g}{\longrightarrow} ~
 \Gamma (u_1 \le e^{2s}) - \Gamma (u_2 \le e^{2s})
  ~ \overset{g}{\longrightarrow} ~\ldots ~.
\end{split}
\end{equation}
Under the inverse map $g^{-1}$, the region $\Gamma (u_i \le a)$ is mapped as
\begin{equation}
\Gamma (u_1 \le a)  ~ \overset{~g^{-1}}{\longrightarrow} ~ \Gamma (u_2 \le a) \,, \quad
\Gamma (u_2 \le a)  ~ \overset{~g^{-1}}{\longrightarrow} ~ \Gamma (u_1 \le e^{-s} a) \,.
\end{equation}
Thus the region $\widetilde{\Gamma}^{(2)}$ is mapped
by a sequence of maps $g^{-1}$ as follows:
\begin{equation}
\begin{split}
\widetilde{\Gamma}^{(2)}
\,=\,\,\,& \Gamma (u_2 \le e^s) - \Gamma (u_1 \le 1)
~ \overset{~g^{-1}}{\longrightarrow} ~
 \Gamma (u_1 \le 1) - \Gamma (u_2 \le 1) \\
~ \overset{~g^{-1}}{\longrightarrow}& ~ \Gamma (u_2 \le 1) - \Gamma (u_1 \le e^{-s})
~ \overset{~g^{-1}}{\longrightarrow} ~ \Gamma (u_1 \le e^{-s}) - \Gamma (u_2 \le e^{-s})
  ~ \overset{~g^{-1}}{\longrightarrow}~ \ldots ~.
\end{split}
\end{equation}

Therefore, any point $(u_1, u_2)\in U^{(2)}$
can be mapped to $\widetilde{\Gamma}^{(2)}$
either by a sequence of maps $g$ or by a sequence of maps $g^{-1}$. Let us denote this map by $\widetilde{G}$. This map
\begin{equation}
    \widetilde{G}:~U^{(2)}~\to~\widetilde{\Gamma}^{(2)}
\end{equation}
is uniquely defined
because the images of $\widetilde{\Gamma}^{(2)}$ under different sequences of either $g$ or $g^{-1}$ do not intersect.
Furthermore, $\widetilde{G}$ is onto because it is the identity map when restricted to $(u_1,u_2)\in\widetilde{\Gamma}^{(2)}$.

We have thus constructed a map $G:\,U^{(2)}\,\to\,\Gamma^{(2)}$ and a map $\widetilde{G}:\,U^{(2)}\,\to\,\widetilde{\Gamma}^{(2)}$. Both maps are onto.
We now define a map $H:\widetilde{\Gamma}^{(2)}\to\Gamma^{(2)}$ that is the restriction of $G$ to $(u_1,u_2)\in\widetilde{\Gamma}^{(2)}$:
\begin{equation}
    H\,=\,G \, \Bigr|_{\widetilde{\Gamma}^{(2)}}\,.
\end{equation}
Similarly, we define $\widetilde{H}:\Gamma^{(2)}\to\widetilde{\Gamma}^{(2)}$ as the restriction of $\widetilde{G}$ to $(u_1,u_2)\in\Gamma^{(2)}$:
\begin{equation}
    \widetilde{H}\,=\,\widetilde{G} \, \Bigr|_{\Gamma^{(2)}}\,.
\end{equation}
The composition of these two maps, $\widetilde{H}\circ H$, is the identity map on $\widetilde{\Gamma}^{(2)}$.
To show this, assume the contrary,
{\em i.e.},
assume $(u'_1,u'_2)\neq(u_1,u_2)$ with
\begin{equation}
    (u'_1,u'_2)=\widetilde{H}\circ H\bigl((u_1,u_2)\bigr)\,.
\end{equation}
As $\widetilde{H}\circ H$ is built from sequences of $g$ and $g^{-1}$, the points $(u_1,u_2)\in\widetilde{\Gamma}^{(2)}$ and  $(u'_1,u'_2)\in\widetilde{\Gamma}^{(2)}$ are related by some sequence of maps $g$ or $g^{-1}$.
However, since
\begin{equation}
\begin{split}
  \underbrace{g\circ\ldots\circ g}_{i-\rm{times}}\bigl(\widetilde{\Gamma}^{(2)}\bigr)\,\cap\,\widetilde{\Gamma}^{(2)}
  \,=\,\emptyset\quad\text{for}\quad i\neq0\,,\\
  \underbrace{g^{-1}\circ\ldots\circ g^{-1}}_{j-\rm{times}}\bigl(\widetilde{\Gamma}^{(2)}\bigr)\,\cap\,\widetilde{\Gamma}^{(2)}
  \,=\,\emptyset\quad\text{for}\quad j\neq0\,,
\end{split}
\end{equation}
we conclude that they cannot be related by a
nontrivial
sequence and thus
\begin{equation}
    (u'_1,u'_2)=(u_1,u_2)\,,
\end{equation}
in contradiction with the assumption.
Thus $\widetilde{H}\circ H$ is
indeed the identity map on $\widetilde{\Gamma}^{(2)}$. Similarly, one can show that $H\circ\widetilde{H}$ is the identity map on $\Gamma^{(2)}$. The maps $H$ and $\widetilde{H}$ are therefore inverses of each other, and in particular they must be one-to-one and onto.
We have thus constructed a map between the integration region $\Gamma^{(2)}$ and the integration region $\widetilde{\Gamma}^{(2)}$,
\begin{equation}
    \widetilde{H}:~\Gamma^{(2)}~\to~\widetilde{\Gamma}^{(2)}\,,
~\text{\,one-to-one\, and\, onto}\,.
\end{equation}
Note that for each fixed $(u_1,u_2)$ this map is either a
 finite\footnote{The maps $g$ and $g^{-1}$ have a fixed-point at the origin in the $(u_1,u_2)$ plane.
However,
as neither
 $\Gamma^{(2)}$ nor $\widetilde{\Gamma}^{(2)}$
contains
the origin, the map $\widetilde{H}$ is perfectly
well defined
and unaffected by this singularity.}
sequence of maps $g$ or a
 finite
sequence of maps $g^{-1}$. Furthermore, recall
that the set of angles $\{\theta_1,\theta_2\}$ in the $\zeta$ frame that the points $(u_1,u_2)$ map to, is invariant under the maps $g$ and $g^{-1}$.
We can thus decompose $\Gamma^{(2)}$ appropriately
and map each piece of the region
to reconstruct $\widetilde{\Gamma}^{(2)}$.
We can explicitly perform this procedure
for a given finite $s$.
We thus conclude that the integration regions $\Gamma^{(2)}$ and $\widetilde{\Gamma}^{(2)}$
are identical:
\begin{equation}
\boxed{\phantom{\Biggl(}
\int_{\Gamma^{(2)}} du_1 du_2 \, V(u_1) \, V(u_2)
= \int_{\widetilde{\Gamma}^{(2)}} du_1 du_2 \, V(u_1) \, V(u_2) \,.
~}
\end{equation}
 This proves~(\ref{Gammanclaim}) for $n=2$.

This proof
can be easily generalized to
 arbitrary $n>2$.
We define a region $U^{(n)}$
by
\begin{equation}
U^{(n)}\,\equiv\,\bigl\{\,(u_1,u_2,\ldots,u_n) ~\bigl|~ u_1 \le u_2 \,,~u_2 \le u_3
\,,\,\ldots\,,~u_{n-1} \le u_n\,,~ u_n \le e^s u_1\,\bigr\}\,.
\end{equation}
The region $\Gamma^{(n)}$ is the subset of $U^{(n)}$ given by
\begin{equation}
\Gamma^{(n)}\,
= \,
\bigl\{\,(u_1,u_2,\ldots,u_n) ~\bigl|~
0 \le u_2 - u_1 \le 1 \,,~ 0 \le u_3 - u_2 \le 1
\,,\,\ldots\,,~1 \le e^s u_1 - u_n \le e^s\,\bigr\}\,.
\end{equation}
If we define $\Gamma (a_1, a_2,\ldots,a_n)\subset U^{(n)}$ by
\begin{equation}
\begin{split}
\Gamma (a_1, a_2,\ldots,a_n)\equiv\,\bigl\{\,(u_1,u_2,\ldots,u_n) ~\bigl|&~
0 \le u_2 - u_1 \le a_1 \,,~ 0 \le u_3 - u_2 \le a_2
\,,\,\ldots\,,\\&
~0 \le u_{n} - u_{n-1} \le a_{n-1} \,,~
0 \le e^s u_1 - u_n \le a_n\,\bigr\}\,,
\end{split}
\end{equation}
the region $\Gamma^{(n)}$ can be written as
\begin{equation}
\Gamma^{(n)} = \Gamma (1,1,\ldots,1,e^s) - \Gamma (1,1,\ldots,1) \,.
\end{equation}
Consider the map $g:~U^{(n)}\,\to\,U^{(n)}$ given by
\begin{equation}
 g\bigl((u_1,u_2,\ldots,u_n)\bigr)=(u_2,u_3,\ldots,u_n,e^s u_1) \,.
\end{equation}
Note that the set of angles $\{\theta_1,\theta_2,\ldots,\theta_n\}$ in the $\zeta$ frame that the points $(u_1,u_2,\ldots,u_n)$ are mapped to via~(\ref{zetaz}) is invariant under $g$.
The region $\Gamma (a_1, a_2,\ldots,a_n)$ is mapped under $g$ as
\begin{equation}
 \Gamma (a_1, a_2,\ldots,a_n) ~ \overset{g}{\longrightarrow} ~ \Gamma (a_2, a_3,\ldots,a_n,e^sa_1)\,.
\end{equation}
Thus the region $\Gamma^{(n)}$ is mapped
by a sequence of maps $g$ as follows:
\begin{equation}
\begin{split}
\Gamma^{(n)}
\,=\,\,\,& \Gamma (1,1,\ldots,1,e^s) - \Gamma (1,1,\ldots,1) \\
 ~ \overset{g}{\longrightarrow}& ~ \Gamma (1,1,\ldots,1,e^s,e^s) - \Gamma (1,1,\ldots,1,e^s) \\
 ~ \overset{g}{\longrightarrow}& ~ \Gamma (1,1,\ldots,1,e^s,e^s,e^s)
- \Gamma (1,1,\ldots,1,e^s,e^s)
~ \overset{g}{\longrightarrow} ~\ldots ~.
\end{split}
\end{equation}
 The map  $g$  is invertible and its inverse is given by
\begin{equation}
 g^{-1}\bigl((u_1,u_2,\ldots,u_n)\bigr)=(e^{-s}u_n, u_1,u_2,\ldots,u_{n-1}) \,.
\end{equation}
Under $g^{-1}$, the region $\Gamma (a_1, a_2,\ldots,a_n)$ is mapped as
\begin{equation}
\Gamma (a_1, a_2,\ldots,a_n)
 ~ \overset{~g^{-1}}{\longrightarrow} ~
 \Gamma (e^{-s}a_n,a_1, a_2,\ldots,a_{n-1}) \,.
\end{equation}
Thus the region $\Gamma^{(n)}$ is mapped
by a sequence of maps $g^{-1}$ as follows:
\begin{equation}
\begin{split}
\Gamma^{(n)} \, =\,\,\,&
\Gamma (1,1,\ldots,1,e^s) - \Gamma (1,1,\ldots,1) \\
 ~ \overset{~g^{-1}}{\longrightarrow}& ~ \Gamma (1,1,\ldots,1) - \Gamma (e^{-s},1,1,\ldots,1) \\
 ~ \overset{~g^{-1}}{\longrightarrow}& ~ \Gamma (e^{-s},1,1,\ldots,1)
- \Gamma (e^{-s},e^{-s},1,1,\ldots,1)
 ~ \overset{~g^{-1}}{\longrightarrow} ~ \ldots ~.
\end{split}
\end{equation}
Therefore, any point $(u_1, u_2,\ldots,u_n)\in U^{(n)}$
can be mapped to $\Gamma^{(n)}$
either by a sequence of the map $g$ or by a sequence of the map $g^{-1}$. Let us denote this map by $G$. This map
\begin{equation}
    G:~U^{(n)}~\to~\Gamma^{(n)}
\end{equation}
is uniquely defined
because the images of $\Gamma^{(n)}$ under different sequences of either $g$ or $g^{-1}$ do not intersect.
Furthermore, $G$ is onto because it is the identity map when restricted to $(u_1,u_2,\ldots,u_n)\in\Gamma^{(n)}$.

The region $\widetilde{\Gamma}^{(n)}$
for the path-ordered exponential at ${\cal O} (\lambda^n)$ is the subset of $U^{(n)}$ given by
\begin{equation}
\widetilde{\Gamma}^{(n)}\,\equiv\,\bigl\{\,(u_1,u_2,\ldots,u_n) ~\bigl|~ 1 \le u_1 \le u_2\le\ldots \le u_n \le e^s\,\bigr\}\,.
\end{equation}
If we define $\Gamma (u_i \le a)\subset U^{(n)}$ by
\begin{equation}
\Gamma (u_i \le a)\,\equiv\,\bigl\{\,(u_1,u_2,\ldots,u_n) ~\bigl|~ u_1 \le u_2 \,,~u_2 \le u_3
\,,\,\ldots\,,~u_{n-1} \le u_n\,,~ u_n \le e^s u_1\,,~u_i\leq a\,\bigr\}\,,
\end{equation}
the region $\widetilde{\Gamma}^{(n)}$ can be written as
\begin{equation}
\widetilde{\Gamma}^{(n)}
= \Gamma (u_n \le e^s) - \Gamma (u_1 \le 1) \,.
\end{equation}
Under the map $g$,
the region $\Gamma (u_i \le a)$ is mapped as
\begin{equation}
\begin{split}
& \Gamma (u_1 \le a) ~ \overset{g}{\longrightarrow} ~ \Gamma (u_n \le e^s a) \,, \\
& \Gamma (u_2 \le a) ~ \overset{g}{\longrightarrow} ~ \Gamma (u_1 \le a) \,, \quad
\Gamma (u_3 \le a) ~ \overset{g}{\longrightarrow} ~ \Gamma (u_2 \le a) \,, \quad \ldots \quad
\Gamma (u_n \le a) ~ \overset{g}{\longrightarrow} ~ \Gamma (u_{n-1} \le a) \,.
\end{split}
\end{equation}
Thus the region $\widetilde{\Gamma}^{(n)}$ is mapped
by a sequence of maps $g$ as follows:
\begin{equation}
\begin{split}
\widetilde{\Gamma}^{(n)}
\, =\,\,\,& \Gamma (u_n \le e^s) - \Gamma (u_1 \le 1)
 ~ \overset{g}{\longrightarrow} ~  \Gamma (u_{n-1} \le e^s) - \Gamma (u_n \le e^s) \\
  ~ \overset{g}{\longrightarrow}& ~\Gamma (u_{n-2} \le e^{s}) - \Gamma (u_{n-1} \le e^s)
 ~ \overset{g}{\longrightarrow} ~\ldots \\
  ~ \overset{g}{\longrightarrow}& ~~~~\Gamma (u_{1} \le e^{s}) - \Gamma (u_{2} \le e^s)~~~
 ~ \overset{g}{\longrightarrow} ~\Gamma (u_{n} \le e^{2s}) - \Gamma (u_{1} \le e^s) \\
  ~ \overset{g}{\longrightarrow}& \,\Gamma (u_{n-1} \le e^{2s}) - \Gamma (u_{n} \le e^{2s})~\,
 ~ \overset{g}{\longrightarrow} ~\ldots \\
 \,.
\end{split}
\end{equation}
Under the inverse map $g^{-1}$, the region $\Gamma (u_i \le a)$ is mapped as
\begin{equation}
\begin{split}
& \Gamma (u_1 \le a)  ~ \overset{~g^{-1}}{\longrightarrow} ~ \Gamma (u_2 \le a) \,, \quad
\Gamma (u_2 \le a)  ~ \overset{~g^{-1}}{\longrightarrow} ~ \Gamma (u_3 \le a) \,, \quad \ldots \quad
\Gamma (u_{n-1} \le a)  ~ \overset{~g^{-1}}{\longrightarrow} ~ \Gamma (u_n \le a) \,, \\
& \Gamma (u_n \le a)  ~ \overset{~g^{-1}}{\longrightarrow} ~ \Gamma (u_1 \le e^{-s} a) \,.
\end{split}
\end{equation}
Thus the region $\widetilde{\Gamma}^{(n)}$ is mapped
by a sequence of maps $g^{-1}$ as follows:
\begin{equation}
\begin{split}
\widetilde{\Gamma}^{(n)}
\,=\,\,\,& \Gamma (u_n \le e^s) - \Gamma (u_1 \le 1)
~ \overset{~g^{-1}}{\longrightarrow} ~\Gamma (u_1 \le 1) - \Gamma (u_2 \le 1) \\
~ \overset{~g^{-1}}{\longrightarrow}& ~\Gamma (u_2 \le 1) - \Gamma (u_3 \le 1)
~ \overset{~g^{-1}}{\longrightarrow} ~\ldots \\
~ \overset{~g^{-1}}{\longrightarrow}& ~\Gamma (u_{n-1} \le 1) - \Gamma (u_{n} \le 1)
~ \overset{~g^{-1}}{\longrightarrow} ~\Gamma (u_{n} \le 1) - \Gamma (u_1 \le e^{-s})\\
~ \overset{~g^{-1}}{\longrightarrow}& ~\Gamma (u_1 \le e^{-s}) - \Gamma (u_2 \le e^{-s})
~ \overset{~g^{-1}}{\longrightarrow} ~\ldots ~.
\end{split}
\end{equation}
Therefore, any point $(u_1, u_2,\ldots,u_n)\in U^{(n)}$
can be mapped to $\widetilde{\Gamma}^{(n)}$
either by a sequence of maps $g$ or by a sequence of maps $g^{-1}$. Let us denote this map by $\widetilde{G}$. This map
\begin{equation}
    \widetilde{G}:~U^{(n)}~\to~\widetilde{\Gamma}^{(n)}
\end{equation}
is uniquely defined
because the images of $\widetilde{\Gamma}^{(n)}$ under different sequences of either $g$ or $g^{-1}$ do not intersect.
Furthermore, $\widetilde{G}$ is onto because it is the identity map when restricted to $(u_1,u_2,\ldots,u_n)\in\widetilde{\Gamma}^{(n)}$.

We have thus constructed a map $G:\,U^{(n)}\,\to\,\Gamma^{(n)}$ and a map $\widetilde{G}:\,U^{(n)}\,\to\,\widetilde{\Gamma}^{(n)}$. Both maps are onto.
We now define a map $H:\widetilde{\Gamma}^{(n)}\to\Gamma^{(n)}$ that is the restriction of $G$ to $(u_1,u_2,\ldots,u_n)\in\widetilde{\Gamma}^{(n)}$:
\begin{equation}
    H\,=\,G \, \Bigr|_{\widetilde{\Gamma}^{(n)}}\,.
\end{equation}
Similarly, we define $\widetilde{H}:\Gamma^{(n)}\to\widetilde{\Gamma}^{(n)}$ as the restriction of $\widetilde{G}$ to $(u_1,u_2,\ldots,u_n)\in\Gamma^{(n)}$:
\begin{equation}
    \widetilde{H}\,=\,\widetilde{G} \, \Bigr|_{\Gamma^{(n)}}\,.
\end{equation}
The composition of these two maps, $\widetilde{H}\circ H$, is the identity map on $\widetilde{\Gamma}^{(n)}$.
To show this, assume the contrary,
{\em i.e.},
assume $(u'_1,u'_2,\ldots,u'_n)\neq(u_1,u_2,\ldots,u_n)$ with
\begin{equation}
    (u'_1,u'_2,\ldots,u'_n)=\widetilde{H}\circ H\bigl((u_1,u_2,\ldots,u_n)\bigr)\,.
\end{equation}
As $\widetilde{H}\circ H$ is built from sequences of $g$ and $g^{-1}$, the points $(u_1,u_2,\ldots,u_n)\in\widetilde{\Gamma}^{(n)}$ and  $(u'_1,u'_2,\ldots,u'_n)\in\widetilde{\Gamma}^{(n)}$ are related by some sequence of maps $g$ or $g^{-1}$.
However, since
\begin{equation}
\begin{split}
  \underbrace{g\circ\ldots\circ g}_{i-\rm{times}}\bigl(\widetilde{\Gamma}^{(n)}\bigr)\,\cap\,\widetilde{\Gamma}^{(n)}
  \,=\,\emptyset\quad\text{for}\quad i\neq0\,,\\
  \underbrace{g^{-1}\circ\ldots\circ g^{-1}}_{j-\rm{times}}\bigl(\widetilde{\Gamma}^{(n)}\bigr)\,\cap\,\widetilde{\Gamma}^{(n)}
  \,=\,\emptyset\quad\text{for}\quad j\neq0\,,
\end{split}
\end{equation}
we conclude that they cannot be related by a
nontrivial
sequence and thus
\begin{equation}
    (u'_1,u'_2,\ldots,u'_n)=(u_1,u_2,\ldots,u_n)\,,
\end{equation}
in contradiction with the assumption.
Thus $\widetilde{H}\circ H$ is
indeed the identity map on $\widetilde{\Gamma}^{(n)}$. Similarly, one can show that $H\circ\widetilde{H}$ is the identity map on $\Gamma^{(n)}$. The maps $H$ and $\widetilde{H}$ are therefore inverses of each other, and in particular they must be one-to-one and onto.
We have thus constructed a map between the integration region $\Gamma^{(n)}$ and the integration region $\widetilde{\Gamma}^{(n)}$,
\begin{equation}
    \widetilde{H}:~\Gamma^{(n)}~\to~\widetilde{\Gamma}^{(n)}\,,
~
\text{\,one-to-one\, and\, onto}\,.
\end{equation}
Note that for each fixed $(u_1,u_2,\ldots,u_n)$ this map is either a
 finite
sequence of maps $g$ or a
 finite
sequence of maps $g^{-1}$. Furthermore,
 recall
that the set of angles $\{\theta_1,\theta_2,\ldots,\theta_n\}$ in the $\zeta$ frame that the points $(u_1,u_2,\ldots,u_n)$ map to, is invariant under the maps $g$ and $g^{-1}$.
We can thus decompose $\Gamma^{(n)}$ appropriately
and map each piece of the region
to reconstruct $\widetilde{\Gamma}^{(n)}$.
We can explicitly perform this procedure
for a given finite $s$.
We thus conclude that the integration regions $\Gamma^{(n)}$ and $\widetilde{\Gamma}^{(n)}$
are identical:
\begin{equation}
\boxed{\phantom{\Biggl(}
\int_{\Gamma^{(n)}} du_1 du_2\ldots du_n \, V(u_1) \, V(u_2)\ldots V(u_n)
= \int_{\widetilde{\Gamma}^{(n)}} du_1 du_2\ldots du_n  \, V(u_1) \, V(u_2)\ldots V(u_n) \,.
~}
\end{equation}
This
completes
our proof of
the
claim~(\ref{Gammanclaim}).

\section{The solution $\Psi_L$}
\setcounter{equation}{0}

\subsection{Ghost sector}\label{KOghost}

Let us consider
the ghost sector
of $\ket{\mB^{(k)}(\Psi_L)}$ in the natural $z$ frame.
The value of $a_0$ is
\begin{equation}\label{a0PsiL}
a_0 = \frac{1}{e^s-1} \sum_{i=1}^k \, n_i \, e^{s_i} \,.
\end{equation}
The ghost sector of the term~(\ref{mBorderk})
in the natural $z$ frame
can be written as
\begin{equation}
\prod_{i=1}^k \,
\biggl[ \, -
\int_{C(s_i)}
\frac{dz}{2 \pi i} \,
(z-\ell_i) \, b(z) \,
c (e^{s_i} + \ell_i)
- c (e^{s_i} + \ell_i)
\int_{C(s_i)}
\frac{dz}{2 \pi i} \, (z-\ell_{i+1}) \, b(z) \, \biggr] \,,
\end{equation}
where
\begin{equation}
\ell_i = \sum_{j=1}^{i-1} n_j \, e^{s_j} + a_0 \,, \quad
\ell_1 = a_0 \,.
\end{equation}
This can be calculated as
\begin{equation}
\begin{split}
& \prod_{i=1}^k \,
\biggl[ \, -
\int_{C(s_i)}
\frac{dz}{2 \pi i} \,
(z-\ell_i) \, b(z) \,
c (e^{s_i} + \ell_i)
- c (e^{s_i} + \ell_i)
\int_{C(s_i)}
\frac{dz}{2 \pi i} \, (z-\ell_{i+1}) \, b(z) \, \biggr] \\
& = \prod_{i=1}^k \,
\biggl[ \, \oint \frac{dz}{2 \pi i} \,
(z-\ell_i) \, b(z) \,
c (e^{s_i} + \ell_i)
+ n_i \, e^{s_i} \, c (e^{s_i} + \ell_i) \, {\cal B}^+_R \,
\biggr] \\
& = \prod_{i=1}^k \, e^{s_i} \, \Bigl[ \,
1 + n_i \, c (e^{s_i} + \ell_i) \, {\cal B}^+_R \, \Bigr] \,.
\end{split}
\end{equation}
 Using the same manipulations as in~(\ref{BcBcBc}) we find
\begin{equation}
c(t_1) \, {\cal B}^+_R \,
c(t_2) \, {\cal B}^+_R \, \ldots
c(t_m) \, {\cal B}^+_R
= (-1)^{m-1} \, c(t_1) \, {\cal B}^+_R
= \frac{(-1)^{m-1}}{e^s-1} \,.
\end{equation}
Therefore, we have
\begin{equation}
\prod_{i=1}^k \, e^{s_i} \, \Bigl[ \,
1 + n_i \, c (e^{s_i} + \ell_i) \, {\cal B}^+_R \, \Bigr]
 =  \Delta_k \, \prod_{i=1}^k \, e^{s_i}
\end{equation}
with
\begin{equation}
\Delta_k = 1 + \frac{1}{e^s-1}
- \frac{1}{e^s-1} \prod_{i=1}^k \, (1 - n_i) \,.
\end{equation}

\subsection{Measure}\label{KOmeasure}
In this appendix we calculate the Jacobian
\begin{equation}
\frac{\partial (t_1^{(1)}, t_1^{(2)}, \ldots , t_1^{(k)})}
{\partial (s_1, s_2, \ldots , s_k)} \,,
\end{equation}
where
\begin{equation}
t_1^{(i)} = \sum_{j=1}^{i-1} n_j \, e^{s_j} + e^{s_i} + a_0
\end{equation}
with
 $a_0$ given in~(\ref{a0PsiL}).
The derivative of $a_0$ with respect to $s_j$ is given by
\begin{equation}
\frac{\partial a_0}{\partial s_j}
= \frac{n_j}{e^s-1} \, e^{s_j} = b_j \, e^{s_j} \,,
\end{equation}
where
\begin{equation}
b_j \equiv \frac{n_j}{e^s-1} \,.
\end{equation}
We define $\widetilde{\Delta}_k$ by
\begin{equation}
\frac{\partial (t_1^{(1)}, t_1^{(2)}, \ldots , t_1^{(k)})}
{\partial (s_1, s_2, \ldots , s_k)}
= \widetilde{\Delta}_k \, \prod_{i=1}^k e^{s_i} \,.
\end{equation}
It follows from
\begin{equation}\label{dtidsj}
\frac{\partial t_1^{(i)}}{\partial s_j} =
\begin{cases}
~ ( n_j + b_j ) \, e^{s_j} \quad
& \text{for} \quad j < i \,, \\
~ ( 1 + b_j ) \, e^{s_j} \quad
& \text{for} \quad j = i \,, \\
~ b_j \, e^{s_j} \quad
& \text{for} \quad j > i
\end{cases}
\end{equation}
that
\begin{equation}
\begin{split}
\widetilde{\Delta}_k & = \left|
\begin{array}{cccccc}
1 + b_1 & b_2 & b_3 & \ldots & b_{k-1} & b_k \\
n_1 + b_1 & 1 + b_2 & b_3 & \ldots & b_{k-1} & b_k \\
n_1 + b_1 & n_2 + b_2 & 1 + b_3  & \ldots & b_{k-1} & b_k \\
\vdots & \vdots & \vdots & \ddots & \vdots & \vdots \\
n_1 + b_1 & n_2 + b_2 & n_3 + b_3  & \ldots & 1 + b_{k-1} & b_k \\
n_1 + b_1 & n_2 + b_2 & n_3 + b_3  & \ldots & n_{k-1} + b_{k-1} & 1 + b_k \\
\end{array}
\right| \,.
\end{split}
\end{equation}
Let us prove that
\begin{equation}
\widetilde{\Delta}_k
= \Delta_k = 1 + \frac{1}{e^s-1}
- \frac{1}{e^s-1} \, \prod_{i=1}^k \, (1-n_i) \,.
\end{equation}
For $k=1$, we have
\begin{equation}
\widetilde{\Delta}_1 = 1 + b_1
= 1 + \frac{n_1}{e^s-1} = \Delta_1 \,.
\end{equation}
For $k > 1$, we find that
\begin{equation}
\begin{split}
\widetilde{\Delta}_k &
= \left|
\begin{array}{cccc}
1 + b_1 & \ldots & b_{k-1} & 0 \\
\vdots & \ddots & \vdots & \vdots \\
n_1 + b_1 & \ldots & 1 + b_{k-1} & 0 \\
n_1 + b_1 & \ldots & n_{k-1} + b_{k-1} & 1 \\
\end{array}
\right| + b_k \left|
\begin{array}{cccc}
1 + b_1 & \ldots & b_{k-1} & 1 \\
\vdots & \ddots & \vdots & \vdots \\
n_1 + b_1 & \ldots & 1 + b_{k-1} & 1 \\
n_1 + b_1 & \ldots & n_{k-1} + b_{k-1} & 1 \\
\end{array}
\right| \,.
\end{split}
\end{equation}
The first term on the right-hand side is $\widetilde{\Delta}_{k-1}$.
The determinant in the second term can be calculated as follows:
\begin{equation}
\begin{split}
& \left|
\begin{array}{cccccc}
1 + b_1 & b_2 & b_3 & \ldots & b_{k-1} & 1 \\
n_1 + b_1 & 1 + b_2 & b_3 & \ldots & b_{k-1} & 1 \\
n_1 + b_1 & n_2 + b_2 & 1 + b_3  & \ldots & b_{k-1} & 1 \\
\vdots & \vdots & \vdots & \ddots & \vdots & \vdots \\
n_1 + b_1 & n_2 + b_2 & n_3 + b_3  & \ldots & 1 + b_{k-1} & 1 \\
n_1 + b_1 & n_2 + b_2 & n_3 + b_3  & \ldots & n_{k-1} + b_{k-1} & 1 \\
\end{array}
\right| \\
& = \left|
\begin{array}{cccccc}
1 - n_1 & -n_2 & -n_3 & \ldots & -n_{k-1} & 0 \\
0 & 1 - n_2 & -n_3 & \ldots & -n_{k-1} & 0 \\
0 & 0 & 1 - n_3  & \ldots & -n_{k-1} & 0 \\
\vdots & \vdots & \vdots & \ddots & \vdots & \vdots \\
0 & 0 & 0  & \ldots & 1 - n_{k-1} & 0 \\
n_1 + b_1 & n_2 + b_2 & n_3 + b_3  & \ldots & n_{k-1} + b_{k-1} & 1 \\
\end{array}
\right|
= \prod_{i=1}^{k-1} \, (1 - n_i ) \,.
\end{split}
\end{equation}
We therefore have
\begin{equation}
\widetilde{\Delta}_k
= \widetilde{\Delta}_{k-1}
+ b_k \, \prod_{i=1}^{k-1} \, (1 - n_i )
= \widetilde{\Delta}_{k-1}
+ \frac{n_k}{e^s-1} \, \prod_{i=1}^{k-1} \, ( 1 - n_i ) \,.
\end{equation}
On the other hand, it is easy to see that
\begin{equation}
\Delta_k - \Delta_{k-1}
= \frac{n_k}{e^s-1} \, \prod_{i=1}^{k-1} \, ( 1 - n_i )
\end{equation}
for $k > 1$.
We thus conclude that
\begin{equation}
\widetilde{\Delta}_k = \Delta_k \,.
\end{equation}

\subsection{Proof of~(\ref{toproveinappendix})}\label{appKO3}
In this appendix we prove the claim~(\ref{toproveinappendix}). We consider
 an arbitrary
point $\{t^{(1)}_1,\ldots,t^{(k)}_{n_k}\}$ in the integration region $\Gamma(\vec n)$ of a partition $\vec n=(n_1,\ldots,n_k)$
contributing to $\ket{\mB(\Psi_L)}$ at $\ord{\lambda^n}$.
The insertion points $\{t^{(1)}_1,\ldots,t^{(k)}_{n_k}\}$ in the $z$ frame
 are mapped
to the unit circle in the $\zeta$ frame as
\begin{equation}
    \{t^{(1)}_1,\ldots,t^{(k)}_{n_k}\}~\to~ \{e^{i\theta_1},\dots,e^{i\theta_{\tilde q-1}},e^{i\theta_{\tilde q}},e^{i\theta_{\tilde q+1}},\dots,e^{i\theta_n}\}\,
       \,.
\end{equation}
We will show that for any
\begin{equation}\label{thetaqregion}
    \theta_{\tilde q-1}\leq \tilde\theta\leq \theta_{\tilde q+1}\,,
\end{equation}
we can find a partition $\tilde{\vec n}$ such that a point in its integration region $\Gamma(\tilde{\vec n})$
 is mapped
to the positions
\begin{equation}
     \{e^{i\theta_1},\dots,e^{i\theta_{\tilde q-1}},e^{i\tilde\theta},e^{i\theta_{\tilde q+1}},\dots,e^{i\theta_n}\}
\end{equation}
in the $\zeta$ frame.
We will prove this claim by showing that we can determine the required partition $\tilde{\vec n}$ starting from the
original partition $\vec n$ as we continuously vary $\tilde \theta$ away from $\theta_{\tilde q}$.
 For the following analysis it is convenient to recall that the integration region of an arbitrary partition $\vec n$.
 The position of the first $V$ insertion associated with $\Psi^{(n_i)}$
in the natural $z$ frame
is given by
 \begin{equation}\label{firstregion}
    t_1^{(i)} = \sum_{j=1}^{i-1} n_j \, e^{s_j} + e^{s_i} + a_0\,,
 \end{equation}
 while the integration region for the remaining $n_i-1$ insertions, which we will denote as \emph{internal insertions} in the following, are given by
 \begin{equation}\label{internalregion}
    t_{j-1}^{(i)}\leq t_{j}^{(i)}\leq t_{1}^{(i)}+e^{s_j}(j-1)\quad\text{ for }j\geq2\,.
 \end{equation}
Let us denote the insertion position in the original partition $\vec n$ which
 is mapped
to the angle $\theta_{\tilde q}$ by
$t_{j}^{(i)}$. We need to distinguish several cases.
\subsubsection*{$V(t^{(i)}_j)$ is an internal insertion ($j\geq2$)}

Consider first a variation
which
decreases $\tilde\theta$,
{\em i.e.},
$\tilde\theta\leq\theta_{\tilde q}$.
 If $\theta_{\tilde q-1}=\theta_{\tilde q}$ then we are already at the lower end of the
 interval~(\ref{thetaqregion}) and we are done.
 If $\theta_{\tilde q-1}<\theta_{\tilde q}$ then
$t^{(i)}_j>t^{(i)}_{j-1}$, so
the internal insertion is not yet at the lower boundary of its integration
 region~(\ref{internalregion}).
 Varying $t^{(i)}_j$
does not affect any other positions, so we conclude that we can find a configuration within the exact same partition
 $\vec n$ for any
\begin{equation}
    \theta_{\tilde q-1}\leq\tilde\theta\leq\theta_{\tilde q}\,.
\end{equation}
Eventually,
 the internal insertion $V(t^{(i)}_j)$ collides with the previous insertion $V(t^{(i)}_{j-1})$,
which is precisely one of the boundaries in the
 interval~(\ref{thetaqregion}) that we expected to find.

Now consider increasing $\tilde \theta$ into the range $\tilde\theta>\theta_{\tilde q}$.
 When increasing $\tilde\theta$ by increasing $t^{(i)}_j$,
eventually two things can happen:
if
$j<n_i$ and $t^{(i)}_{j+1}\leq t^{(i)}_{1}+e^{s_i}(j-1)$,
 it follows from~(\ref{internalregion}) that $V(t^{(i)}_j)$
can collide with the next $V$ insertion at $t^{(i)}_{j+1}$
which corresponds to the upper boundary of the interval~(\ref{thetaqregion}) and we are done.
Otherwise, $t^{(i)}_j$ eventually hits its upper limit of integration at
$t^{(i)}_j=t^{(i)}_{1}+e^{s_i}(j-1)$. Within this partition
 $\vec n$,
the integration region stops although we have not encountered an operator collision. So this integration region must smoothly match
with
another integration region in a different partition
 $\tilde{\vec n}$.
This is indeed the case. Intuitively, this can be understood as a ``breaking'' of the solution insertion $\Psi^{(n_i)}$ into two pieces when $t^{(i)}_j$
becomes
too large. We now have two solution insertions, $\Psi^{(j-1)}$ and $\Psi^{(n_i-j+1)}$, with a new half-propagator strip opening up between them.
The relevant partition is thus obtained by replacing
\begin{equation}
    \cdots\PP(s_{i-1},s_i)\Psi^{(n_i)}\PP(s_i,s_{i+1})\cdots
    \,\to\,
    \cdots\PP(s_{i-1},s_i)\Psi^{(j-1)}\PP(s_i,\tilde s)\Psi^{(n_i-j+1)}\PP(\tilde s,s_{i+1})\cdots\,.
\end{equation}
 We thus have
 \begin{equation}
    \vec n=\{\ldots,n_{i-1},n_i,n_{i+1},\ldots\}\,\to\,
    \tilde{\vec n}=\{\ldots,n_{i-1},(j-1),(n_{i-1}-j+1),n_{i+1},\ldots\}\,.
 \end{equation}
It is easy to see that for $t^{(i)}_j=t^{(i)}_{1}+e^{s_i}(j-1)$ in the first partition, the operator insertions match
smoothly to $\tilde s=s_i$ in the second partition. The boundary of the integration region
 $\Gamma(\vec n)$
 at $t^{(i)}_j=t^{(i)}_{1}+e^{s_i}(j-1)$ thus matches smoothly
to the boundary
of the integration region
$\Gamma(\tilde{\vec n})$.
In the new partition
 $\tilde{\vec n}$,
the $V$ insertion at angle $\tilde \theta$ originates from the first $V$ insertion in the solution $\Psi^{(n_i-j+1)}$. To see that we can then still continue varying $\tilde \theta$, we turn to the second
case.

\subsubsection*{$V(t^{(i)}_j)$ is the first insertion in a solution ($ j=1$)}
The insertion position  $t^{(i)}_1$ is not independently integrated over in our parameterization using integrals over $s_i$.
In fact, to move $t^{(i)}_1$ while keeping all positions fixed, we generically need to vary \emph{all} the $s_m$.
  Recalling~(\ref{dtidsj}), we find
  \begin{equation}
   \frac{\partial s_m}{\partial t_1^{(i)}}>0
   \quad\text{ for all }\quad 1\leq m,i\leq k\,.
  \end{equation}
Furthermore, we note that
 the integral regions
  (\ref{internalregion})
  of \emph{internal} insertions depend on the $s_m$.
As we vary $\tilde \theta$ and thus $t^{(i)}_1$, the following things can happen.
\begin{itemize}
  \item  {\bf $V(t^{(i)}_1)$ can collide with $V(t^{(i)}_2)$}\\
        This can happen if $n_i\geq2$ and constitutes
          the upper boundary of the interval~(\ref{thetaqregion}).
  \item {\bf An internal insertions hits the upper boundary of its integration}\\
        As mentioned above,
this is caused by the change in integration regions for internal insertions as we vary $t^{(i)}_1$. We have encountered such a situation before in the previous subsection.
         Denote the position of the insertion which reaches its upper limit of integration by $t^{(m)}_j$ with $ j\geq 2$.
        Just as above we can match this configuration smoothly by breaking the affected solution into two pieces:
        \begin{equation}
            \PP(s_{m-1},s_m)\Psi^{(n_m)}\PP(s_m,s_{m+1})
            \,\to\,
            \PP(s_{m-1},s_m)\Psi^{(j-1)}\PP(s_m,\tilde s)\Psi^{(n_m-j+1)}\PP(\tilde s,s_{m+1})\,.
        \end{equation}
         This corresponds to the change of partition
         \begin{equation}
                \vec n=\{\ldots,n_{m-1},n_{m},n_{m+1},\ldots\}\,\to\,
            \tilde{\vec n}=\{\ldots,n_{m-1},(j-1),(n_{m}-j+1),n_{m+1},\ldots\}\,.
         \end{equation}
        For $t^{(m)}_j=t^{(m)}_{1}+e^{s_m}(j-1)$ in the
        partition
         $\vec n$,
        the operator insertions match
        smoothly to $\tilde s=s_m$ in the
         partition $\tilde{\vec n}$.
  \item {\bf $s_m\to s_{m+1}$ for some $1\leq m<k$}\\
        This is in a sense the opposite case to the ones we have encountered so far. Instead of breaking solutions, in
        this case solutions merge. We match smoothly to a new
         configuration
        \begin{equation}\label{mergeit}
            \ldots\PP(s_{m-1},s_{m})
              \Psi^{(n_m)}
            \PP(s_{m},s_{m+1})\Psi^{(n_{m+1})}\ldots
            \,\to\,
            \ldots\PP(s_{m-1},s_{m+1})\Psi^{(n_{m}+n_{m+1})}\ldots\,.
        \end{equation}
          This corresponds to the change of partition
          \begin{equation}
                \vec n=\{\ldots,n_{m-1},n_{m},n_{m+1},n_{m+2},\ldots\}\,\to\,
             \tilde{\vec n}=\{\ldots,n_{m-1},(n_{m}+n_{m+1}),n_{m+2},\ldots\}\,.
          \end{equation}
        Thus there is no longer an integral over  $s_{m}$ present -- instead the solution insertion $\Psi^{(n_{m}+n_{m+1})}$ carries one more internal integral than the previous solution insertions $\Psi^{(m)}$ and $\Psi^{(n_{m+1})}$ combined.
        Note that if $m+1=i$, then the insertion
         $V(t_1^{(i)})$
        which we are varying
         corresponds to an internal insertion in the new partition $\tilde{\vec n}$
        and we have to keep track of the variation of its position as we did in the previous subsection. If not, we continue the analysis of this subsection with the new partition
         $\tilde{\vec n}$.
           \item {\bf $s_k\to s$}\\
        As we increase $\tilde\theta$ and thus increase $t^{(i)}_1$, $s_k$ also increases because
        \begin{equation}
            \frac{\del s_k}{\del t^{(i)}_1}>0\,.
        \end{equation}
        Thus eventually we can hit its upper limit of integration $s_k=s$, and  the last half-propagator $\PP(s_k,s)$ in the partition~(\ref{mBorderk}) collapses. As we generically do not have an operator collision in this configuration, we need to match it smoothly to a different partition. But in fact, using the cyclic property of $\oints s$, we can rewrite this configuration as
        \begin{equation}
        \begin{split}
            &[\dots]\oints{s}\PP(0,s_1)\dots\PP(s_{k-1},s_k)\Psi^{(n_k)}\PP(s_k,s)\biggr|_{s_k=s}\\
            \,\to\,~&
            [\dots]\oints{s}\PP(0,s_0)\Psi^{(n_k)}\PP(s_0,s_1)\dots\PP(s_{k-1},s)\biggr|_{s_0=0}\,.
        \end{split}
        \end{equation}
        After cyclic index relabeling $i\rightarrow i+1$, we again obtain a partition of the form~(\ref{mBorderk})
        with $s_1=0$.
                   This corresponds to the change of partition
          \begin{equation}
                \vec n=\{n_1,\ldots,n_{k-1},n_k\}\,\to\,
             \tilde{\vec n}=\{n_k,n_1,\ldots,n_{k-1}\}\,.
          \end{equation}
        We can now continue to increase $\tilde\theta$, which is now represented
        by the
        position $t^{(i+1)}_1$
         in $\tilde{\vec n}$.
        As we increase $\tilde \theta$, the new
        Schwinger parameter $s_1$ leaves its lower boundary $s_1=0$ and enters its allowed region $s_1>0$, because
        \begin{equation}
            \frac{\del s_1}{\del t^{(i+1)}_1}>0\,.
        \end{equation}
  \item {\bf $s_1\to 0$}\\
        We can hit $s_1=0$ when decreasing $\tilde \theta$. This is reversed situation of the collision $s_k\to s$ and can be dealt with in the exact same way.
\end{itemize}
We conclude that we can continue to vary $\tilde \theta$
 throughout the interval~(\ref{thetaqregion})
 while keeping all other insertion angles fixed.
This completes the proof of~(\ref{toproveinappendix}).

\end{document}